%% file: main.tex
\documentclass[english,american,superscriptaddress,prd,a4paper,nofootinbib,11pt,eqsecnum,secnumarabic,aps, preprintnumbers]{revtex4-2}
\usepackage{tabularx}
\usepackage{amsmath}  
\usepackage{amsfonts,amssymb}
\usepackage{enumerate}
\usepackage{xcolor}
\usepackage{subfigure}
\usepackage{graphicx} 
\usepackage{xspace}
\usepackage[final]{hyperref} 
\hypersetup{
	colorlinks=true,       
	linkcolor=blue,        
	citecolor=blue,       
	filecolor=magenta,     
	urlcolor=blue         
}
\usepackage[normalem]{ulem}
\usepackage{braket}

\usepackage[top=1.2in, bottom=1in, left=0.8in, right=0.8in]{geometry}

\begin{document}
\preprint{MSUHEP-24-002}

\title{ A generalized statistical model for fits to parton distributions}

\author{Mengshi Yan}
\affiliation{Department of Physics and State Key Laboratory of Nuclear Physics and Technology, Peking University, Beijing 100871, China.}
\email{msyan@pku.edu.cn}

\author{Tie-Jiun Hou}
\affiliation{School of Nuclear Science and Technology, University of South China, Hengyang 421001, China.}
\email{tjhou@msu.edu}

\author{Zhao Li}
\affiliation{Institute of High Energy Physics, Chinese Academy of Sciences, Beijing 100049, China; \\
	School of Physics Sciences, University of Chinese Academy of Sciences, Beijing 100039, China}
 \affiliation{Center for High Energy Physics, Peking University, Beijing 100871, China.}
 \email{zhaoli@ihep.ac.cn}

 \author{Kirtimaan Mohan}
 \email{kamohan@msu.edu}
 \author{C.--P. Yuan}
 \affiliation{Department of Physics and Astronomy,
	Michigan State University, East Lansing, MI 48824, U.S.A.}
 \email{yuanch@msu.edu}

\begin{abstract}
Parton distribution functions (PDFs) form an essential part of particle physics calculations. Currently, the most precise predictions for these non-perturbative functions are generated through fits to global data. A problem that several PDF fitting groups encounter is the presence of tension in data sets that appear to pull the fits in different directions. In other words, the best fit depends on the choice of data set. Several methods to capture the uncertainty in PDFs in presence of seemingly inconsistent fits have been proposed and are currently in use. These methods are important to ensure that uncertainty in PDFs are not underestimated. 
Here we propose a novel method for estimating the uncertainty  by introducing a generalized statistical model based on Bayesian Hierarchical models which is implemented via the Gaussian Mixture Model (GMM). The methodology is inspired by unsupervised machine learning techniques  
and is closely related to the statistical methods of Ensemble Learning and Bayesian Model Averaging. 
Using a toy model of PDFs, we demonstrate how the GMM can be used to faithfully reconstruct the likelihood associated with PDF fits, which can in turn be used to accurately determine the uncertainty on PDFs, especially in the presence of tension in the fitted data sets.
We further show how this statistical model reduces to the usual chi-squared likelihood function for a consistent data set and provide measures to optimize the number of Gaussians in the GMM.
\end{abstract}

\date{\today}

\maketitle
\flushbottom
\newpage
\tableofcontents
\newpage
\section{Introduction}
\label{sec:introduction}

Parton distribution functions (PDFs) are non-perturbative functions quantifying the probability of finding a quark or gluon with certain momentum fraction $x$ from a nucleon.
Taking advantage of a wide variety of experimental data and advanced computing packages for perturbative QCD (and electroweak) calculations, various groups~\cite{Hou:2019efy, Bailey:2020ooq, NNPDF:2021njg, Alekhin:2017kpj, ATLAS:2021vod} provide a sophisticated determination of PDFs. The recently released NNPDF~\cite{NNPDF:2021uiq} and \texttt{xFitter}~\cite{Alekhin:2014irh} fitting codes open opportunities for general users to also conduct global QCD analyses.

These PDFs form a critical component of the theory community's infrastructure~\cite{Amoroso:2022eow} and developing the next generation of PDFs are crucial for achieving the physics goals of the Large Hadron Collider (LHC) and the Electron-Ion Collider (EIC). 
An unbiased determination of PDFs are essential for precision phenomenology and the search for new physics, such as the recent measurement of $W$ boson mass by the CDF collaboration~\cite{CDF:2022hxs}. 
A robust understanding of the the uncertainties arising from PDFs is necessary to provide precise predictions. PDF uncertainties constitute one of the dominant 
systematic errors both for direct searches of new physics~\cite{Cepeda:2019klc} and for precision tests of the Standard Model (SM), such as determining the properties of the Higgs boson~\cite{LHCHiggsCrossSectionWorkingGroup:2016ypw}. 
A faithful determination of PDFs and its uncertainties are thus necessary in order to make precise theory predictions at hadron-hadron and lepton-hadron colliders. 
Improving PDFs will require new experimental data, new theoretical (higher-order) calculations and a determination of fits to global data in a statistically robust way~\cite{Amoroso:2022eow}.

Many issues could be encountered when trying to perform global fits to data sets from a variety of experiments. One of these issues, which is the primary focus of this paper, is in dealing with data that appear to be in tension with each other, leading to inconsistencies within the global fit.
For example, suppose we have two data sets that pull the fit parameters to different values. In this scenario, the data are in tension with each other. The source of such tension could arise due to deficiencies in theoretical models and calculations or even from incorrect estimation of experimental uncertainties. To keep the discussion simple, we work with an example of the latter, although, the method presented here is agnostic to the source of tension. 

An example of tension between data sets would be the measurement of the ATLAS 8 TeV $W$ and $Z$ rapidity distribution~\cite{ATLAS:2016nqi}, which was found to be inconsistent with other data included in the PDF global analysis and favored an enhanced strange quark distribution function~\cite{Hou:2019efy, Bailey:2020ooq, Hou:2022sdf}. Although a non-vanishing strangeness asymmetry $(s-\bar{s})(x)$ can partially release this tension~\cite{Bailey:2020ooq, Hou:2022sdf}, it is still inadequate to accommodate the ATLAS 8 TeV $W/Z$ data~\cite{ATLAS:2016nqi} consistently in a PDF global analysis.

Another example is the inconsistency between the E866 NuSea and E906 SeaQuest measurements of the cross-section ratio $\sigma(pd)/2\sigma(pp)$, which approximates the ratio of anti-quark PDFs $\sigma(pd)/2\sigma(pp) \approx \big{(} 1 + \bar{d}_p(x_2) / \bar{u}_p(x_2) \big{)}/2$. The E866 NuSea and E906 SeaQuest measurements  provide valuable information on the PDF ratio $(\bar{d}/\bar{u})(x)$ in the large-$x$ region. However, they are found to be in tension at large-$x$~\cite{Guzzi:2021fre,Hou:2022ajg,Hou:2022sdf}.
The effect of tension between certain data sets generally becomes negligible for large data sets, when outliers appear with small probability. However, in this case, E866 NuSea and E906 SeaQuest, the measurements probe a specific kinematic region that is otherwise weakly constrained and hence the tension between them cannot be ignored.
It is thus necessary to develop a prescription to accommodate such inconsistencies in a PDF global analysis in order to properly estimate the PDF uncertainty.

Several methods have been proposed and are currently in use to provide an estimate of uncertainties especially in the presence of tension\cite{Kovarik:2019xvh}.
One naive approach would be to simply omit one of the troublesome data sets or perhaps attach a significantly lower weight to this data set. This approach introduces biases in the fitting procedure that needs to be justified with some prior knowledge. In some cases, such a justification may be possible. For example, if perhaps the methodology of the omitted experiment is analyzed and found to be unsatisfactory. In such a scenario, the prior bias in  omitting the data set from this experiment is justified, but this is not always the case, since it is not always straightforward to reanalyze particle physics experiments and their data.

Modern PDFs perform fits to data by defining a likelihood or loss function. For PDFs produced by CTEQ-TEA (CT)~\cite{Hou:2019efy} and MSHT~\cite{Bailey:2020ooq}, the likelihood is the usual $\Delta \chi^2$. Under normal circumstances with a consistent set, the uncertainty can be defined by the $68\%$ confidence level limit of the $\chi^2$ distribution, which amounts to varying the fit and determining boundaries where $\Delta \chi^2 = 1$. 

The convention used by the PDF global analysis groups, such as CT and MSHT, 
is to introduce a tolerance $T$ such that the fits should be varied within $\Delta\chi^2 = T^2$. Values of $T$ are chosen either dynamically~\cite{Bailey:2020ooq} or set to a fixed value~\cite{Hou:2019efy}. Typical values of $T$ are much larger than $1$, which increases the uncertainties in the PDFs. 
For example, $T^2=37$ corresponds to the CT tolerance at the 68\% CL. 

NNPDF uses the Monte-Carlo method for determining uncertainties, where they produce representative sample of PDFs by sampling their likelihood or loss function with the help of pseudo-data,which might be one of the reasons for the uncertainties from NNPDF's analysis being notably different~\cite{Ball:2022qks}. Although, as pointed out in Refs.~\cite{Stegeman:2022wrn,lambri2020thesis,thesis:talon},
the level of inconsistency with data sets is found to be negligible in the NNPDF3.1 analyses, further study is needed to understand differences between NNPDF and the other fitting groups. One such task was performed in Ref.~\cite{Cridge:2021qjj}, which details benchmark comparisons of three global PDF sets - CT18, MSHT20 and NNPDF3.1
- and their similarities and differences that have been observed. The end result of that study
was a new PDF4LHC21 ensemble of combined PDFs suitable for a wide range of LHC applications.

In this paper we describe how to apply a more general statistical framework, namely that of Bayesian Hierarchical Models~\cite{gelman2013bayesian,mcglothlin2018bayesian,Erler:2020bif}, to accommodate potential inconsistencies in PDF data. 
This rather non-trivial task is accomplished with the aid of 
the Gaussian Mixture Model (GMM), which is commonly used as an unsupervised machine learning technique as well as a method of combining models.
The approach used here can be interpreted as an implementation of the statistical method of Ensemble Learning~\cite{Murphy:ML, bishop2007} and the related method of Bayesian Model Averaging (BMA)~\cite{leamer1978specification,george1993variable,kass1995bayes,hoeting1999,raftery2003discussion}.  BMA is a statistical method used to account for uncertainties arising from model selection. Although it is most often used in the context of linear regression, here
we apply it to highly non-linear fits.\footnote{Application of Bayesian Model Averaging to analyze Lattice Field Theory results can be found in Refs.~\cite{Jay:2020jkz,Neil:2022joj,Neil:2023pgt}.} 
On the other hand, Ensemble Learning, also known as  Ensemble Averaging or Bayesian Model Combination, is a method to combine predictions of multiple statistical models.
Although there are some subtle differences between Ensemble Learning and Bayesian Model Averaging, which we discuss briefly in Appendix~\ref{app:sec:Bayesian view of the GMM likelihood}, we do not dwell on these differences as they are subject to differences in the interpretation of priors.
Ensemble Learning is primarily used to combine different theoretical models and their point estimates. On the other hand, Bayesian Hierarchical Models are primarily applied to cases where there are variations across groups of data and its use has also been suggested in estimating uncertainty when there is tension in data sets~\cite{Erler:2020bif}. Here, we demonstrate how Ensemble Learning can be used to construct a Bayesian Hierarchical Model and how it can be used to provide an estimate of the uncertainty in the scenario where there are tensions among data sets.
We will not discuss combining or averaging over different theoretical models which is beyond the scope of this paper.

Our major findings are summarized below for the benefit of the reader.

\begin{itemize}
\item When fitting inconsistent data sets with tension, the usual Least Squares (LS) method, that employs the $\chi^2$ likelihood, underestimates uncertainty. Specifically, the posterior probability distribution for fitted theory parameters should reflect the tension in the data sets and appropriately span the range of possible theory values. We demonstrate how the LS method and its adaptations to address  tension in data sets do not appropriately span the range of possible theory values. Another drawback of these methods is that they introduce ad hoc criteria to expand uncertainties. These deficiencies are overcome by the use of the GMM.
 \item We introduce a new method to fitting inconsistent data, utilizing the Gaussian Mixture Model. The method is closely related to  Ensemble Learning and Bayesian Model Averaging, that are used to estimate uncertainties due to selection of theoretical models. Here, we show how these methods can be repurposed to help determine uncertainties by constructing a Bayesian Hierarchical Model when there is tension between data sets. 
 \item The main idea is a modification of the likelihood function used to determine the posterior probability distribution of the theory parameters. This modification, given in Eq.~\eqref{eq:GMM},  accounts for the possibility of fitting with a multi-modal distribution rather than a uni-modal distribution assumed in the LS method. Estimation of theory parameters and their associated uncertainties  proceed through the usual Maximum Likelihood estimation. We describe methods and provide formulae for the estimate of the mean and variance  in Eq.~\eqref{eq:GMM_mean} and Eq.~\eqref{eq:cov_GMM_2}, respectively. We demonstrate the use and characteristics of the GMM method using various toy examples where we try to combine various pseudo-data sets with varying degrees of tension. 

 \item The fit presented in Fig.~\ref{fig:GMM_case1:d} and Fig.~\ref{fig:GMM_case1:e} is an example of how the GMM method provides a more accurate representation of the uncertainty compared to the LS method by spanning the range of possible theory predictions, given the tension in the data sets. Specifically, using the GMM, we find that the uncertainty bands are larger when the tension is more prominent and smaller otherwise. On the other hand, more traditional methods of addressing tension tend to expand uncertainties more uniformly, resulting in larger uncertainty bands in regions where the tension is not large.

 \item The GMM introduces the number of modes ($K$) of the distribution as a hyperparameter. We demonstrate the use of  Akaike Information Criterion (AIC) and Bayesian Information Criterion (BIC) not only to determine $K$ in order to avoid  over-fitting or under-fitting, but also to be used  as quantitative tests of the consistency of data sets. 
\end{itemize}

The rest of the paper is organized as follows.
In Sec.~\ref{sec:tension}, we set up a toy PDF model and describe different pseudo-data sets, some of which have obvious tension between them.  In Appendix~\ref{app:sec:pseudo_data} we describe how the pseudo-data sets we use in the fits are generated. Section~\ref{sec:GMM} introduces the Gaussian Mixture Model as a statistical model that can be used to tackle  inconsistencies in data sets. In Sec.~\ref{sec:result_GMM}, the results of the GMM fit are compared with those of conventional fits using the usual least squares fit. Consistency between the GMM and the commonly-used Least Squares (LS) method is evaluated using a data set with no tension in Sec.~\ref{sec:consistency}.
We discuss our results and conclude in Sec.~\ref{sec:conclusion}. 
Additionally, we describe a more intuitive Bayesian perspective of the GMM in Appendix~\ref{sec:appendix:Bayesian} and discuss some of the differences and intricacies related to BMA and Ensemble Learning in Appendix~\ref{sec:app:BMA_and_BMC}. In Appendix~\ref{sec:app:W_boson},  we provide a much simpler and more accessible example using the real world data of $W$-boson mass measurements.

\section{Tension in a toy model}
\label{sec:tension}

In this section, we introduce a toy model of PDFs and the pseudo-data sets to fit the PDFs. We perform LS fits to illustrate the problems encountered when there is tension between data sets.


\begin{figure}[htbp]  
\centering
\subfigure[]
{{\includegraphics[width=0.4\columnwidth]{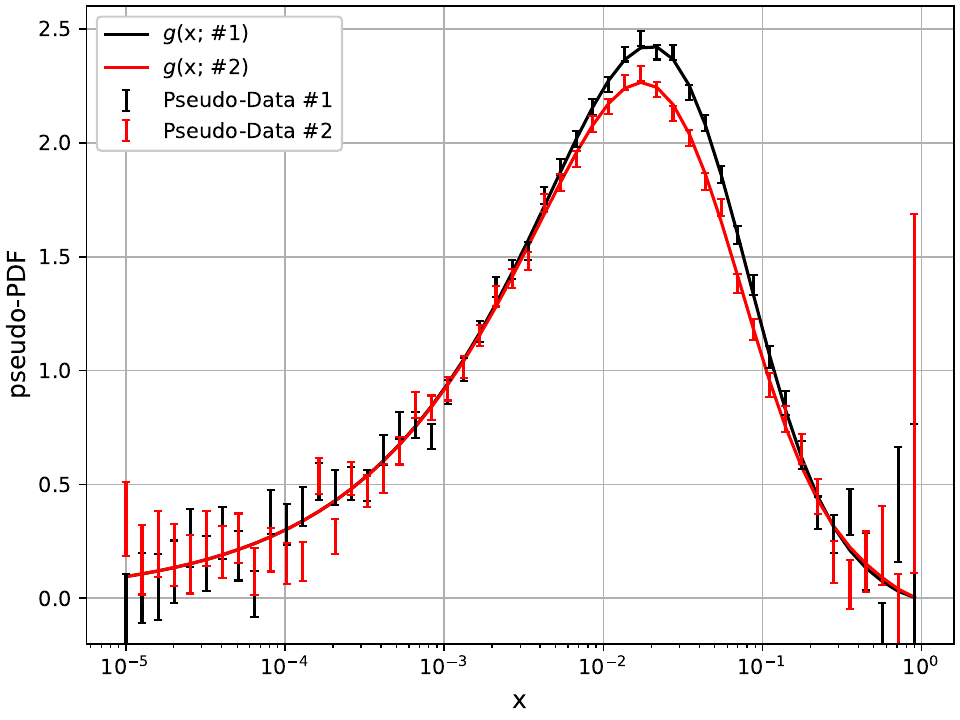}}
\label{fig:pData:a}}
\subfigure[]
{{\includegraphics[width=0.4\columnwidth]{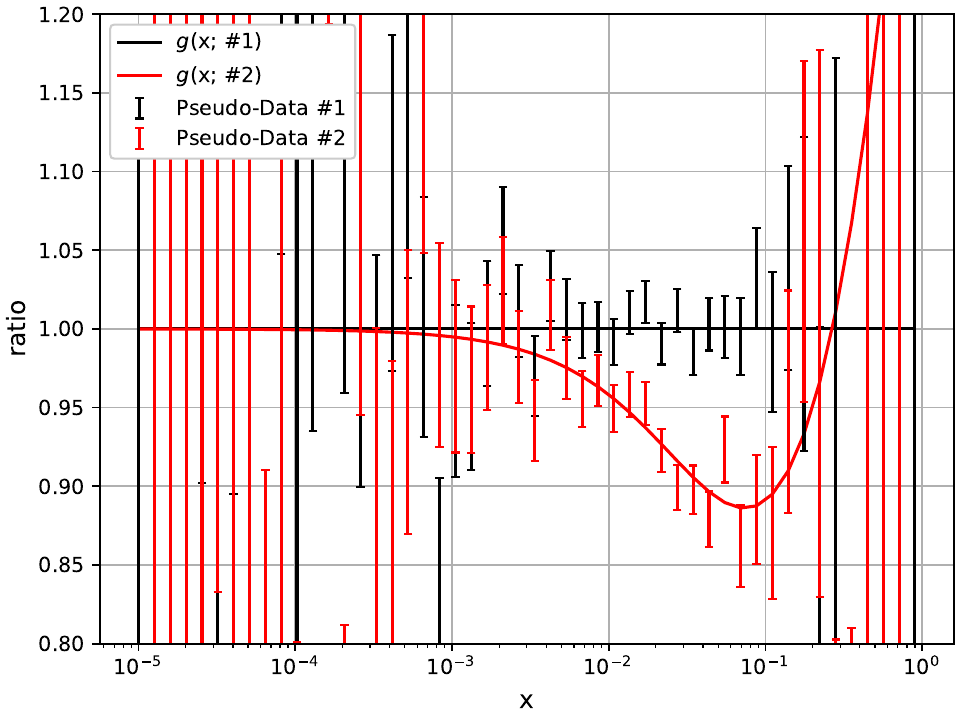}}
\label{fig:pData:b}}
\caption{\label{fig:pData} 
Two pseudo-data sets,  Pseudo-Data \#1 (black) and Pseudo-Data \#2 (red) are shown to be in tension with each other. 
The solid curves which are referred to as $g(x; \# 1)$ and $g(x; \# 2)$, correspond to the curves given by the function in Eq.~\eqref{eq:func}, with parameter values given in Table~\ref{tab:pData_truth_2}. The plot on the left shows the pseudo-PDFs whereas the plot on the right shows their ratio with respect to $g(x; \# 1)$.
}
\end{figure}


    We begin by considering two sets of data that we refer to as pseudo-data \#1 and pseudo-data \#2, containing fifty data points each. These two data sets are not real, but are simulated. However, we treat them as though they are measurements that likely came from two separate experiments. Details of how these pseudo-data sets simulated are relegated to Appendix~\ref{app:sec:pseudo_data}. 
    The two pseudo-data sets are presented in Fig.~\ref{fig:pData}, showing measured values of a pseudo-PDF as a function of $x$. Both data sets have a larger uncertainty at low and high values of $x$, but smaller uncertainties at intermediate values of $x\sim 10^{-2}$, to mimic general features of real experimental data. This choice represents the possibility that larger values of the PDF may correspond to larger number of events and therefore a reduced statistical uncertainty. The data sets have an obvious tension between them in the region $10^{-2}\lesssim x \lesssim 10^{-1}$.
    Such a tension might be due to underestimated systematics or even due to underestimated theoretical uncertainties. 

    Given the two pseudo-data sets, we would like to perform a combined fit to both data sets. We fit the data to a  function of the form
\begin{equation}
g(x) = a_0 \ x^{a_1} (1-x)^{a_2} e^{x a_3} ( 1 + x e^{a_4} )^{a_5}.
\label{eq:func}
\end{equation}
Here $\left\{a_0,a_1,a_2,a_3,a_4,a_5\right\}$ are the theory parameters that we would like to determine from the pseudo-data. In addition, we would also like to estimate the uncertainty on these fitted parameters from the given data. To keep the discussion simple we only perform fits to a subset of these parameters, while keeping others fixed. We would like to emphasize that although we choose an explicit functional form, the starting point could have also  been a deep neural network or a polynomial parametrization.\footnote{The use of a deep neural network is supported by the Universal Approximation Theorem~\cite{Cybenko:1989} whereas a polynomial is supported by the Stone-Weierstrass Theorem~\cite{Stone:1937,Stone:1948}. These theorems ensure that any smooth function can be well approximated either by neural networks or by polynomials.} Here, we restrict ourselves to use only the functional form in Eq.~\eqref{eq:func} for clarity of presentation.   

Our purpose is to determine point estimates, i.e. the mean and variance, of the theoretical parameters $a$, given the experimentally measured data. Note, in the process of generating the two pseudo-data sets, we do have knowledge the values of these parameters, c.f. Appendix~\ref{app:sec:pseudo_data}. However, when performing the fits we do so by excluding any prior knowledge of the theory parameters $a$. Our purpose is not to try to recover parameter values that were used to generate each pseudo-data set separately, instead, we would like to determine an estimate of the theory parameters $a$ given a combination of both data sets. Furthermore,
to be consistent with QCD, we demand that we have only one set of theoretical parameters, so that there is only one estimate of the mean.  We now proceed to first perform LS fits to the two pseudo-data sets in Fig.~\ref{fig:pData}.

\subsection{Least Squares Fits and Tension}
\label{subsec:chi2_fit_case_1}

Before introducing the GMM method we first illustrate the method of handling inconsistent data in the Least Squares fitting framework. Fits are performed by optimizing the log likelihood, which in our case is the  $\chi^2$ function defined as
\begin{equation}
 \chi^2(\theta) = \sum_{j=1}^{N_{\text{pt}}} \bigg{(} \frac{y_i - T(\theta)}{\Delta y_i} \bigg{)}^2.
 \label{eq:LS_method}
\end{equation}
Given the known $N_{\text{pt}}$ measurements $y_i$ associated with uncertainties $\Delta y_i$, the $\chi^2$ provides a normalized distance from theoretical predictions $T(\theta)$ to the measurements. Here $\theta$ are the fit parameters, which are some subset of the parameters $a_i$, and are varied in order to optimize the $\chi^2$ function.

To illustrate the problem with the LS fit when tension exists, three LS fits are performed, which we refer to as SG-A (performed only with pseudo-data \#1), SG-B (performed only with pseudo-data \#2) and SG-C (performed by combining both  pseudo-data \#1  and pseudo-data \#2).  
We label these fits as ``SG" to refer to the single-Gaussian nature of the likelihood in the LS method, in contrast to the multi-modal distribution that we will introduce later.
Results of these fits are shown in Table~\ref{tab:chi2_fit_def_results}.

We fit the data to the function given  in Eq.~\eqref{eq:func}, by varying and determining the parameters of the function that minimize the $\chi^2$ in Eq.~\eqref{eq:LS_method}. To keep the discussion and fits simple in this toy example, we fix parameters $\{ a_0, a_1, a_2, a_3 \}$ to values which can be found in Table~\ref{tab:pData_truth_2}. Only two parameters $a_4$ and $a_5$ are varied and the values that minimize the $\chi^2$ function are shown in Table~\ref{tab:chi2_fit_def_results}. 
The values of $\chi^2_{\# 1}/N_{\text{pt}}$, $\chi^2_{\# 2}/N_{\text{pt}}$, and $\chi^2_{\# 1 + \# 2}/N_{\text{pt}}$, which are approximately equal to the reduced $\chi^2$, indicate the goodness of fit to individual pseudo-data sets \#~1, \#~2 and the combined pseudo-data set, respectively. As is well known, a good fit will have values of a $\chi^2/N_{\text{pt}} \sim 1$, whereas a poor fit  has $\chi^2/N_{\text{pt}} $ values significantly larger than 1.

The best-fit of SG-A, where only pseudo-data set \# 1 is included, is found with a  minimum of $\chi^2_{\# 1}/N_{\text{pt}} = 0.88$, with values of $a_4=2.32$ and $a_5=-3.22$. Also shown in the same row of Table~\ref{tab:chi2_fit_def_results} for SG-A are values of  $\chi^2_{\# 2}/N_{\text{pt}}$ and $\chi^2_{\# 1 + \# 2}/N_{\text{pt}}$. These indicate the values of each of the likelihood functions $\chi^2_{\# 2}/N_{\text{pt}}$ and $\chi^2_{\# 1 + \# 2}/N_{\text{pt}}$  determined at the best fit values of $a_4$ and $a_5$ for fits SG-A. We see that for SG-A the values of  $\chi^2_{\# 2}/N_{\text{pt}} = 6.55$ and $\chi^2_{\# 1 + \# 2}/N_{\text{pt}} = 3.72$ are significantly larger than one, indicating that there is significant tension between both the combined data set and data set \#2. 

Likewise, the best-fit of SG-B, where only pseudo-data set \# 2 is included, is found with a minimum $\chi^2_{\# 2}/N_{\text{pt}} = 1.02$, indicating a reasonably good fit. For SG-B, the values $\chi^2_{\# 1}/N_{\text{pt}}$ and $\chi^2_{\# 1 + \# 2}/N_{\text{pt}}$ are determined by using the best fit value of $a_4 = 2.63$ and $a_5 =2.72$. Once again large values of $\chi^2_{\# 1}/N_{\text{pt}} = 7$ and $\chi^2_{\# 1 + \# 2}/N_{\text{pt}} = 4.01$ indicate tension with both the combined data set and data set \#1.\footnote{    
The two pseudo-data sets, \#1 and  \#2, are generated by starting with different choices of $a_4$ and $a_5$ as given in Table~\ref{tab:pData_truth_2}. As expected, these values of $a_4$ and $a_5$ are similar to the fitted values in Table~\ref{tab:chi2_fit_def_results} for SG-A and SG-B and lie within the 1-$\sigma$ confidence interval of the fit.
   
}

The fit, SG-C, to the the combined data set where both pseudo-data sets \#~1 and \#~2 are included, is found with a minimum $\chi^2_{\# 1 + \# 2}/N_{\text{pt}} =2.42$, indicating, as expected, a poor fit to the combined data set. In this case, the values of  $\chi^2_{\# 1}/N_{\text{pt}}=2.27$, $\chi^2_{\# 2}/N_{\text{pt}} = 2.42$ indicate that the combined fit, still has tension with both individual data sets.

\begin{table}[htpb]
\begin{center}
\begin{tabular}{ccc|cccccc} \hline
  fits & pseudo-data & $N_{\text{pt}}$ & best-fit $a_4$ & best-fit $a_5$ & $\chi^2_{\# 1}/N_{\text{pt}}$ & $\chi^2_{\# 2}/N_{\text{pt}}$ & $\chi^2_{\# 1 + \# 2}/N_{\text{pt}}$ \\ \hline \hline
 SG-A & \# 1 & 50 & 2.32 & -3.22 & 0.88 & 6.55 & 3.72 \\
 SG-B & \# 2 & 50 & 2.63 & -2.73 & 7.00 & 1.02 & 4.01 \\
 SG-C & \# 1 and \# 2 & 100 & 2.48 & -2.94 & 2.27 & 2.56 & 2.42 \\ \hline
\end{tabular}
\end{center}
\caption{
The table shows the best-fit values of $a_4$ and $a_5$, as well as a measure of the goodness of fit $\chi^2/N_{\text{pt}}$ for three Least Square fits SG-A, SG-B, and SG-C, performed using pseudo-data sets \#~1 and \#~2. The columns $\chi^2_{\#1}/N_{\text{pt}}$, $\chi^2_{\#2}/N_{\text{pt}}$ and $\chi^2_{\#1 + \#2}/N_{\text{pt}}$ indicate the values of these log-likelihood functions evaluated at values of $a_4$ and $a_5$ in the given rows. Also shown, for comparison, in the last two rows are the true values of $a_4$ and $a_5$ that were used to generate the two psuedo-data sets. }
\label{tab:chi2_fit_def_results}
\end{table}

\begin{figure}[htbp]  
\centering
\subfigure[]
{{\includegraphics[width=0.4\columnwidth]{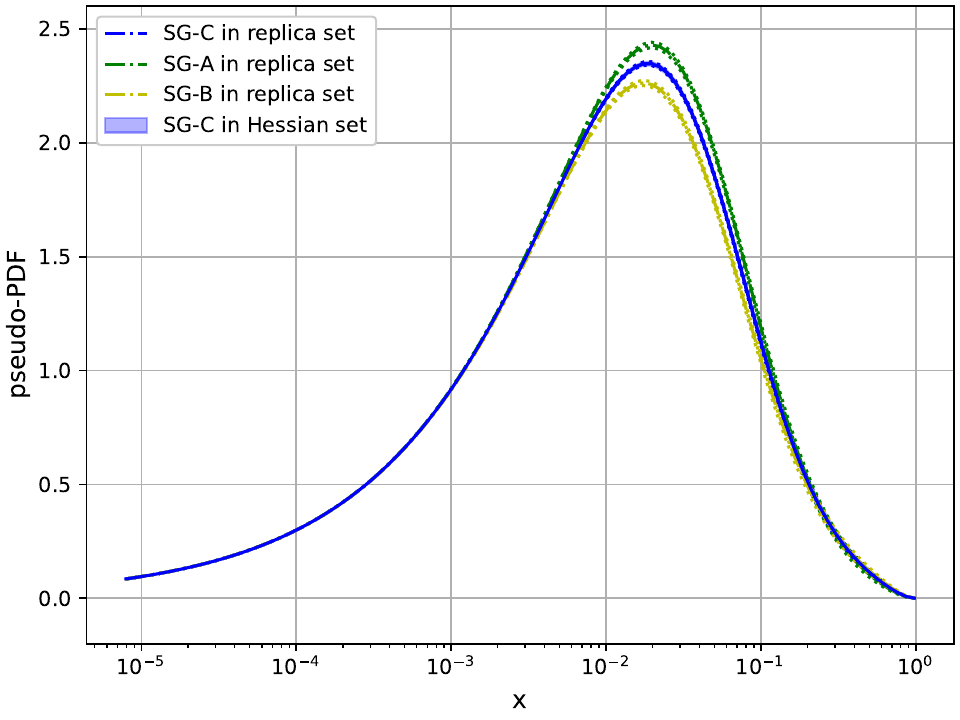}}
\label{fig:problem_of_usual_chi2:a}}
\subfigure[]
{{\includegraphics[width=0.4\columnwidth]{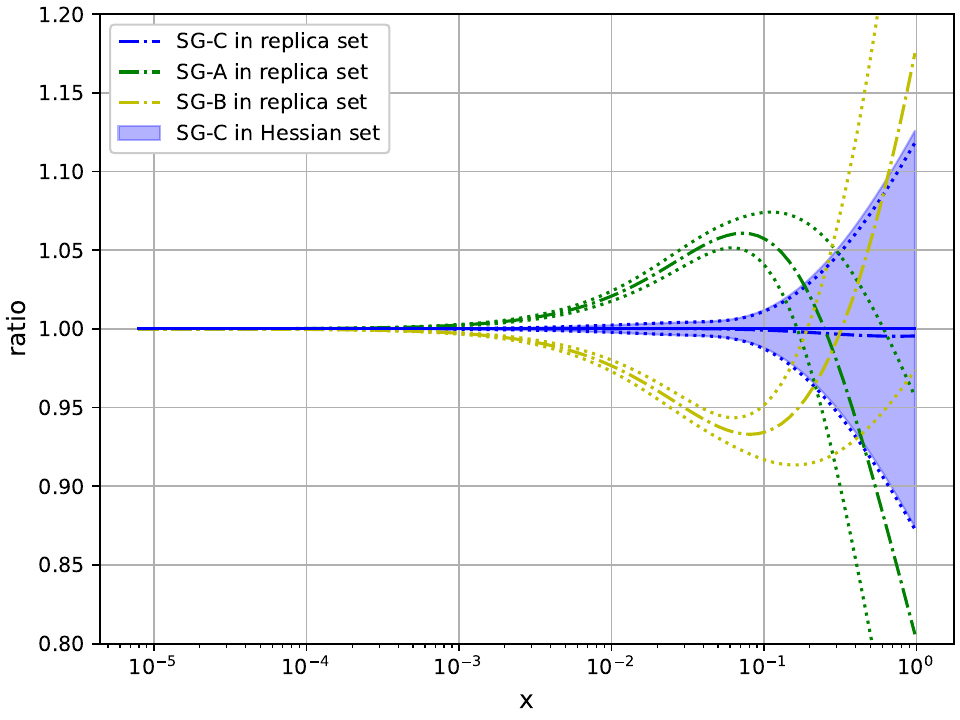}}
\label{fig:problem_of_usual_chi2:b}}
\subfigure[]
{{\includegraphics[width=0.4\columnwidth]{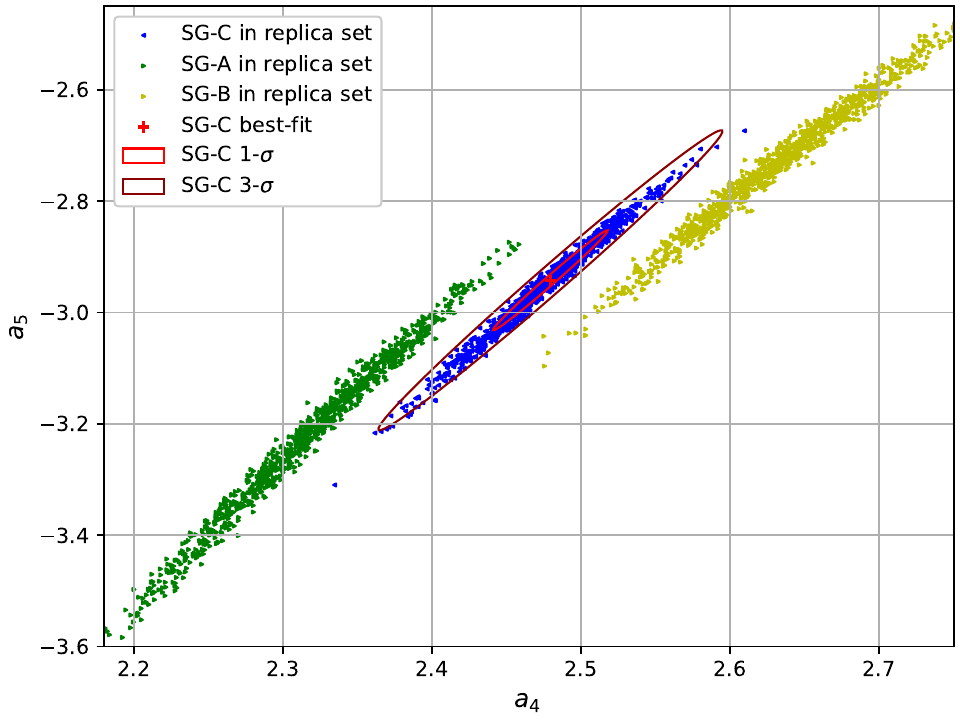}}
\label{fig:problem_of_usual_chi2:c}}
\caption{\label{fig:problem_of_usual_chi2} 
Panel (a): the pseudo-PDFs determined from each of the fits are shown. The dot-dashed lines refer to the central PDFs for SG-A (green), SG-B (yellow) and the combination SG-C (blue). Uncertainties at 1 standard deviation in the replica sets are shown by the dotted lines for each fit. For the LS fit C, the blue shaded region represents the Hessian uncertainty with $\Delta \chi^2 = 1$, which deviates from the Monte Carlo uncertainty by a negligible amount.
The sizes of uncertainties are small with respect to  the magnitude of the pseudo-PDFs.
Panel (b): similar to panel (a), but for ratios to the central value of the Hessian set of the SG-C fit. 
Panel (c): the values of parameters $a_4$ and $a_5$ in replica sets for SG-A, SG-B and SG-C are shown along with the 1 and 3 standard deviation uncertainties for SG-C given by the solid curves.
} 
\end{figure}

Although the fit to the combined data is poor, the main problem arises when attempting to determine the relevant uncertainty associated with such a fit. This is illustrated in  Fig.~\ref{fig:problem_of_usual_chi2}, which presents
results of the LS fits SG-A, SG-B and SG-C. For all  three fits, results are presented in the form of replica sets, where the probability distribution of the pseudo-PDF is sampled with the Monte Carlo method as described in~\cite{NNPDF:2021uiq}. 
For the purpose of comparison, we also estimate the uncertainty of the SG-C fit using the Hessian method~\cite{Hou:2019efy}, which is consistent with the Monte Carlo method. Here we only consider the Hessian uncertainty generated by the variation of $\chi^2$ and determine the uncertainty with the criterion that $\Delta\chi^2=1$. We do not include any additional tolerance criterion, such as the two-tier penalty for the CT uncertainty~\cite{Hou:2019efy} and the dynamic tolerance for the MSHT uncertainty~\cite{Bailey:2020ooq}. Later, we will include such tolerance criteria for comparison with the GMM.

In panels~\ref{fig:problem_of_usual_chi2:a} and~\ref{fig:problem_of_usual_chi2:b}, pseudo-PDFs from fits SG-A, SG-B and SG-C are found to be in disagreement in the range $10^{-3}<x<0.2$. Even the $1\sigma$ uncertainty bands do not overlap in this region. For $x>0.2$, the tension is weak due to larger uncertainty in the data which is represented by the overlapping uncertainty bands in this region. Even though the uncertainty bands overlap at large $x$, the impact of significant tension in the range $10^{-3}<x<0.2$ still affects the pseudo-PDF at large $x$ as can be observed by the different shapes of the uncertainty bands for the different LS fits.
For the combined LS fit SG-C, the Hessian uncertainty with $\Delta \chi^2 = 1$ is also provided, which deviates from the Monte Carlo uncertainty by a negligible amount.
The tension between the different fits, SG-A, SG-B and SG-C can also be observed by looking at Fig.~\ref{fig:problem_of_usual_chi2:c} that shows uncertainty on the parameters $a_4$ and $a_5$. Monte Carlo samples of 1000 replicas are generated for each of the fits SG-A, SG-B and SG-C. 

None of the replicas generated for either of the fits overlap, once again indicating the strong tension among the psuedo-data sets.  

We would like to emphasize that either or both experiments may have underestimated systematic uncertainties and in the absence of further information about the source of systematics, the range of possible theory values should span the theory values predicted by each experimental data set individually. For example, an uncertainty band 
that looks like some combination of the uncertainty bands of both SG-A and SG-B,
should envelope  the central values of the fits SG-A and SG-B, or at least come close to doing so. Such an uncertainty band would be more representative of the tension in the data. Fig.~\ref{fig:problem_of_usual_chi2:c} shows that this is not the case for the SG-C fit.
The tolerance criteria used by the PDF fitting groups increases the uncertainty, but as will be shown in Section~\ref{sec:result_GMM}, this approach is not as precise as using the GMM. Further, in Sec.~\ref{sec:GMM}, we implement Ensemble Learning with the use of the GMM and show how this allows for the combined likelihood of the pseudo-data to be multi-modal and results in a uncertainty band that overlaps with the uncertainty bands of both SG-A and SG-B. 
Before we conclude this section, we comment on alternate methods to deal with such inconsistencies in fits. For example, as described in Appendix~\ref{sec:app:W_boson}, the Particle Data Group (PDG)~\cite{ParticleDataGroup:2022pth}, recommends increasing the uncertainty on data sets
by a scale factor $S = \sqrt{\chi^2/(N_{\text{pt}}-1)}$. For SG-C, this implies scaling the uncertainty on all the data by $S\simeq 1.56$, which, for our simple example, translates to setting a tolerance $T=S\simeq1.56$. Increasing the error with the scale factor, or equivalently the tolerance, does not shift the central value of the fit and only increases the uncertainty marginally. In Sec.~\ref{subsec:tolerance}, we explore even larger values of the tolerance and we will see that  such a small value 
of $T \simeq 1.56$ is not sufficient to reproduce uncertainty bands that represent the tension between the data sets. 

It is crucial to understand why the uncertainty band of SG-C, even with the increased tolerance ($T=S\simeq 1.56$), is unsatisfactory. It is easy to imagine, that the true theory value could easily be close to the central values of SG-A or SG-B, i.e. there is a large range of possible values for the true theory parameters, given the data. On the other hand, the central values of SG-A and SG-B lie well outside the uncertainty band of the combined LS fit SG-C, even with $T\simeq1.56$. Hence, the uncertainty bands of SG-C, do not provide an accurate representation of the situation. In other words, precision is lost  and the uncertainties must be increased in the region of tension in order to avoid biasing the fits and to accommodate our lack of information about the source of tension. We describe a simpler example using $W$-boson mass in Appendix~\ref{sec:app:W_boson}, where we show how a naive combination of data with tension can bias the fits even after increasing the scale factor $S$.

The method using the GMM, to be presented in the next section, provides an alternate way to combine inconsistent data sets  with large tension and provides a more accurate determination of the uncertainties. Furthermore, if desired, it can also provide an alternate estimate for the PDG scale factor $S$ required to make data sets consistent. We describe some of these ideas briefly in Sec.~\ref{subsec:GMM_overfit} and Appendix~\ref{sec:app:W_boson}.

\section{Gaussian mixture model in PDF fit}
\label{sec:GMM}

As mentioned in Sec.~\ref{sec:tension}, the likelihood from each of the data sets (pseudo-data \#1 and pseudo-data \#2) are significantly different and need to be combined in a meaningful way. We propose the use of the Gaussian Mixture Model to  construct the combined likelihood for data with tension.

 The GMM was first used by Karl Pearson in 1894~\cite{pearson1894contributions}. It is now used extensively as an unsupervised machine learning technique to classify data~\cite{Murphy:ML}. 
Here, we use it to determine the uncertainty estimate on theoretical fit parameters given a set of discrepant data. In Appendix~\ref{sec:appendix:Bayesian}, we present the more intuitive Bayesian description of the GMM. Here, we present the frequentist approach by starting with an ansatz for the likelihood function.
In general, the GMM is a mixture of $K$ base Gaussian distributions.
Given the data set $Y=\{ (y_1, \Delta y_1), (y_2, \Delta y_2), \cdots, (y_{N_{\text{pt}}}, \Delta y_{N_{\text{pt}}}) \}$, where $\Delta y_j$ is the uncertainty associated with the measured value $y_j$,\footnote{ For clarity of presentation, we use an uncorrelated data set and present formulas for this scenario. These can easily be generalized when correlated uncertainties are present.} we can define the GMM likelihood as
\begin{eqnarray}
L(Y|\vec{\theta}) &=& \prod_{j=1}^{N_{\text{pt}}} \pi(y_j, \Delta y_j|\vec{\theta}) = \prod_{j=1}^{N_{\text{pt}}} \left(\sum_{i=1}^K \omega_i \mathcal{N}(y_j, \Delta y_j|\theta_i)\right)\ , 
 \label{eq:GMM} \\
 &0& \leq \omega_i \leq 1 \quad \text{and} \quad \sum_i \omega_i = 1, \nonumber
\end{eqnarray}
where the $i$'th base distribution $\mathcal{N}(y_j, \Delta y_j|\theta_i)$ is associated with a normalized weight $\omega_i$.

    Here, the index $j$ labels each experimental data point, index $i$ labels each Gaussian with theory parameters $\theta_i$  that are estimated by maximizing the likelihood above. 
    In the example fit that we use $\theta_i = \left\{a_{4_i},a_{5_i}\right\}$ where $i =1,2,...K$. For example, if $K=2$, then  $\theta_1 = \left\{a_{4_1},a_{5_1}\right\}$ and $\theta_2 = \left\{a_{4_2},a_{5_2}\right\}$ and we have effectively introduced four parameters to fit.  If $K=1$, then $\theta_1 = \left\{a_{4_1},a_{5_1}\right\}=\left\{a_{4},a_{5}\right\}$ and the log likelihood reduces to the usual $\chi^2$ likelihood used in Sec.~\ref{subsec:chi2_fit_case_1}. For $K>1$, in addition to the $(K-1)$ independent weight parameters $w_i$, we have  $N_{\text{parm}}\times K$ parameters that are varied to maximize the likelihood, where $N_{\text{parm}}$ is the number of parameters in the fitting function $g(x)$ that are being estimated.
    Since the number of original parameters $N_{\text{parm}}$ has increased in the GMM by a factor $K$, 
we need to determine not only the central value of $K\times N_{\text{parm}}$ parameters but also the uncertainty on each of them. Later, we will describe how this fit to $K\times N_{\text{parm}}$ parameters can be combined in a meaningful way to produce a fit and uncertainty on only $N_{\text{parm}}$ parameters.\footnote{ For example, there were two fit parameters ($a_4$ and $a_5$) in  the LS fits  in Sec.~\ref{sec:tension}, whereas in the GMM there would be $2\times K$ parameters. Ultimately, we want to be able determine the uncertainty on two parameters and not $2\times K$ parameters.}

To simplify the following expressions, we will abbreviate $\mathcal{N}$ and $\pi$ for the $i$'th base distribution and the $j$'th measurement as follows 
\begin{equation}
\mathcal{N}(y_j, \Delta y_j|\theta_i) = \mathcal{N}_{ij}  , \quad \pi(y_j, \Delta y_j|\vec{\theta}) = \pi_j \ .
\end{equation}
In the GMM, the logarithm of the likelihood function is
\begin{eqnarray}
 - \ln L &=& - \ln \bigg{(} \prod_{j=1}^{N_{\text{pt}}} L(y_j, \Delta y_j|\vec{\theta}) \bigg{)} = - \sum_{j=1}^{N_{\text{pt}}} \ln \bigg{(} \sum_{i=1}^K \omega_i \mathcal{N}_{ij} \bigg{)}\nonumber \\
 &=& - \sum_{j=1}^{N_{\text{pt}}} \ln \bigg{(} \sum_{i=1}^K \frac{\omega_i}{\sqrt{2\pi} \Delta y_j} \text{exp} \Big{[} -\frac{1}{2} \Big{(} \frac{y_j - T(\theta_i)}{\Delta y_j} \Big{)}^2 \Big{]} \bigg{)}.
 \label{eq:likelihood_GMM}
\end{eqnarray}
For a single Gaussian distribution with $K=1$, the  log-likelihood in Eq.~\eqref{eq:likelihood_GMM} is proportional to $\chi^2$, which reproduces the Least Square method as shown in the equation below
\begin{equation}
 - \ln L = -\sum_{j=1}^{N_{\text{pt}}} \ln \bigg{(} \mathcal{N}(y_j, \Delta y_j|\theta) \bigg{)} = \frac{1}{2} \chi^2 + \sum_{j=1}^{N_{\text{pt}}} \ln \Big{(} \sqrt{2\pi} \Delta y_j \Big{)} \ .
\end{equation}
Notice, in Eq.~\eqref{eq:likelihood_GMM}, we have chosen to define the likelihood, such that each Gaussian has information about all experimental data points, i.e. for each Gaussian we sum over all values of $j=1$ to $j=N_{\text{pt}}$. An alternate way of defining the likelihood would be to distribute the experimental data among the various Gaussians, i.e. each data set would have its own Gaussian.  These two definitions have different interpretations. In the latter, we are trying to assign different theoretical predictions to different experiments, whereas in the former we are trying to find the theoretical values that different combinations of all experiments predict. Since we are trying to combine all experiments into a single global fit, we focus on using the definition as given in Eq.~\eqref{eq:likelihood_GMM}. 

    See Appendix~\ref{app:sec:Bayesian view of the GMM likelihood}, for a more intuitive explanation of this idea using a Bayesian formalism.

The GMM method allows us to extend the statistical model assumed in the optimization procedure, from a single Gaussian distribution to a more generalized configuration, where a multi-modal feature is allowed.
Importantly, for a consistent data set we will see that the multiple Gaussian modes will overlap and yield a result that is equivalent to a single Gaussian, i.e. the Least Squares fit. We will discuss the consistency between the GMM and LS methods in Sec.~\ref{sec:consistency}.

\subsection{Performing parameter fits with the GMM}
\label{subsec:GMM_best-fit}
In order to determine the best fit parameters, we minimize the the log-likelihood defined in Eq.~\eqref{eq:likelihood_GMM}. We remark here that, conventionally the Expectation Maximization algorithm~\cite{Murphy:ML} is used to learn the best fit parameters of the mixture model.\footnote{We remark here that it is possible to use the Expectation Maximization algorithm if we produce a large pseudo-data set and apply the Gaussian mixture model directly to the theory parameter space.} The advantage of this algorithm is that it provides an iterative method that does not require information about the derivative of the likelihood. 
For the purpose of demonstration, we use methods that rely on information about the derivative of the likelihood. Specifically, we use {\tt TensorFlow}~\cite{TensorFlow:2016} to determine derivatives of the likelihood and minimize the log-likelihood.
As described in the previous section, the number of parameters have increased and for $K$ Gaussians we  have 
\begin{align}
    N_{\text{parm}}^{\text{GMM}} = K\times N_{\text{parm}} + (K-1) \ ,
\end{align}

where, the last term, $K-1$, come from the constraint on the weights $\omega_i$ that are required to sum to unity.
We carry out minimization of the likelihood by varying all $ N_{\text{parm}}^{\text{GMM}} $ parameters.\footnote{It is possible to determine the value of the weights $\omega_i$ in other more complicated ways since they indicate the degree of importance to each Gaussian. Here we take the approach that we learn the values of $\omega_i$ directly from the data and hence it is also varied during the minimization process.}
The gradient of the log-likelihood with respect to the fit parameters that belong to $i$-th Gaussian is
\begin{eqnarray}
 \frac{\partial}{\partial \theta_i} \ln L &=& \sum_{j=1}^{N_{\text{pt}}} \omega_i \frac{\mathcal{N}_{ij}}{\pi_{j}} \frac{y_j - T(\theta_i)}{\Delta y^2_j} \frac{\partial T(\theta_i)}{\partial \theta_i}, \label{eq:GMM_gradient} 
\end{eqnarray}
 and that with respective to weight $\omega_i$ is 
\begin{eqnarray}
     \frac{\partial}{\partial \omega_i} \ln L &=& \sum_{j=1}^{N_{\text{pt}}} \frac{\mathcal{N}_{ij}}{ \big{(} \sum_{i=1}^K \omega_i \mathcal{N}_{ij} \big{)}} 
     =\sum_{j=1}^{N_{\text{pt}}} \frac{\mathcal{N}_{ij}}{ \pi_j} 
     \, .
      \label{eq:GMM_gradient_wrt_wgt}
\end{eqnarray}
Optimal values, $\hat{\theta}_i$, for the parameters  ${\theta}_i$ are found by extremizing the log-likelihood by solving~\cite{mclachlan2000finite}
\begin{eqnarray}
    \frac{\partial}{\partial \theta_i} \ln L \bigg{|}_{\theta_i=\hat{\theta}_i} &=& 0 
    \label{eq:best_fit_GMM}
\end{eqnarray}
along with a similar equation involving the weights $\omega_i$,
\begin{eqnarray}
 \frac{\partial}{\partial w_i} \ln L \bigg{|}_{w_i=\hat{w}_i} &=& 0.
\end{eqnarray}
In this work, we extremize the log-likelihood function by using the \texttt{Adam} optimizer with learning rate being $10^{-6}$ provided by the \texttt{TensorFlow} API~\cite{TensorFlow:2016}.
Note that, in Eq.~\eqref{eq:GMM_gradient}, the best-fit value of the prediction $y_j(\theta_i)$ is weighted not only by the variance in measurements ($\Delta y_i$), but also by the weights $\omega_i$ and the factor $\mathcal{N}(y_j|\theta_i)/\pi(y_j, \Delta y_j|\vec{\theta})$ of the $i$-the base distribution, which becomes trivial for a single Gaussian.
The Jacobian $\partial y_j(\theta_i)/\partial \theta_i$, transforms  information from the data space to the PDF space. The presence of this transformation makes it difficult to use the Expectation Maximum algorithm directly.

In this work we use the above equation as a constraint that determines the value of $\omega_i$. This is generally the approach taken  for other unsupervised machine learning applications. In general, this strategy tends to assign higher weights to experiments with higher precision. If desired, it is possible to control the weights by increasing or decreasing the value of the uncertainties associated with a certain experimental data set.\footnote{One method of method of managing inconsistent data sets is to artificially inflate uncertainties until  we arrive at a consistent data set~\cite{Kovarik:2019xvh}.  

In Appendix~\ref{sec:app:W_boson}, we describe how the GMM, in conjunction with certain information criteria, provides a way to determine the extent to which it is necessary to inflate the uncertainty on data until the fits become consistent.

}
Thus along with $\theta_i$, the weights $\omega_i$ are also varied in order to optimize the log-likelihood.

\subsection{Mean and Uncertainty in the GMM}
\label{subsec:GMM_Hessian}

It is straightforward to show that the expectation value of parameters is the weighted sum of the optimal parameters for individual base distributions~\cite{raftery2005using},
\begin{equation}
 \mathbb{E}[\theta] = \sum_{i=1}^K \omega_i \hat{\theta}_i.
 \label{eq:GMM_mean}
\end{equation}

    Here $\hat{\theta}_i$ are the estimated values of the $N_{\text{parm}}$ parameters of the fitted function $g(x)$ and there are $K$ such $\hat{\theta}_i$.

Consequently, an experiment with higher precision would be associated with a higher weight and, as a result, the central value of the parameters is pulled towards those values that better agree with higher precision experiments. 

There are several ways to determine the uncertainty with the modified log-likelihood. Below, we describe the use of the observed Fisher Information Matrix to estimate uncertainty~\cite{mclachlan2000finite} on the fit parameters. We note that this is only one way to estimate the uncertainty and there are other possible methods which will be discussed in detail elsewhere.

When using a single Gaussian or when all the Gaussians have the same mean and covariance, the observed Fisher information matrix reduces to the usual Hessian matrix used to determine the covariance of the fit parameters in the LS method.\footnote{ For the GMM with multiple Gaussians, the likelihood function as defined in Eq.~\eqref{eq:likelihood_GMM} includes all experimental points for each Gaussian. Thus, if the means of each Gaussian are identical, then this implies that the covariance, which is related to the second order derivative about the minimum of the likelihood, must also be identical, c.f. Eq.~\eqref{eq:cov_GMM_2}. In general, the covariance matrices for Gaussians with identical means may be different in the GMM if the likelihoods were to be defined differently. 
} 
Like the Hessian matrix, the information matrix can also be found with numerical methods. Here we use algorithmic differentiation~\cite{Bucker2005ABo} within {\tt TensorFlow} to determine derivatives of the likelihood function. We note that the information matrix is only an estimate of the covariance of the GMM and it is possible to determine the covariance matrix as well, which will be described in detail elsewhere.

Elements of the information matrix ($\mathcal{I}$) are determined by taking a further derivative on Eq.~\eqref{eq:best_fit_GMM} as follows, 
\begin{align}
 \left(\mathcal{I}\right)_{mk}&=- \frac{\partial^2}{\partial \theta_m \partial \theta_k} \ln L = \frac{\partial}{\partial \theta_m} \left( - \sum_{j=1}^{N_{\text{pt}}} \frac{\omega_k}{\pi_j} \frac{\partial \mathcal{N}_{kj}}{\partial \theta_k}\right)
& \nonumber \\
&=
 \sum_{j=1}^{N_{\text{pt}}} \left[ - 
 \frac{\omega_k}{\pi_j} \frac{\partial^2 \mathcal{N}_{mj}}{\partial \theta_m \partial \theta_m} \delta_{mk} + \frac{\omega_k \omega_m}{\pi_j^2} \frac{\partial \mathcal{N}_{mj}}{\partial \theta_m}\frac{\partial \mathcal{N}_{kj}}{\partial \theta_k} 
 \right]
&
 \label{eq:wgt_Hess_GMM}
\end{align}

Notice that for a fit with $N_{\text{parm}}$ 
parameters per Gaussian, the information matrix is a $(K N_{\text{parm}}) \times (K N_{\text{parm}})$ symmetric matrix, i.e. it spans a space that is a direct product of an $N_{\text{parm}}$ parameter and a $K$ parameter space.
The second term contributes  to the elements of the information matrix ($\mathcal{I}$) that are off-diagonal in $K$-space but not in $N_{\text{parm}}$ space. In contrast, the off-diagonal terms in $N_{\text{parm}}$ space contribute to the covariance matrix. On the other hand, the $K$-space off-diagonal terms are small since they must add up to the left hand side of Eq.~\eqref{eq:best_fit_GMM}, which vanishes at $\theta_i=\hat{\theta}_i$ by definition. This is also verified numerically for all of the fits performed here. For simplicity, we will neglect these particular $K$-space off-diagonal terms and only consider the $N_{\text{parm}} \times N_{\text{parm}}$ block diagonal terms of the information matrix. Although, it is not necessary to neglect these off diagonal terms, this allows us to quickly approximate the probability distribution function in parameter space ($\theta$) as a mixture of Gaussians. In other words, we are trying to use the Likelihood function defined in Eq.~\eqref{eq:likelihood_GMM}, to approximate the probability distribution function of $\theta$. 
Thus, we assume that, when evaluated at the mean ($\theta_i = \hat{\theta}_i$) the information matrix takes on an approximate block diagonal form, with each $N_{\text{parm}} \times N_{\text{parm}}$ block belonging to the subspace of the $i$'th Gaussian $\mathcal{N}(y_j, \Delta y_j|\theta_i)$. 
We denote the information matrix for individual base distributions of the GMM as  $\mathcal{I}_{\text{GMM},i} = - (\partial^2 \ln L / \partial \theta_i \partial \theta_i) / \omega_i$, which are the unweighted $N_{\text{parm}} \times N_{\text{parm}}$ diagonal blocks of the full information matrix $\mathcal{I}$.
An estimate of the covariance of individual base Gaussian distributions of the GMM is then,
\begin{equation}
 \text{cov}_{\text{GMM},i} \simeq \bigg{(} \mathcal{I}_{\text{GMM},i} \bigg{)}^{-1} = \Bigg{(} 
  \sum_{j=1}^{N_{\text{pt}}} \left[ - 
 \frac{1}{\pi_j} \frac{\partial^2 \mathcal{N}_{ij}}{\partial \theta_i^2}  + \frac{\omega_i }{\pi_j^2} \left(\frac{\partial \mathcal{N}_{ij}}{\partial \theta_i}\right)^2 
 \right]\Bigg{)}^{-1}
 \label{eq:cov_GMM_i}
\end{equation}
We can now define the probability distribution of the parameters of the fit as mixture of Gaussians as follows:
\begin{align}
\mathcal{P}(x|\theta_i, \text{cov}_{\text{GMM,i}}) &= \sum_{i=1}^K \frac{\omega_i}{(2\pi)^{N_{\text{parm}}/2} \det (\text{cov}_{\text{GMM,i}})^{1/2}}& \nonumber \\  &\times \text{exp} \Big{[} -\frac{1}{2} \Big{(} (x - \theta_i)^T\cdot\left( \text{cov}_{\text{GMM,i}} \right)^{-1} \cdot (x - \theta_i)  \Big{)} \Big{]} \ .&  \end{align}
With this definition, we can now determine estimates of the uncertainty on the parameters $\theta$. We first note that when $K=1$, we have a single Gaussian and Eq.~\eqref{eq:wgt_Hess_GMM} reduces to the Hessian matrix of the LS method,
\begin{equation}
- \frac{\partial^2}{\partial \theta \partial \theta} \ln L \bigg|_{\theta = \hat{\theta}}= \frac{1}{2} \frac{\partial^2}{\partial \theta \partial \theta} \chi^2 \bigg|_{\theta = \hat{\theta}} = \sum_{j=1}^{N_{\text{pt}}} \frac{1}{\Delta y^2_j} \frac{\partial^2 T(\theta)}{ \partial \theta ^2}\bigg|_{\theta = \hat{\theta}},
\end{equation}
and the covariance is 
\begin{equation}
 \text{cov}_{\chi^2} = \bigg{(} \sum_{j=1}^{N_{\text{pt}}} \frac{1}{\Delta y^2_j} \frac{\partial^2 T(\theta)}{ \partial \theta ^2} \bigg|_{\theta = \hat{\theta}} \bigg{)}^{-1}.
 \label{eq:cov_chi2}
\end{equation}

For the GMM, in order to provide a useful estimate of uncertainty, we need to reduce the $(K N_{\text{parm}}) \times (K N_{\text{parm}})$ space to a $(N_{\text{parm}}) \times ( N_{\text{parm}})$ space in a meaningful way.  To that end, let us consider the covariance of the GMM which is defined as~\cite{raftery2005using} 
\begin{eqnarray}
 \text{cov}_{\text{GMM}} &=& \mathbb{E}[\theta^2] - \mathbb{E}[\theta]^2 \nonumber \\
 &=& \sum_{i=1}^K \omega_i \hat{\theta}_i^{(2)} - \big{(} \sum_{i=1}^K \omega_i \hat{\theta}_i \big{)}^2,
 \label{eq:cov_GMM_1}
\end{eqnarray}
where $\hat{\theta}_i^{(2)}$ denotes the estimated second moment of parameters for the $i$-th base distribution. It can be re-expressed using the covariance of $i$-th base Gaussian distribution,
\begin{equation}
 \hat{\theta}_i^{(2)} = \text{cov}_{\text{GMM},i} + \hat{\theta}^2_i.
 \label{eq:2nd_mom_GMM_i}
\end{equation}
Combining Eqs.~\eqref{eq:cov_GMM_1} and~\eqref{eq:2nd_mom_GMM_i}, an estimate of the covariance of the GMM is given by~\cite{raftery2005using}
\begin{eqnarray}
 \text{cov}_{\text{GMM}} &=& \sum_{i=1}^K \omega_i \ \text{cov}_{\text{GMM},i} + \sum_{i=1}^K \omega_i (\mathbb{E}[\theta] - \hat{\theta}_i )^2 \nonumber \\
  &\simeq& \sum_{i=1}^K \omega_i \Bigg{(} 
  \sum_{j=1}^{N_{\text{pt}}} \left[ - 
 \frac{1}{\pi_j} \frac{\partial^2 \mathcal{N}_{ij}}{\partial \theta_i^2}  + \frac{\omega_i }{\pi_j^2} \left(\frac{\partial \mathcal{N}_{ij}}{\partial \theta_i}\right)^2 
 \right]\Bigg{)}^{-1}  + \sum_{i=1}^K \omega_i (\mathbb{E}[\theta] - \hat{\theta}_i )^2.
 \label{eq:cov_GMM_2}
\end{eqnarray}
The covariance of the GMM shown in the above equation receives two contributions: the first is the weighted sum of covariances for the individual base distributions, and the second term accounts for differences in the mean values of each of the base distributions. When the means of each Gaussian overlap, the second term vanishes and the first term reproduces the covariance in LS method, Eq.~\eqref{eq:cov_chi2}.

\section{Application of the GMM in a toy model of PDFs}
\label{sec:result_GMM}
In this section, we compare results obtained by the GMM method, using the likelihood defined in Eq.~\eqref{eq:likelihood_GMM}, to those obtained using the LS method. When estimating uncertainty using the LS method,  we first diagonalize the Hessian and determine eigenvalue directions. We then determine values of the parameters that satisfy the  $\Delta \chi^2 = 1$ condition and define the uncertainty on the fit parameters, which corresponds to the 68\% confidence level interval. Here, we also implement the CTEQ-TEA~\cite{Hou:2019efy} and MSHT~\cite{Bailey:2020ooq} tolerance criteria, which provide a more conservative estimate of uncertainty under the potential imperfections of fits.

Below we present the results of three different combinations of pseudo-data sets, including also the two pseudo-data sets that were discussed in Sec.~\ref{sec:tension}.

\subsection{Applying GMM to different scenarios}
\label{subsec:GMM_vs_chi2}

In this section we consider three cases of combining different pairs of pseudo-data sets. Details of how each of these psuedo-data sets are generated can be found in Appendix~\ref{app:sec:pseudo_data}.

\begin{figure}[htbp]  
\centering
\subfigure[Case-2]
{{\includegraphics[width=0.4\columnwidth]{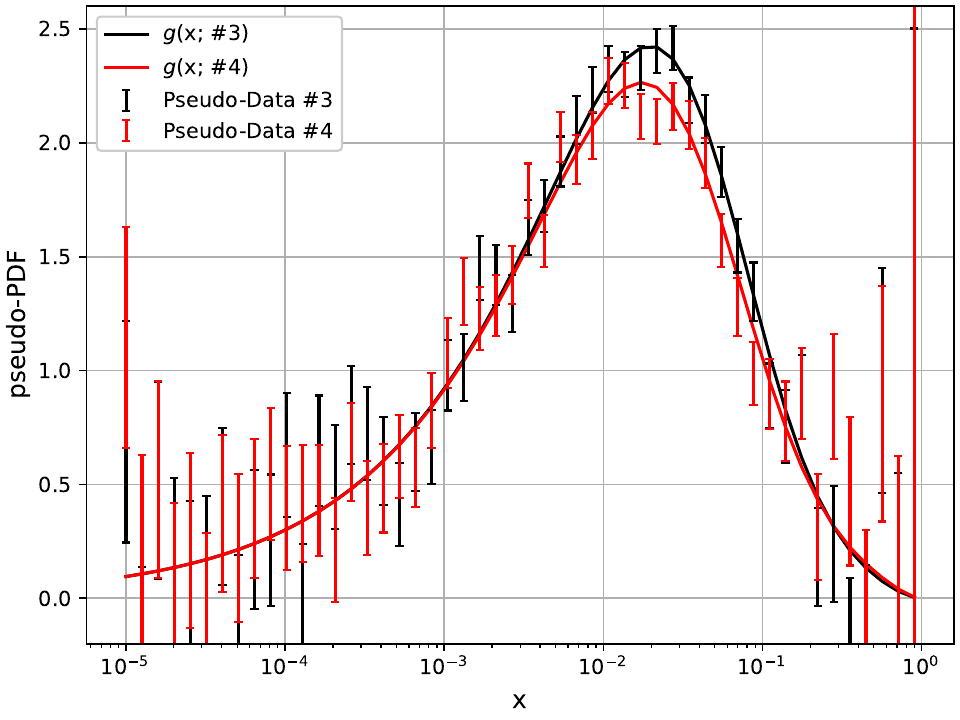}}
\label{fig:pData_case23:a}}
\subfigure[Case-2]
{{\includegraphics[width=0.4\columnwidth]{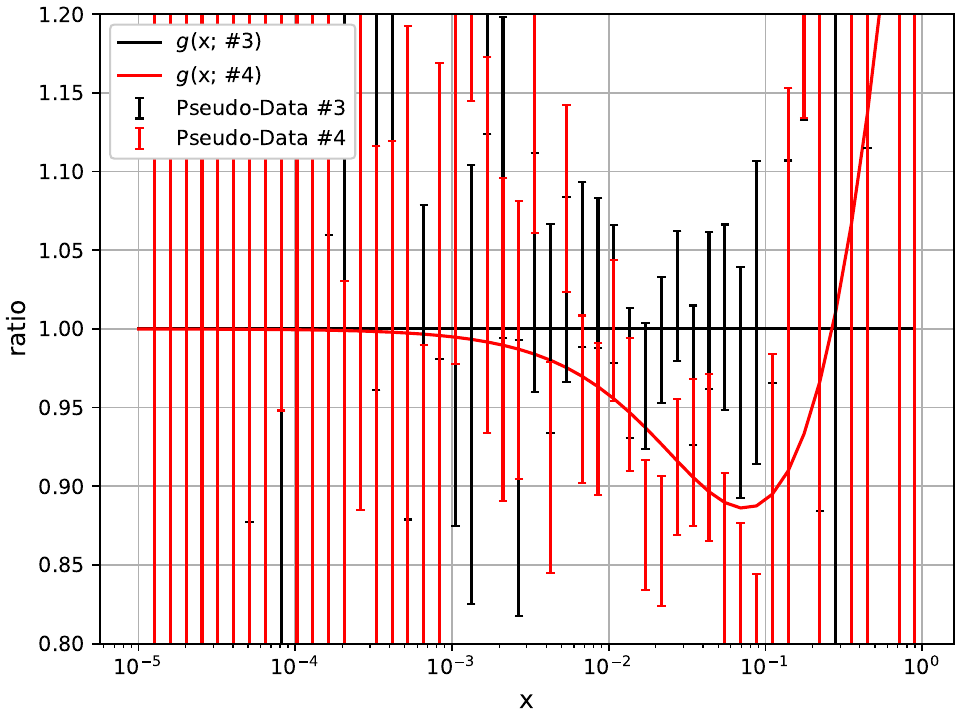}}
\label{fig:pData_case23:b}}
\subfigure[Case-3]
{{\includegraphics[width=0.4\columnwidth]{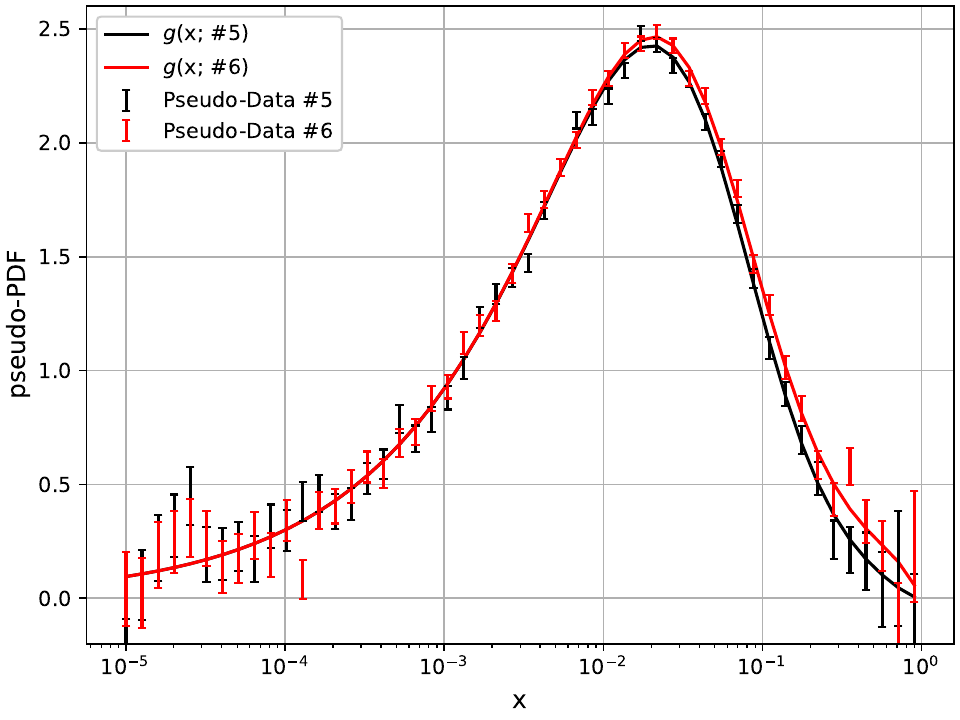}}
\label{fig:pData_case23:c}}
\subfigure[Case-3]
{{\includegraphics[width=0.4\columnwidth]{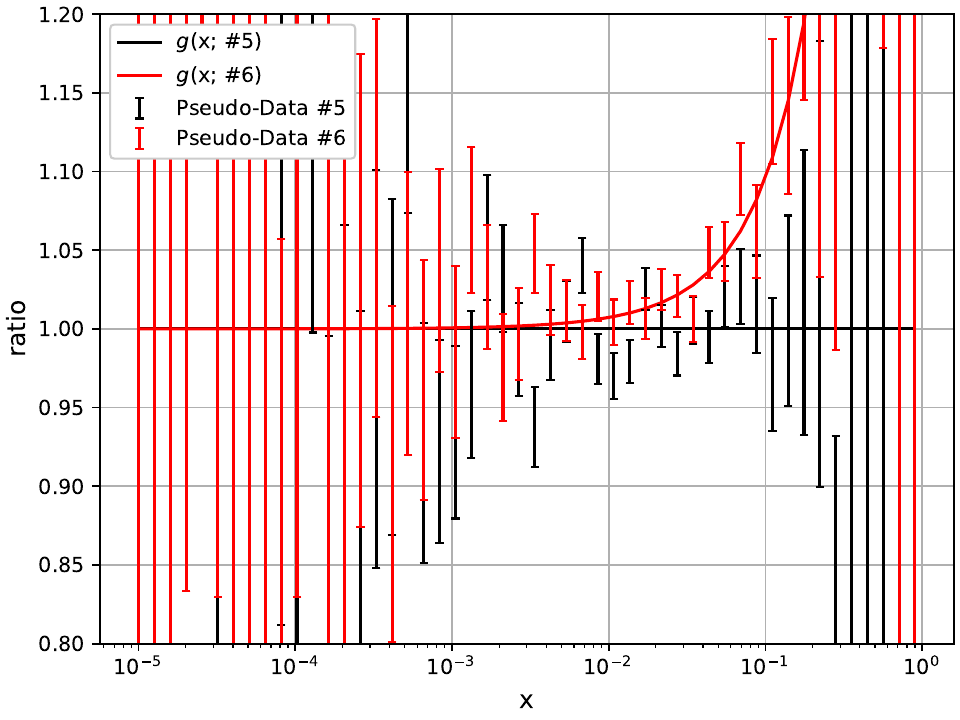}}
\label{fig:pData_case23:d}}
\caption{\label{fig:pData_case23} 
Similar to Fig.~\ref{fig:pData}, but for Case-2 and Case-3.
} 
\end{figure}

Here, we list and describe each of these cases below:
\begin{itemize}
\item Case-1: Here we use pseudo-data sets $\#~1$ and $\#~2$ which were introduced in Sec.~\ref{sec:tension}. The pseudo-data sets are plotted in Fig.~\ref{fig:pData}.This case represents the scenario when there is strong tension between data sets. With this case, we will demonstrate how the GMM can provide a more accurate estimate of uncertainty.
 \item Case-2:  
 
 Here we combine pseudo-data sets $\#~3$ and $\#~4$, which are shown in Fig.~\ref{fig:pData_case23:a} and Fig.~\ref{fig:pData_case23:b}. 
     These two data sets are similar to data sets $\#~1$ and $\#~2$,  except that their uncertainties are tripled and thus have less tension.
 
 With this combination we will be able to show how the GMM behaves when the errors of the data sets are large enough so that there is less tension. We can connect this to the alternative suggestions of dealing with uncertain data by artificially increasing the errors of the data sets~\cite{Kovarik:2019xvh}.
 \item Case-3: 
  
 This case contains pseudo-data sets $\#~5$ and $\#~6$, which are shown in Fig.~\ref{fig:pData_case23:c} and Fig.~\ref{fig:pData_case23:d}. 
  The tension between these two pseudo-data sets resides in the large-$x$ region    
 
where the uncertainty is large. We fit this case with parameters $a_2$ and $a_4$. With this combination we demonstrate overfitting with the GMM method, and later in Sec.~\ref{subsec:GMM_overfit} we demonstrate how to address overfitting.
\end{itemize}

\begin{table}[htpb]
\begin{center}
\begin{tabular}{ccc|ccccccc}
 Case-1 & fits & pseudo-data & $a_4$ & $a_5$ & $\chi^2_{\# 1}/N_{\text{pt}}$ & $\chi^2_{\# 2}/N_{\text{pt}}$ & $\chi^2_{\# 1 + \# 2}/N_{\text{pt}}$ & weight \\ \hline
 & SG-A & \# 1 & 2.32 & -3.22 & 0.88 & 6.55 & 3.72 & - \\
 & SG-B & \# 2 & 2.63 & -2.73 & 7.00 & 1.02 & 4.01 & - \\
 & SG-C & \# 1 and \# 2 & 2.48 & -2.94 & 2.27 & 2.56 & 2.42 & - \\
 & GMM Gauss.-1 & \# 1 and \# 2 & 2.31 & -3.25 & 0.88 & 6.55 & 3.72 & 0.52 \\
 & GMM Gauss.-2 & \# 1 and \# 2 & 2.64 & -2.72 & 7.00 & 1.02 & 4.01 & 0.48 \\ \hline \hline
 Case-2 & fits & pseudo-data & $a_4$ & $a_5$ & $\chi^2_{\# 3}/N_{\text{pt}}$ & $\chi^2_{\# 4}/N_{\text{pt}}$ & $\chi^2_{\# 3 + \# 4}/N_{\text{pt}}$ & weight \\ \hline
  &  SG-A & \# 3 & 2.20 & -3.50 & 0.76 & 1.68 & 1.22 & - \\
  & SG-B & \# 4 & 2.74 & -2.54 & 1.61 & 0.89 & 1.25 & - \\
  & SG-C & \# 3 and \# 4 & 2.48 & -2.98 & 0.96 & 1.11 & 1.04 & - \\
  & GMM Gauss.-1 & \# 3 and \# 4 & 2.40 & -3.10 & 0.81 & 1.36 & 1.09 & 0.69 \\
  & GMM Gauss.-2 & \# 3 and \# 4 & 2.66 & -2.73 & 1.68 & 0.90 & 1.29 & 0.31 \\ \hline \hline
 Case-3 & fits & pseudo-data & $a_2$ & $a_4$ & $\chi^2_{\# 5}/N_{\text{pt}}$ & $\chi^2_{\# 6}/N_{\text{pt}}$ & $\chi^2_{\# 5 + \# 6}/N_{\text{pt}}$ & weight \\ \hline
  & SG-A & \# 5 & 2.27 & 2.40 & 1.16 & 2.33 & 1.75 & - \\
  & SG-B & \# 6 & 1.25 & 2.41 & 2.46 & 0.82 & 1.64 & - \\
  & SG-C & \# 5 and \# 6 & 1.60 & 2.41 & 1.54 & 1.14 & 1.34 & - \\ 
  & GMM Gauss.-1 & \# 5 and \# 6 & 2.06 & 2.40 & 1.19 & 1.96 & 1.58 & 0.58 \\
  & GMM Gauss.-2 & \# 5 and \# 6 & 1.13 & 2.41 & 2.58 & 0.82 & 1.70 & 0.42 \\ 
\end{tabular}
\end{center}
\caption{
The best-fit values of parameters, and $\chi^2/N_{\text{pt}}$ values for individual and combined pseudo-data sets obtained by the LS method and the GMM fit.
}
\label{tab:GMM_best_fits}
\end{table}

\begin{figure}[htbp]  
\centering
\subfigure[]
{{\includegraphics[width=0.37\hsize]{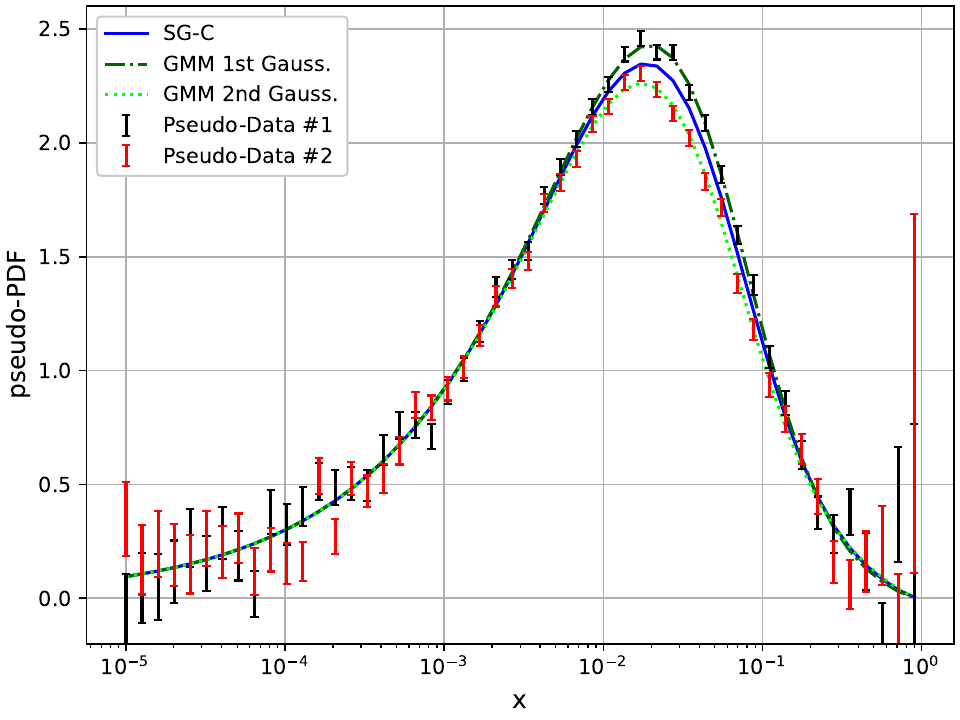}}
\label{fig:GMM_case1:a}}
\subfigure[]
{{\includegraphics[width=0.37\hsize]{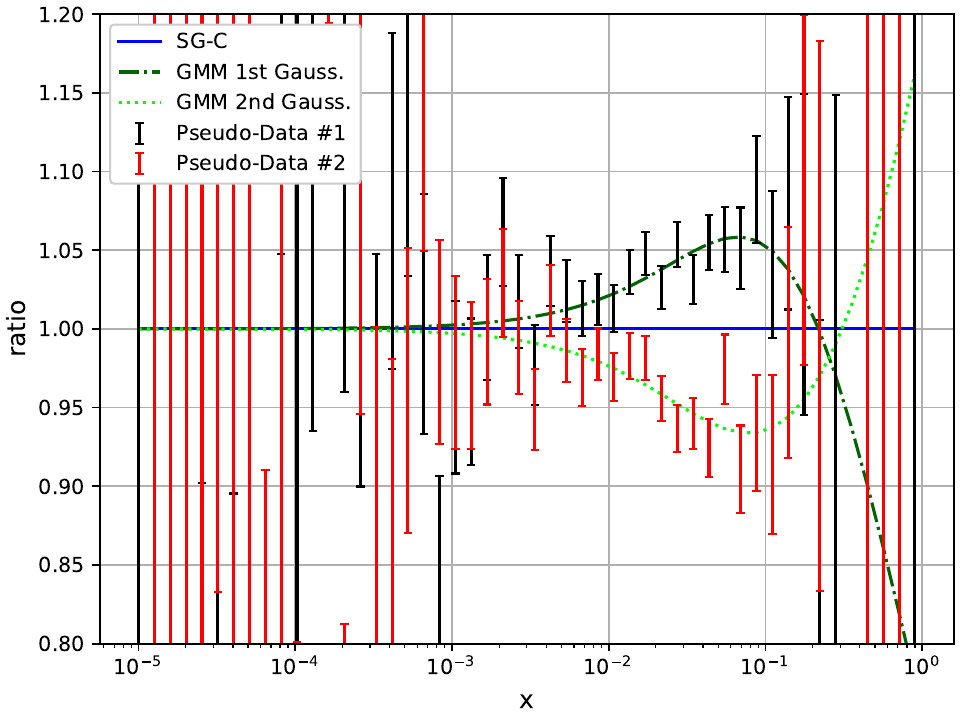}}
\label{fig:GMM_case1:b}}
\subfigure[]
{{\includegraphics[width=0.37\hsize]{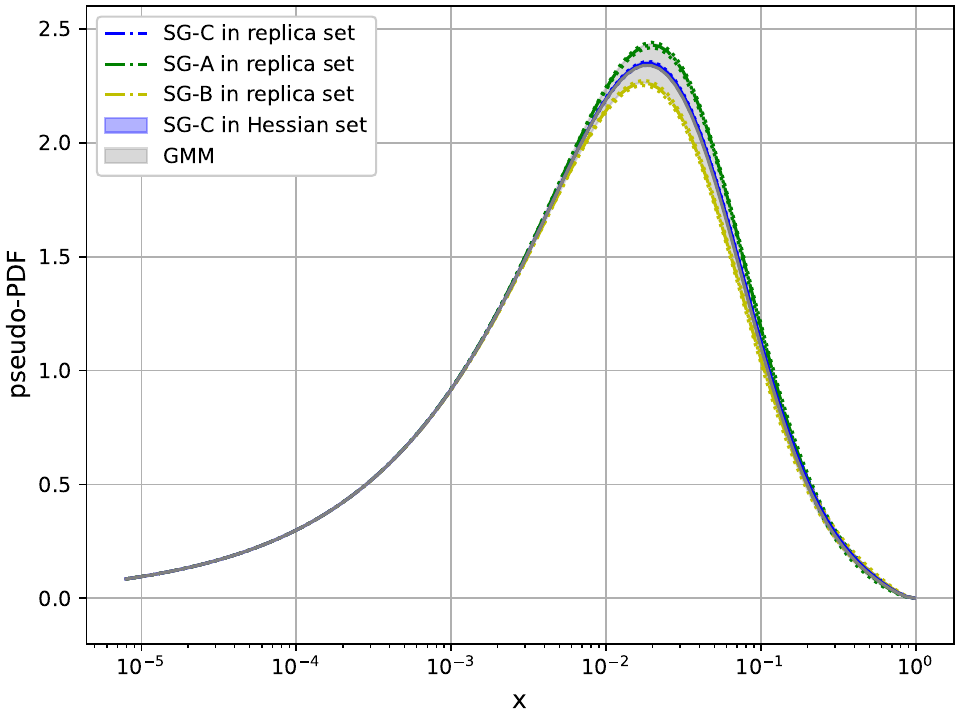}}
\label{fig:GMM_case1:c}}
\subfigure[]
{{\includegraphics[width=0.37\hsize]{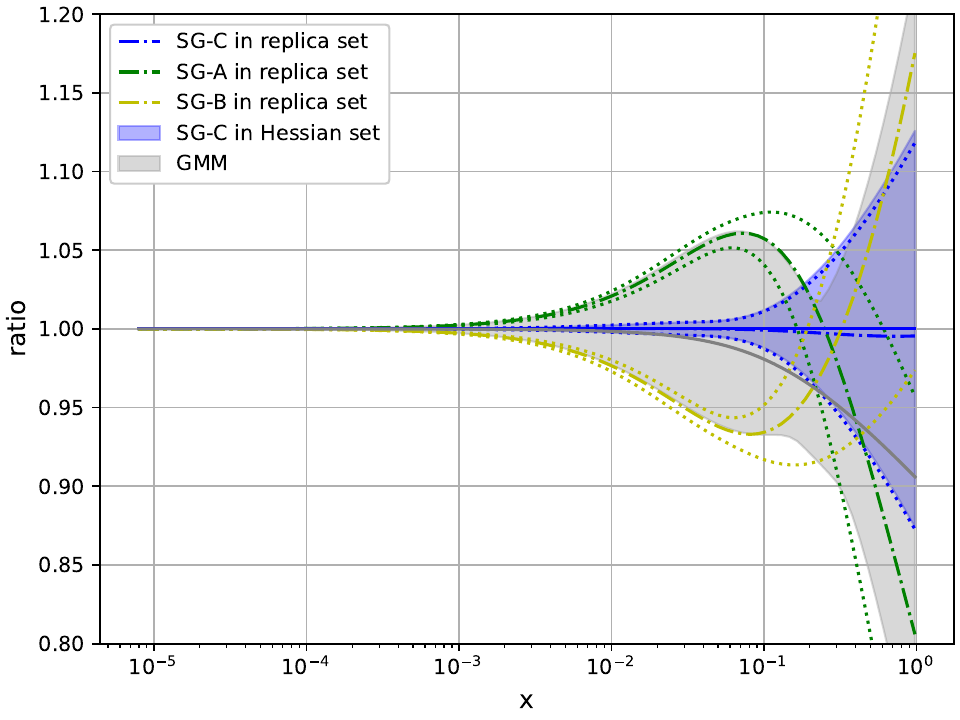}}
\label{fig:GMM_case1:d}}
\subfigure[]
{{\includegraphics[width=0.37\hsize]{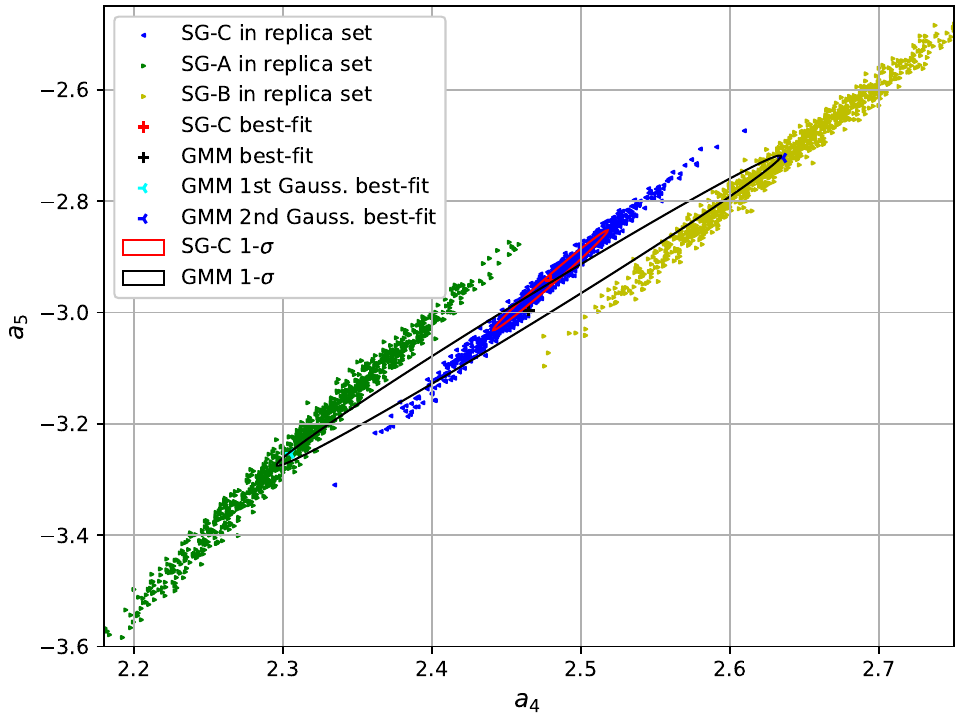}}
\label{fig:GMM_case1:e}}
\caption{\label{fig:GMM_case1} 
Pseudo-PDFs obtained by the GMM method and the LS method for Case-1 are compared.
Panel (a): The pseudo-data sets are compared to best-fit pseudo-PDFs. 
Although shown separately, the two best-fit pseudo-PDFs from the GMM method are eventually combined according to weights shown in Table~\ref{tab:GMM_best_fits}.
Panel (b): Similar to panel (a), but presented as ratios of the best-fit obtained from the SG-C fit.
Panel (c): The absolute values of pseudo-PDFs for all SG-A, -B, -C, and GMM fits.
Panel (d): Similar to panel (c), but presented as ratios to the best-fit obtained from the SG-C Hessian set.
Panel (e): The correlation ellipse from the GMM method is compared to SG-C correlation ellipses, and to replicas from SG-A, -B, and -C fits. The best-fit points are located at their corresponding values in Table~\ref{tab:GMM_best_fits}, while the GMM best-fit corresponds to the weighted average of best-fit of two individual Gaussian distributions as given by Eq.~\eqref{eq:GMM_mean}.
} 
\end{figure}

The best-fit obtained by the GMM method for Case-1 is shown in Figs.~\ref{fig:GMM_case1:a} and~\ref{fig:GMM_case1:b} and in  Table~\ref{tab:GMM_best_fits}. We choose to fit the GMM likelihood function with a mixture of two Gaussian distributions. As a result, the mean values of each Gaussian, can describe the two distinct pseudo-data sets well and are in agreement with the fits SG-A and SG-B. We find the weights $\omega_i$ to be roughly equal, which is a result of the errors in each of the data sets $\# 1$ and $\# 2 $ being almost equal and also having the same number of data points.
In Figs.~\ref{fig:GMM_case1:c}and~\ref{fig:GMM_case1:d}, the pseudo-PDF uncertainty estimated by the GMM is compared to uncertainties predicted using the LS method. For  $10^{-3} < x < 0.1$, the (small) error band of  SG-C pseudo-PDF exclude those of SG-A and SG-B fits, the error band estimated by the GMM is consistent and includes areas covered by both SG-A and SG-B pseudo-PDFs. For $x > 0.1$, where the error in the data sets are large, uncertainties from SG-C and the GMM overlap. Fig.~\ref{fig:GMM_case1:e} shows the results of the different fits in the fit parameter space. The 1-$\sigma$ correlation ellipse estimated by SG-C is small, as has been mentioned before, and does not overlap with SG-A and SG-B replicas. On the other hand, the GMM correlation ellipse is large enough and, importantly, also possesses a different axes orientations, that enables the GMM covariance to overlap with some of the replicas of both SG-A and SG-B.
The essential feature we see here  is that the tension in the pseudo-data sets lies in the region $10^{-2} < x <10^{-1}$ (see for example Fig.~\ref{fig:GMM_case1:d}). The GMM expands errors in this region, whereas the $\chi^2$ method with tolerance displays a smaller uncertainty band in this region and expands the error in the region $x > 10^{-1}$. The GMM therefore provides a better estimate of the probability distribution of the theory parameters given the discrepancies in the data.
Note that we have represented the GMM by a single covariance matrix but this should not be confused with the covariance arising from a single Gaussian, since the shape of the GMM likelihood is multi-modal and not uni-modal.

\begin{figure}[htbp]  
\centering
\subfigure[]
{{\includegraphics[width=0.4\columnwidth]{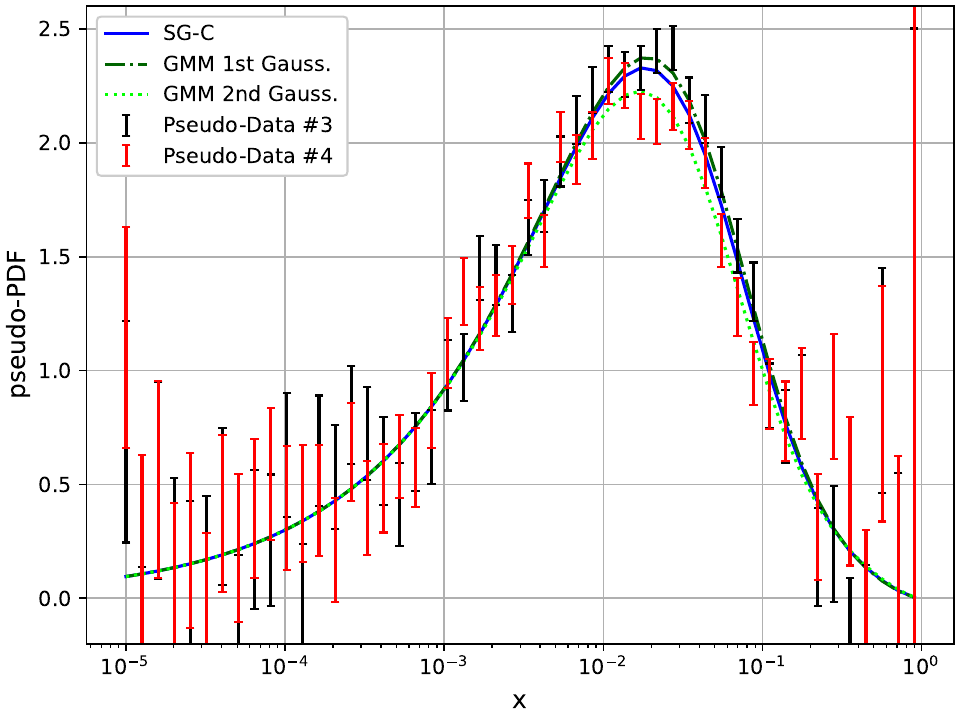}}
\label{fig:GMM_case2:a}}
\subfigure[]
{{\includegraphics[width=0.4\columnwidth]{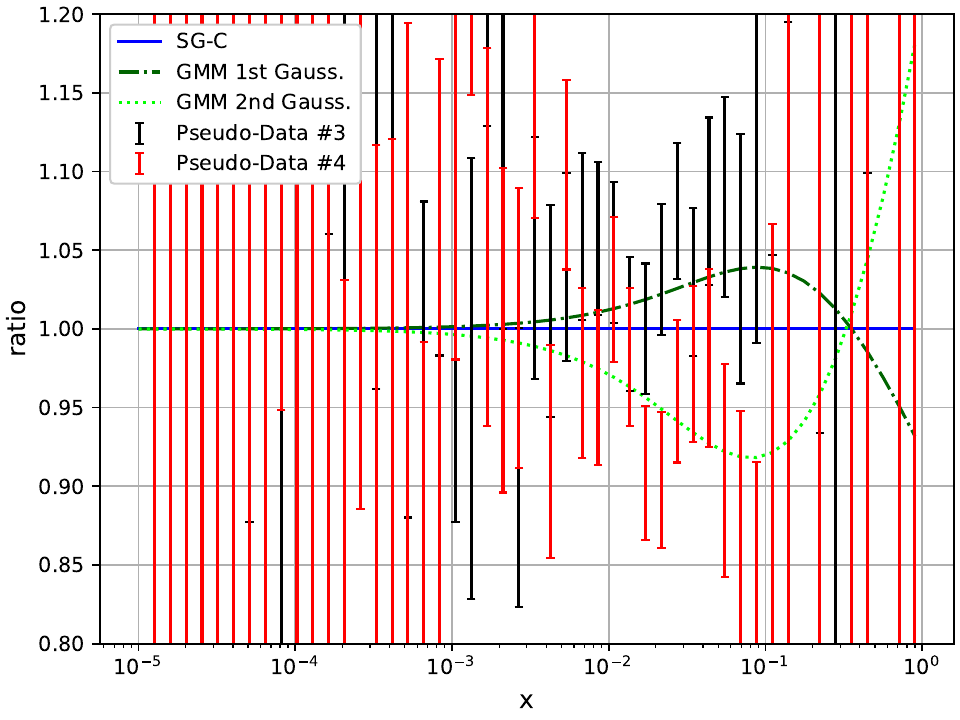}}
\label{fig:GMM_case2:b}}
\subfigure[]
{{\includegraphics[width=0.4\columnwidth]{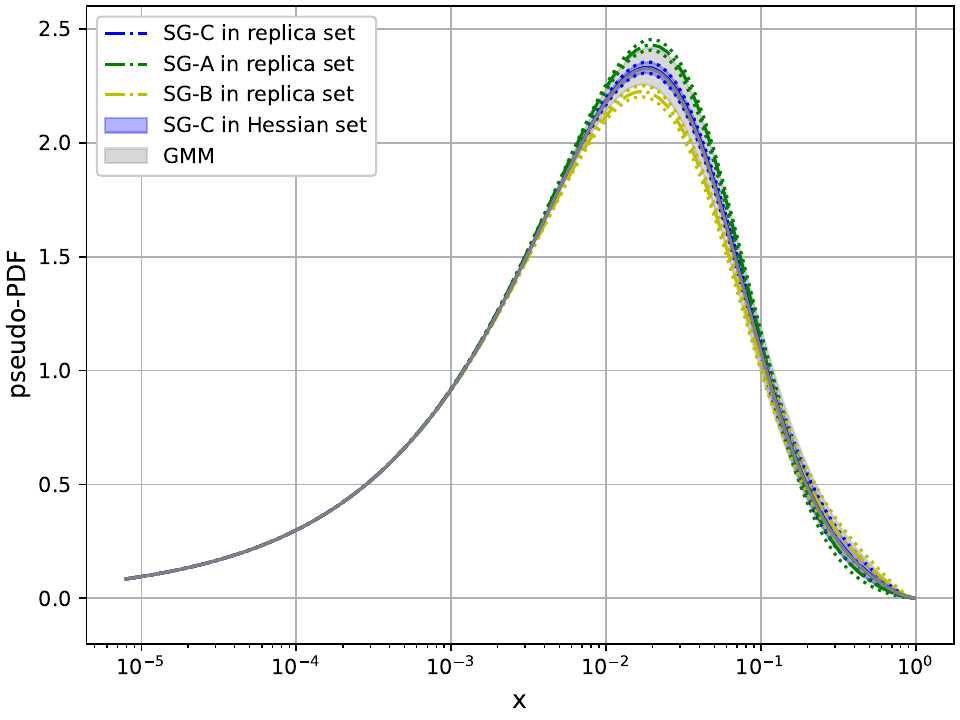}}
\label{fig:GMM_case2:c}}
\subfigure[]
{{\includegraphics[width=0.4\columnwidth]{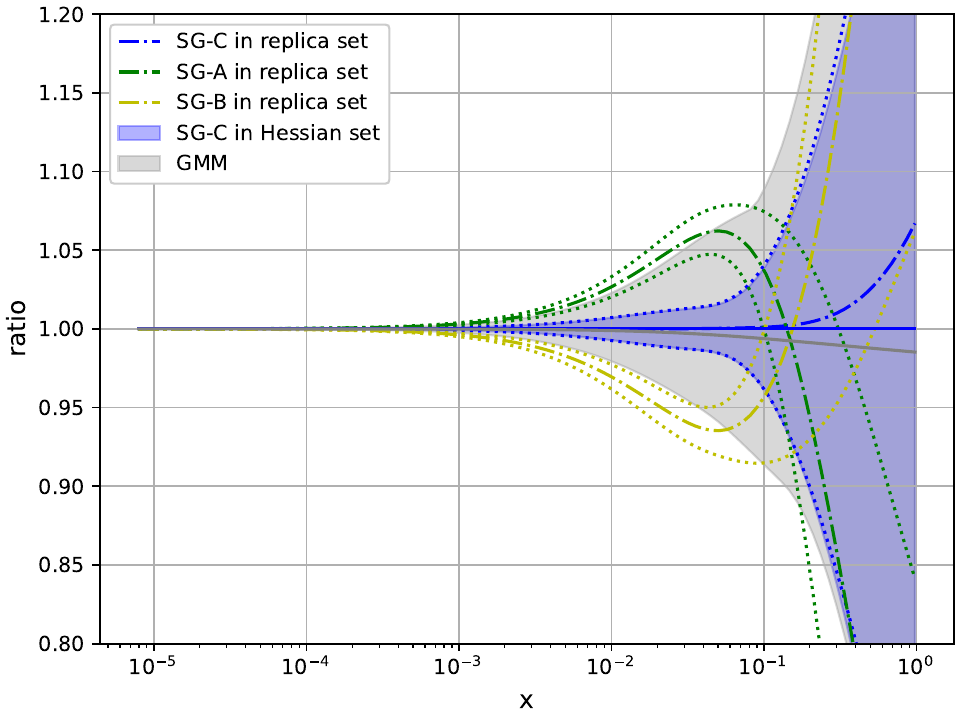}}
\label{fig:GMM_case2:d}}
\subfigure[]
{{\includegraphics[width=0.4\columnwidth]{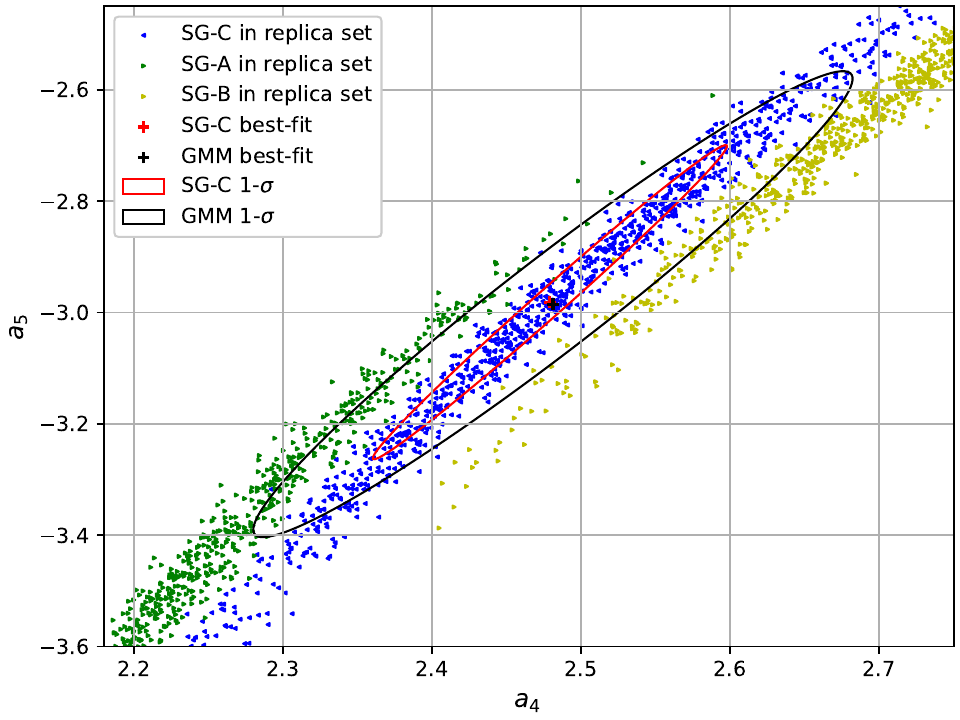}}}
\label{fig:GMM_case2:e}
\caption{\label{fig:GMM_case2} 
Similar to Fig.~\ref{fig:GMM_case1}, but for the Case-2.
} 
\end{figure}

For Case-2, shown in Fig.~\ref{fig:GMM_case2},  the tension in the data set has been reduced by increasing the error on the data, so that SG-C describes the combined pseudo-data set with $\chi^2_{\# 3 + \# 4}/N_{\text{pt}} = 1.04$.
Even for such a mild tension, the fits SG-A and SG-B do not overlap with SG-C. On the other hand, the GMM fit overlaps with both SG-A and SG-B. The difference between the axes orientations of  the correlation ellipses between SG-C and the GMM is small. Although, the GMM method produces a larger uncertainty which is more consistent with SG-A and SG-B, the differences with SG-C are small and we need a way to determine whether we should use SG-C or the GMM method. We will return to a discussion of this in Sec.~\ref{subsec:GMM_overfit}.

\begin{figure}[htbp]  
\centering
\subfigure[]
{{\includegraphics[width=0.4\columnwidth]{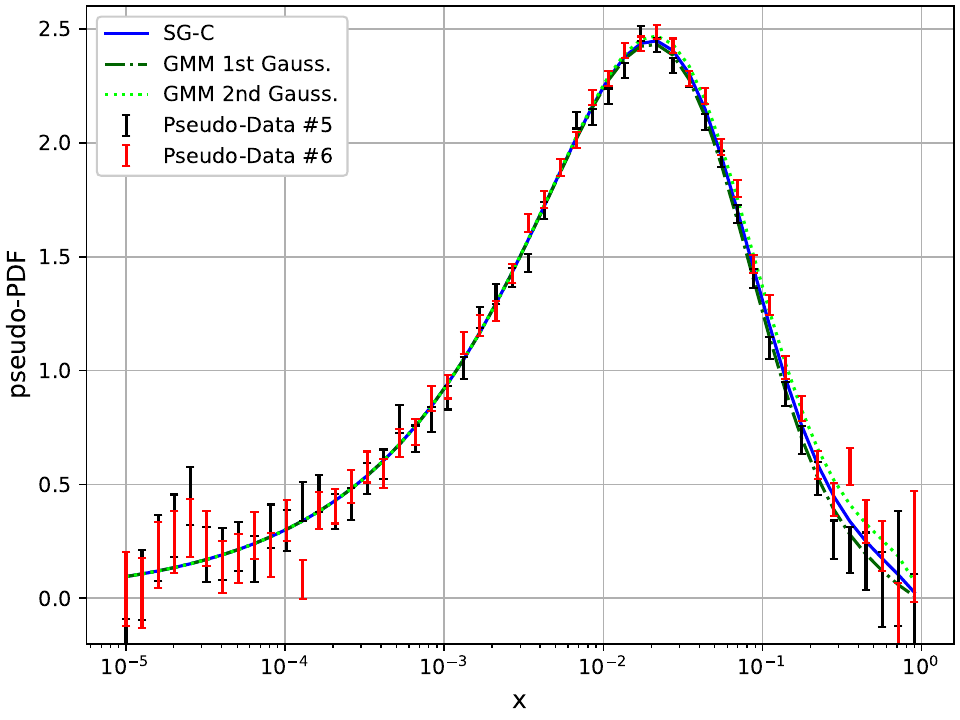}}
\label{fig:GMM_case3:a}}
\subfigure[]
{{\includegraphics[width=0.4\columnwidth]{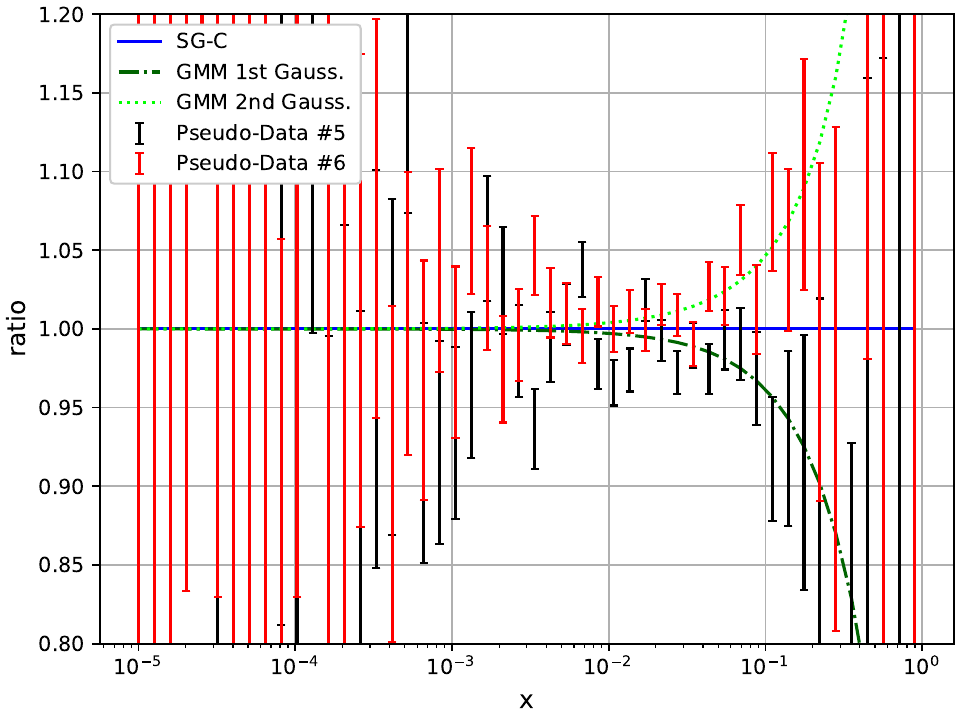}}
\label{fig:GMM_case3:b}}
\subfigure[]
{{\includegraphics[width=0.4\columnwidth]{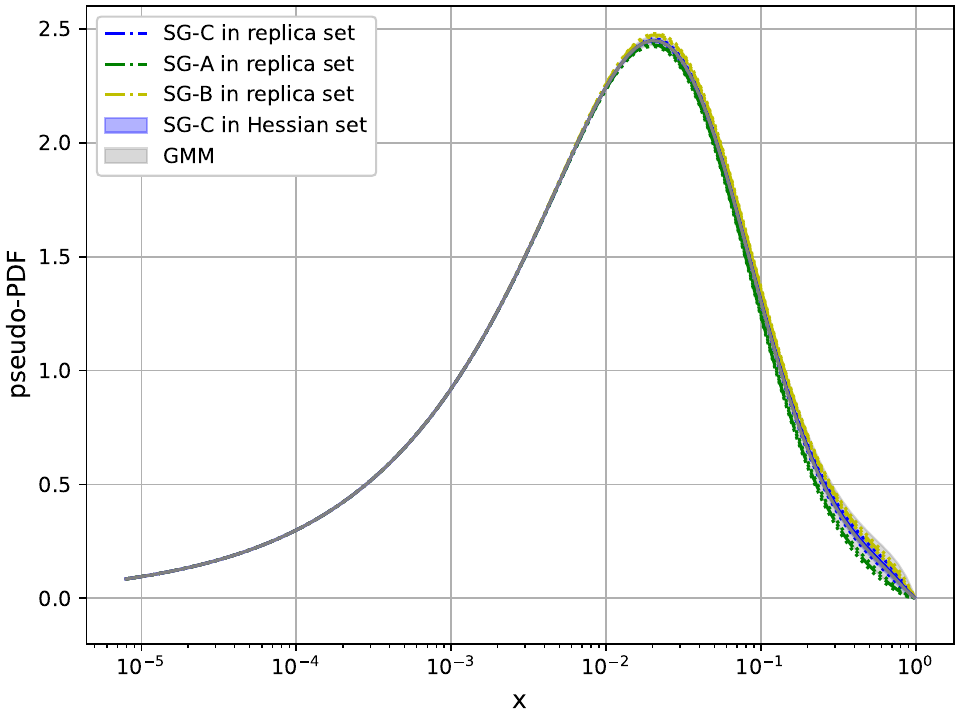}}
\label{fig:GMM_case3:c}}
\subfigure[]
{{\includegraphics[width=0.4\columnwidth]{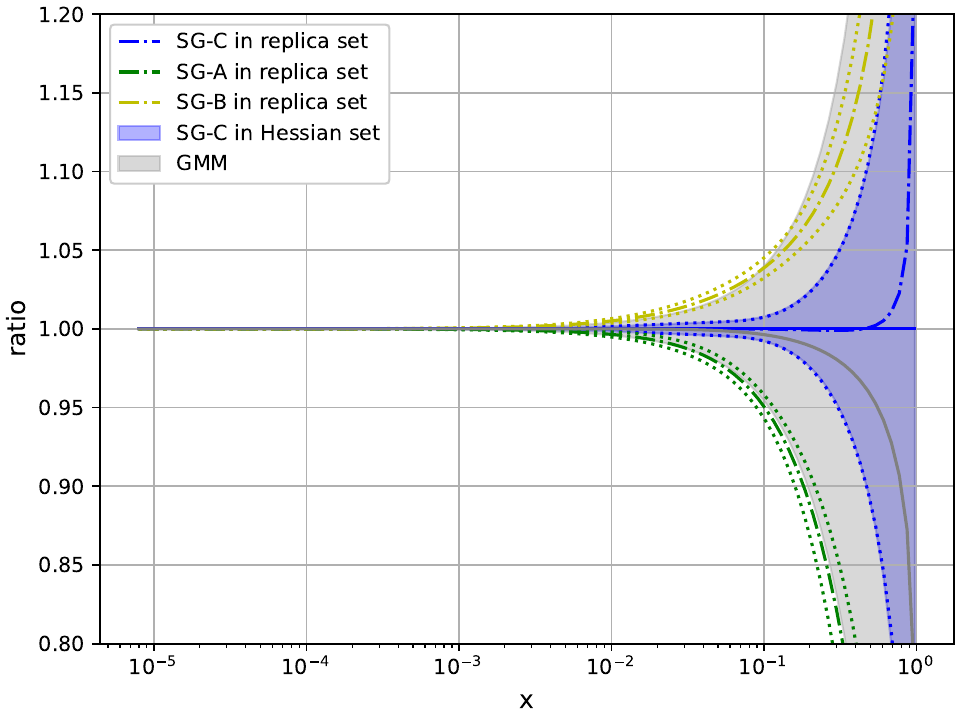}}
\label{fig:GMM_case3:d}}
\subfigure[]
{{\includegraphics[width=0.4\columnwidth]{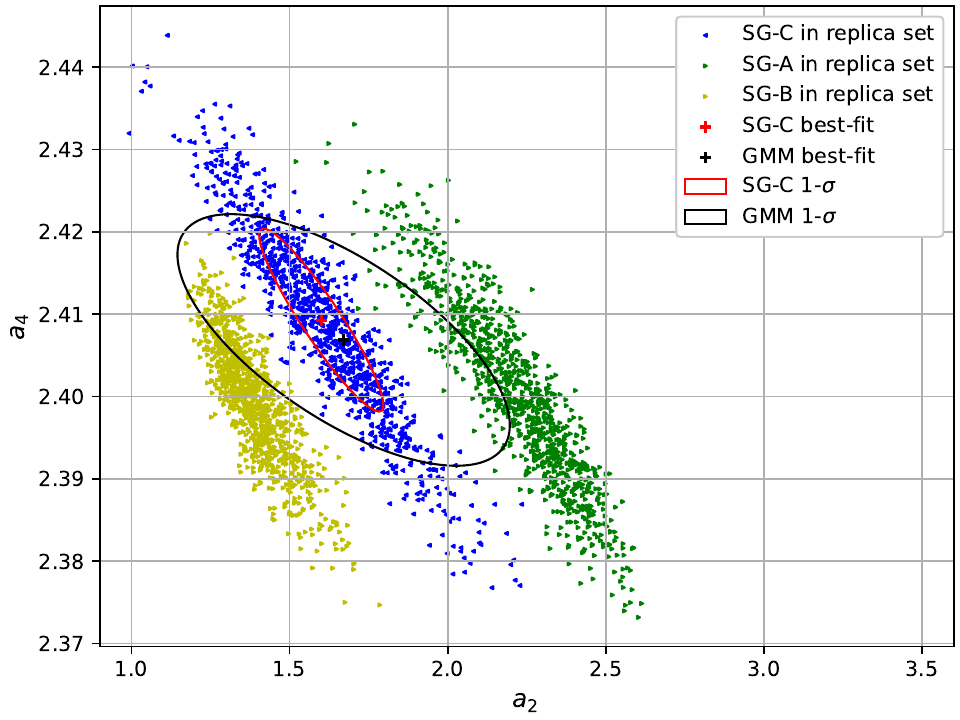}}\label{fig:GMM_case3:e}}
\caption{\label{fig:GMM_case3} 
Similar to Fig.~\ref{fig:GMM_case1}, but for the Case-3.
} 
\end{figure}

The performances for Case-3 of the GMM and LS methods are similar to Case-1. The LS method cannot fit both pseudo-data sets \#~5 and \#~6 well, c.f. the bottom rows of Table~\ref{tab:GMM_best_fits}, even though the uncertainties in both data sets are large. The GMM is able to learn about the tension in these data sets and shows a strong  inclination to two modes.
In terms of the fitted pseudo-PDFs, the GMM in Fig.~\ref{fig:GMM_case3} systematically enlarges the uncertainty and envelopes both data sets \#~5 and \#~6 simultaneously.

\subsection{The GMM method compared to different tolerance criteria}
\label{subsec:tolerance}

Various factors that determine the precision of PDFs include 
theoretical uncertainties, 
tension between experimental data, and the sampling of PDF functional forms and fitting methodologies, often referred to as epistemic uncertainties, c.f. Ref.~\cite{Courtoy:2022ocu}.

Global PDF analysis groups are aware of the shortcomings of using a  fixed tolerance $\Delta \chi^2 = 1$ and have proposed alternate methods. Specifically, CTEQ-TEA~\cite{Hou:2019efy} and MSHT~\cite{Bailey:2020ooq} global PDF analyses, adopt certain tolerance criteria, known as the two-tier tolerance and the dynamic tolerance criteria, respectively, on top of the basic Hessian method. 

In Sec.~\ref{subsec:GMM_vs_chi2}, we compared results of the GMM with the usual LS method of determining 1-$\sigma$ contours by setting $\Delta \chi^2 = 1$.
In this subsection, we compare the GMM results presented in Sec.~\ref{subsec:GMM_vs_chi2} to the Hessian results with the CTEQ-TEA and MSHT tolerance criteria.

We remind the reader that CTEQ-TEA Hessian eigenvector PDFs are constructed using a two-tier prescription~\cite{Lai:2010vv,Gao:2013xoa} that prevents unexpectedly large increments to either the global $\chi^2$ or to the $\chi^2_E$ values of individual experiments. Hence, the CTEQ-TEA PDF error sets are determined using a two-tier (global+dynamic) tolerance ($T$) that accounts for theory, experimental, parametrization, and methodological uncertainties and results in error bands that are
wider than the $\Delta\chi^2=1$ criterion. The CTEQ-TEA Hessian uncertainty, at the 68\% confidence level, nominally corresponds to $T^2 \equiv \Delta \chi^2_{total}\approx 37$, where $\chi^2_{total}$ is the sum of the global $\chi^2$ and the so-called Tier-2 (dynamic) penalty~~\cite{Lai:2010vv,Gao:2013xoa}.
The MSHT PDF fits  by default apply the ``dynamic tolerance'' procedure, as detailed in~\cite{Martin:2009iq,Harland-Lang:2014zoa,Bailey:2020ooq}. This enlarges the uncertainties beyond the $\Delta\chi^2=1$ definition to account for data sets with tension, as well as potential mismatch of data and theory due to imprecise theory or parametrization. 
The impact of different tolerance criteria on the determination of PDF induced uncertainties has been examined in Ref.~\cite{Jing:2023isu}. In this study, we compare the results of applying the CTEQ-TEA and MSHT like tolerance criteria with $T^2 = 37$ to the ones with $\Delta\chi^2 = 1$.

\begin{figure}[htbp]  
\centering
\subfigure[GMM compared to tolerance criteria]
{{\includegraphics[width=0.4\columnwidth]{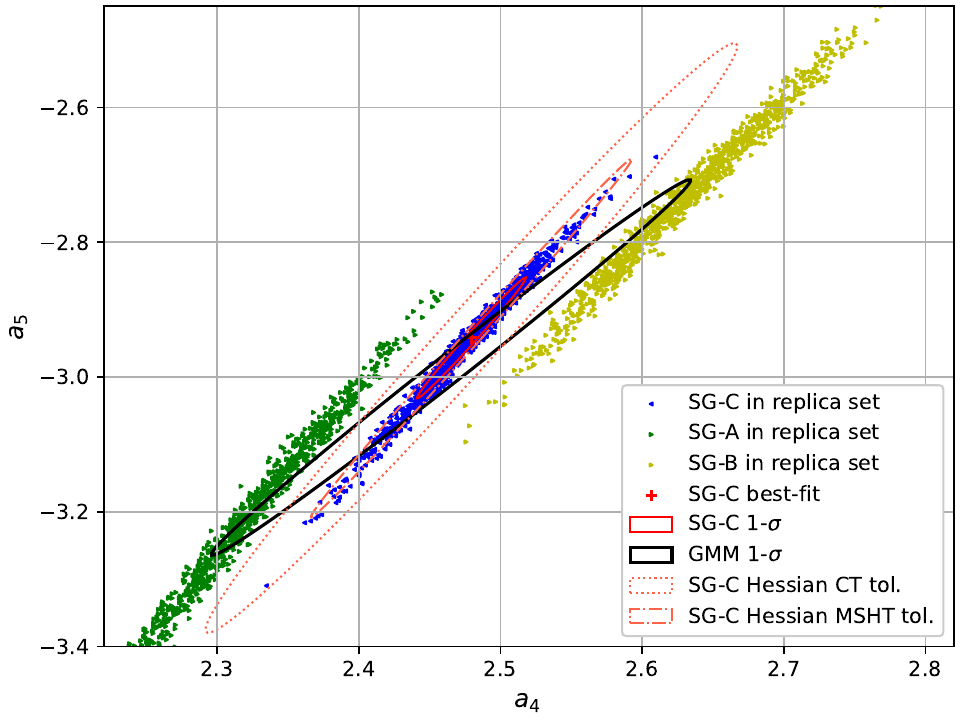}}
\label{fig:GMM_case1_T2:a}}
\subfigure[pseudo-PDF ratio]
{{\includegraphics[width=0.4\columnwidth]{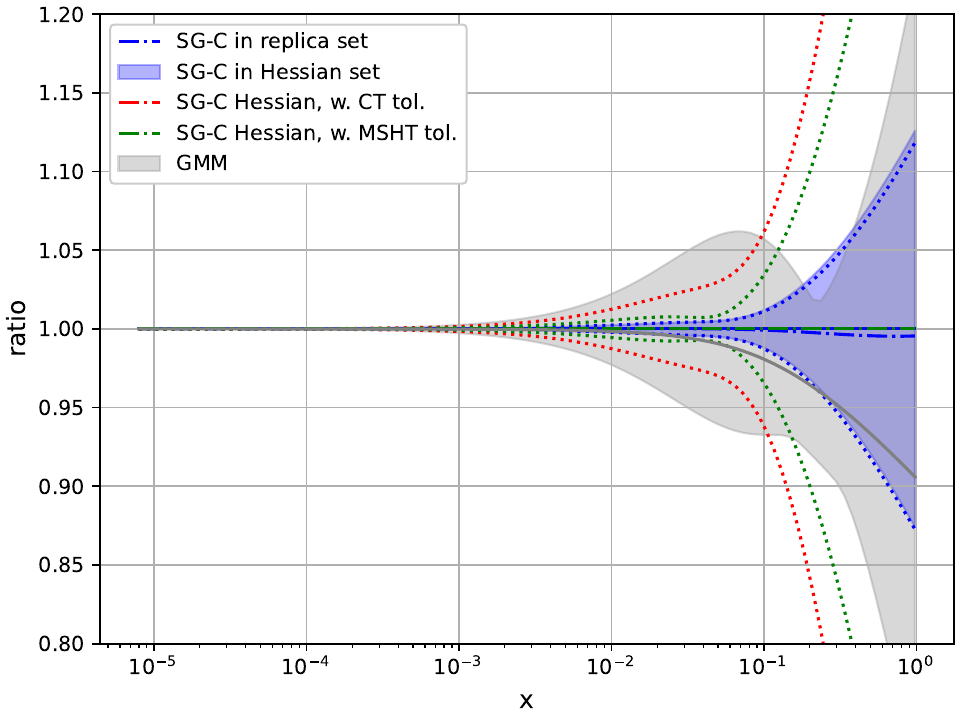}}
\label{fig:GMM_case1_T2:b}}
\caption{\label{fig:GMM_case1_T2} 
The pseudo-PDF uncertainty estimated by the GMM method in Case-1 is compared to pseudo-PDF uncertainties predicted by the LS method along with the CT and MSHT tolerance criteria.
} 
\end{figure}

\begin{figure}[htbp]  
\centering
\subfigure[GMM compared to tolerance criteria]
{{\includegraphics[width=0.4\columnwidth]{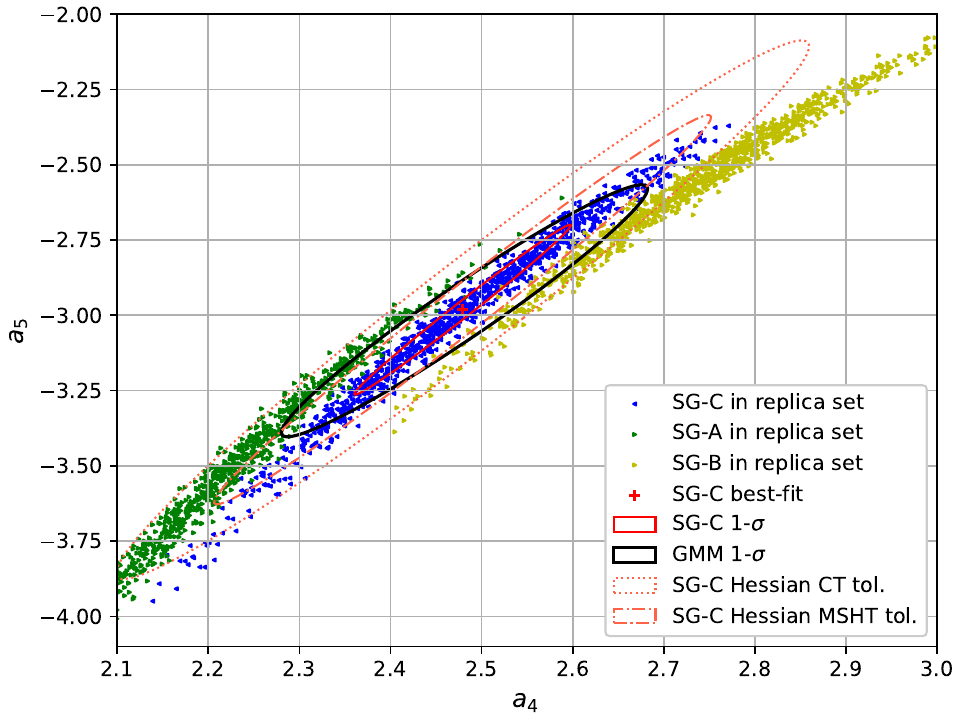}}
\label{fig:GMM_case2_T2:a}}
\subfigure[pseudo-PDF ratio]
{{\includegraphics[width=0.4\columnwidth]{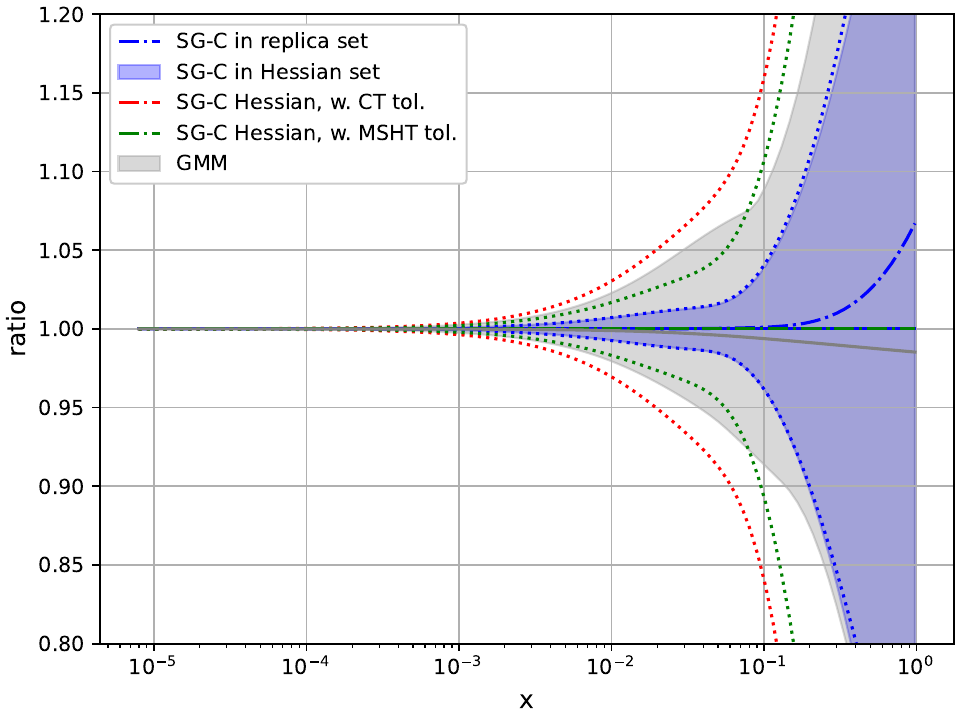}}
\label{fig:GMM_case2_T2:b}}
\caption{\label{fig:GMM_case2_T2} 
Similar to Fig.~\ref{fig:GMM_case1_T2}, but for the Case-2.
} 
\end{figure}

\begin{figure}[htbp]  
\centering
\subfigure[GMM compared to tolerance criteria]
{{\includegraphics[width=0.4\columnwidth]{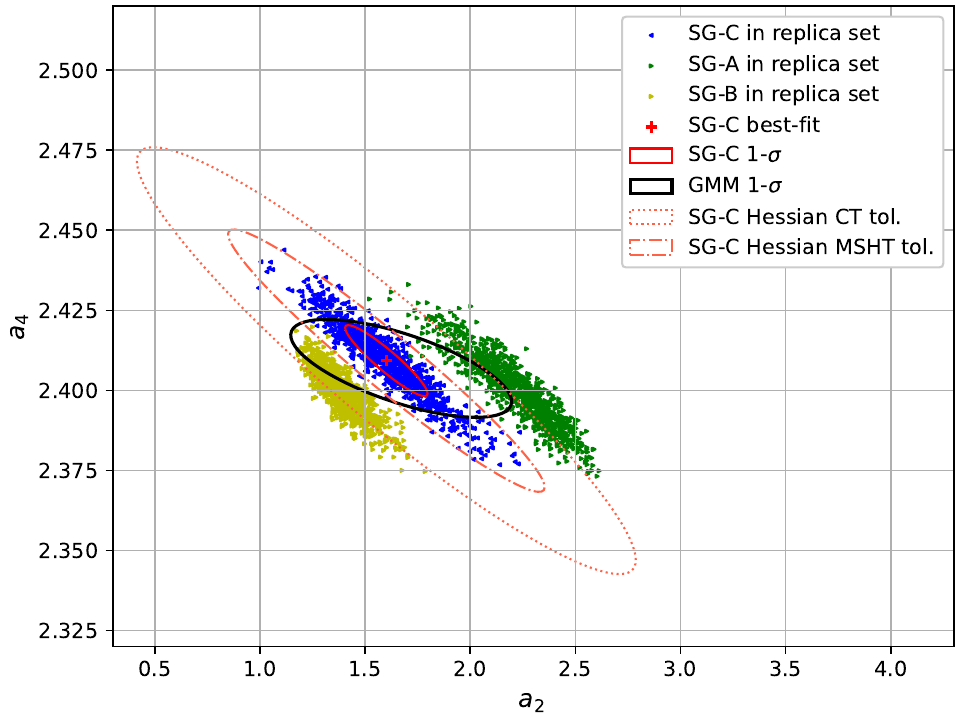}}
\label{fig:GMM_case3_T2:a}}
\subfigure[pseudo-PDF ratio]
{{\includegraphics[width=0.4\columnwidth]{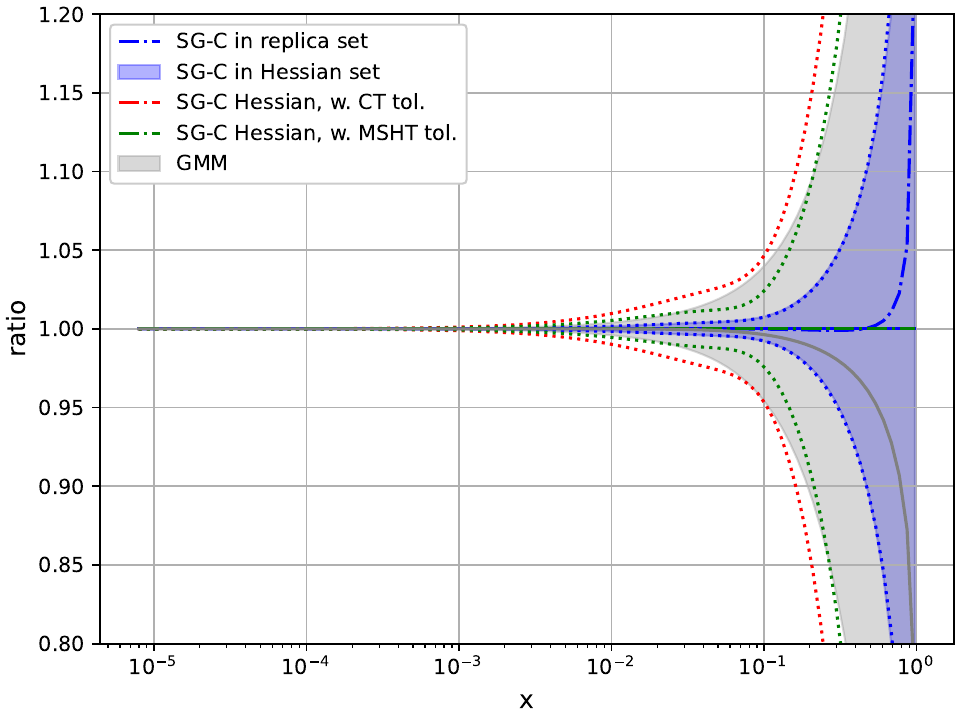}}
\label{fig:GMM_case3_T2:b}}
\caption{\label{fig:GMM_case3_T2} 
Similar to Fig.~\ref{fig:GMM_case1_T2}, but for the Case-3.
} 
\end{figure}

The Hessian uncertainties based on the single Gaussian LS method  with a tolerance criteria  of  $T^2 = 37$ for all three cases are compared to the GMM method in Figs.~\ref{fig:GMM_case1_T2} to~\ref{fig:GMM_case3_T2}.
\begin{itemize}
\item Case-1:
 Both Hessian sets with CTEQ and MSHT tolerance criteria estimate a smaller pseudo-PDF uncertainty for $10^{-3} < x < 0.1$, compared to the GMM. Recall for Case-1, the GMM produced a more faithful representation of uncertainties and overlapped with SG-A and SG-B. We see that the alternate tolerance criteria is not able to account for the larger uncertainties in this $x$ range. Further, the uncertainties given by the modified tolerance criteria are much larger than required when $x>0.1$.  The same can also be observed in the fit parameter space. Here, the SG-C correlation ellipses with tolerance cover larger regions than the default Hessian set, but still cannot predict a correlation that spans regions of parameter space that cover portions of SG-A and SG-B.
\item Case-2:
As shown in  Fig.~\ref{fig:GMM_case2_T2}, the larger tolerance leads to an overestimation of pseudo-PDF uncertainty, particularly for $x>0.1$. In fact, the correlation ellipse for the larger tolerance, seen in the fit parameter space, has regions that are not populated by any of the replica sets.
\item Case-3:
As shown in Fig.~\ref{fig:GMM_case3_T2:b}, the GMM uncertainty (gray shaded regions of the pseudo-PDF ratios) shows better agreement with the uncertainty bands of the higher tolerance criteria. However, when comparing results in the fit parameter space, shown in  Figs.~\ref{fig:GMM_case3_T2:a}, there appears to be significant mismatch between  the higher tolerance ellipse and the GMM ellipse.
\end{itemize}
	
The tolerance criteria, currently in use by PDF collaborations, was introduced to capture larger uncertainties when there is tension in the fits. However, these methods are still based on the assumption that the likelihood behaves like a single Gaussian and is uni-modal. The tolerance criteria are not able to faithfully capture the effect of tension in the data. In some regions, the higher tolerance underestimates the uncertainty, and in others it overestimates the uncertainty. The GMM, on the other hand, is able to provide a more faithful representation of the combined likelihoods of two data sets with tension.

\section{Consistency checks for the GMM}
\label{sec:consistency}
In Sec.~\ref{sec:result_GMM}, we demonstrated how the GMM with two base Gaussian distributions provides a better description of the combined likelihood of two pseudo-data sets with tension and provides more accurate uncertainties. 
In this section, we show how the GMM reduces to the usual single Gaussian LS method when data  are consistent. Further, as with any present day machine learning technique, when implementing the GMM, one must contend with over-fitting. In this section, we also remark on the consequences of over-fitting with the GMM.

\subsection{Consistent results from consistent data}
\label{subsec:consistent_GMM_chi2}
In Sec.~\ref{sec:GMM}, we have shown how the GMM likelihood function reduces to the LS method when the number of Gaussians used in the GMM is  $K=1$. Another possibility for reducing to the usual $\chi^2$ fit is when all base Gaussians end up with the same mean and covariance, i.e. 
\begin{equation}
 \mathbb{E}[\theta] \sim \hat{\theta}_i \sim \hat{\theta}_k, \quad \text{cov}_{\text{GMM},i} \sim \text{cov}_{\text{GMM},k}, \quad \text{and} \quad \mathcal{N}(y_j, \Delta y_j|\theta_i) \sim \mathcal{N}(y_j, \Delta y_j|\theta_k). \nonumber
\end{equation}
Consequently, the second term in Eq.~\eqref{eq:cov_GMM_2}, which accounts for the difference of means, vanishes. 
In this scenario, the covariance of the GMM reduces to
\begin{eqnarray}
  \text{cov}_{\text{GMM}} &\simeq& \sum_{i=1}^K \omega_i \Bigg{(} 
  \sum_{j=1}^{N_{\text{pt}}} \left[ - 
 \frac{1}{\pi_j} \frac{\partial^2 \mathcal{N}_{ij}}{\partial \theta_i^2}  + \frac{\omega_i }{\pi_j^2} \left(\frac{\partial \mathcal{N}_{ij}}{\partial \theta_i}\right)^2 
 \right]\Bigg{)}^{-1}+ \sum_{i=1}^K \omega_i (\mathbb{E}[\theta] - \hat{\theta}_i )^2 \nonumber \\
  &\simeq& \bigg{(} \sum_{j=1}^{N_{\text{pt}}} \frac{1}{\Delta y^2_j} \bigg{)}^{-1},
  \label{eq:GMM_chi2_consistent}
\end{eqnarray}
which is the covariance found by the LS method in Eq.~\eqref{eq:cov_chi2}. 
When fitting a consistent data set, consistency between the  GMM method and LS method are expected as shown analytically in Eq.~\eqref{eq:GMM_chi2_consistent}. In practice, due to overfitting, we will see that the GMM reduces to the LS method only if no local minima exist in the log-likelihood, i.e. there is only one minimum of the log-likelihood.

\begin{figure}[htbp]  
\centering
\subfigure[Truth for Case-4]
{{\includegraphics[width=0.4\columnwidth]{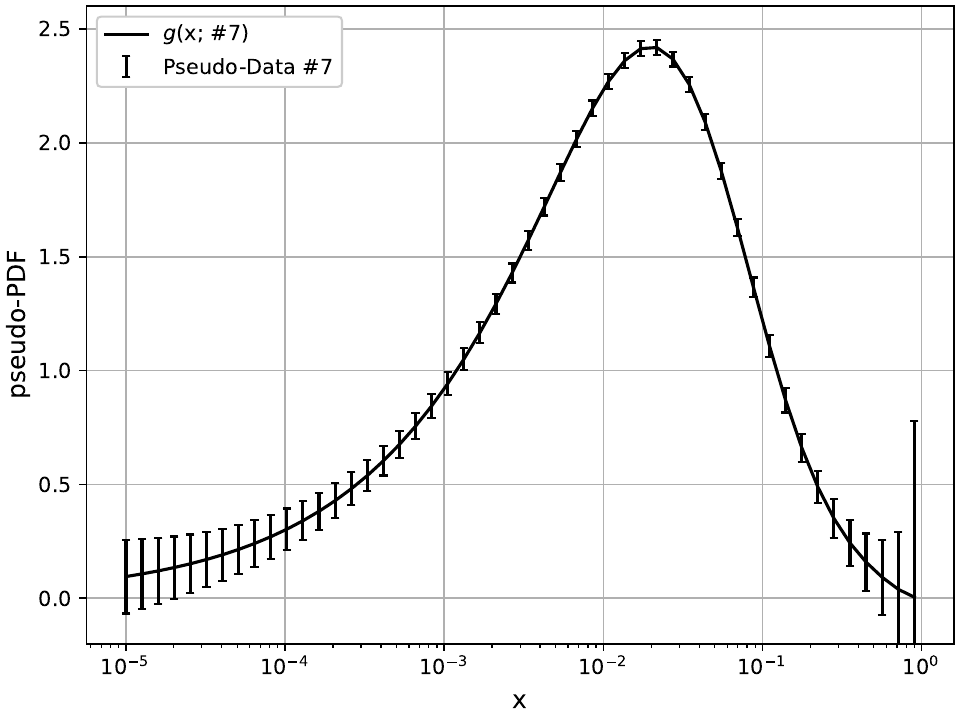}}
\label{fig:consistent_0:a}}
\subfigure[Truth for Case-5]
{{\includegraphics[width=0.4\columnwidth]{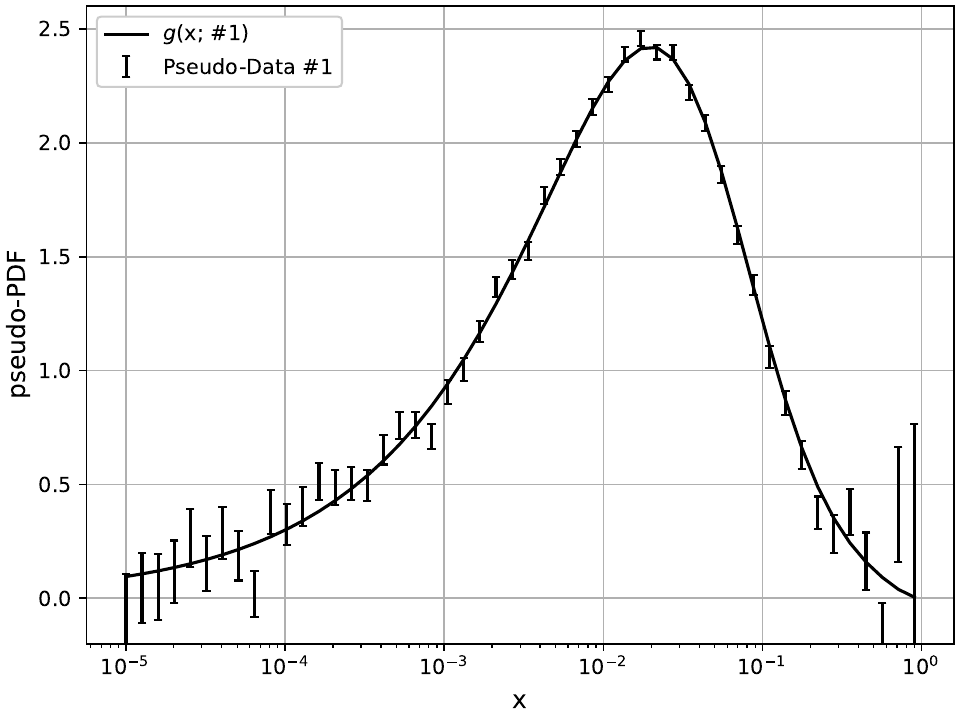}}
\label{fig:consistent_0:b}}
\subfigure[best-fits for Case-4]
{{\includegraphics[width=0.4\columnwidth]{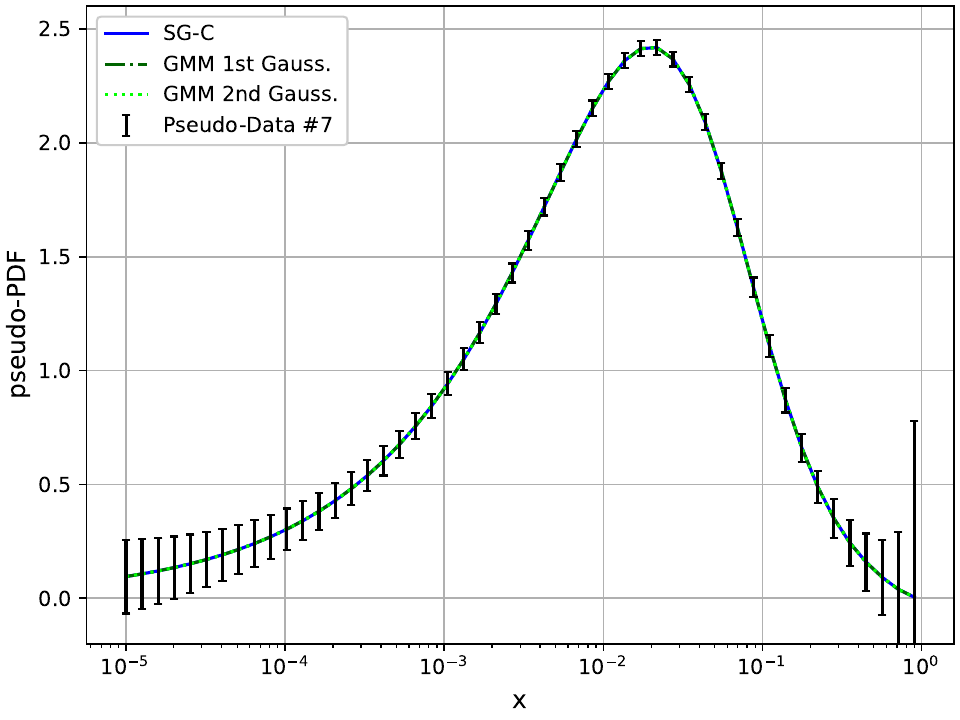}}
\label{fig:consistent_0:c}}
\subfigure[best-fits for Case-5]
{{\includegraphics[width=0.4\columnwidth]{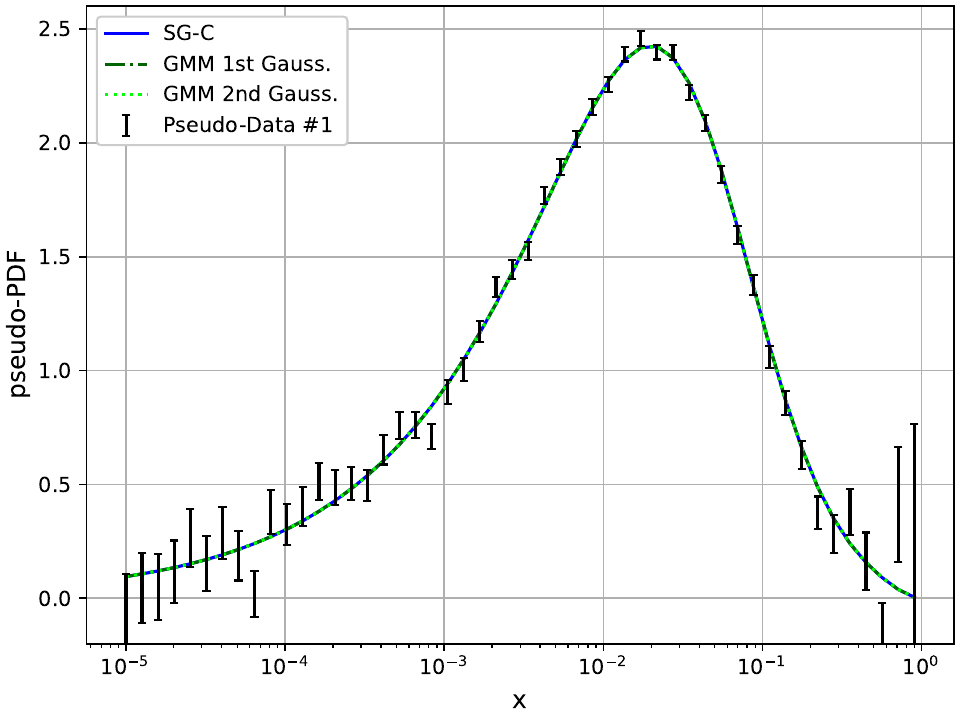}}
\label{fig:consistent_0:d}}
\caption{\label{fig:consistent_0} 
Upper panels: the truths are compared to pseudo-data for Case-4 and Case-5.
Lower panels: the best-fit pseudo-PDFs are compared to pseudo-data for both cases.
}
\end{figure}

Here, we verify Eq.~\eqref{eq:GMM_chi2_consistent} by performing two more fits on pseudo-data sets, as shown in Fig.~\ref{fig:consistent_0}:
\begin{itemize}
 \item Case-4: This case concerns a fit to the pseudo-data $\#~7$,
  
 which is the same as pseudo-data $\#~1$, but without any statistical fluctuations. 
 
 In this case, only a single extremum of the log-likelihood of the GMM exists, i.e. $\theta_i = \theta_k$, and we shall see that numerically as well, the GMM  reduces to the LS method fit.
 \item Case-5:  Here, we fit only to a single data set, namely pseudo-data set $\#~1$. Due to overfitting, we will see that although $\theta_i \simeq \theta_k$, the GMM tends to find multiple possible values of $\theta_k$. This in turn implies that the results from the  GMM method are different from the LS method.
\end{itemize}

\begin{figure}[htbp]  
\centering
\subfigure[Correlations for Case-4]
{{\includegraphics[width=0.4\columnwidth]{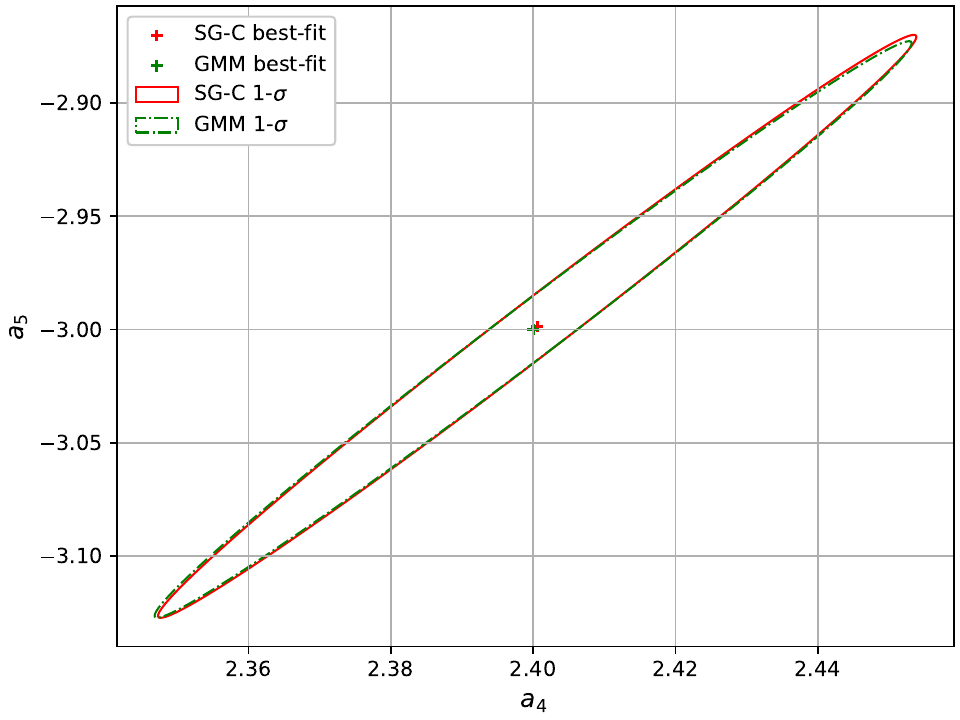}}
\label{fig:consistent_1:a}}
\subfigure[Correlations for Case-5]
{{\includegraphics[width=0.4\columnwidth]{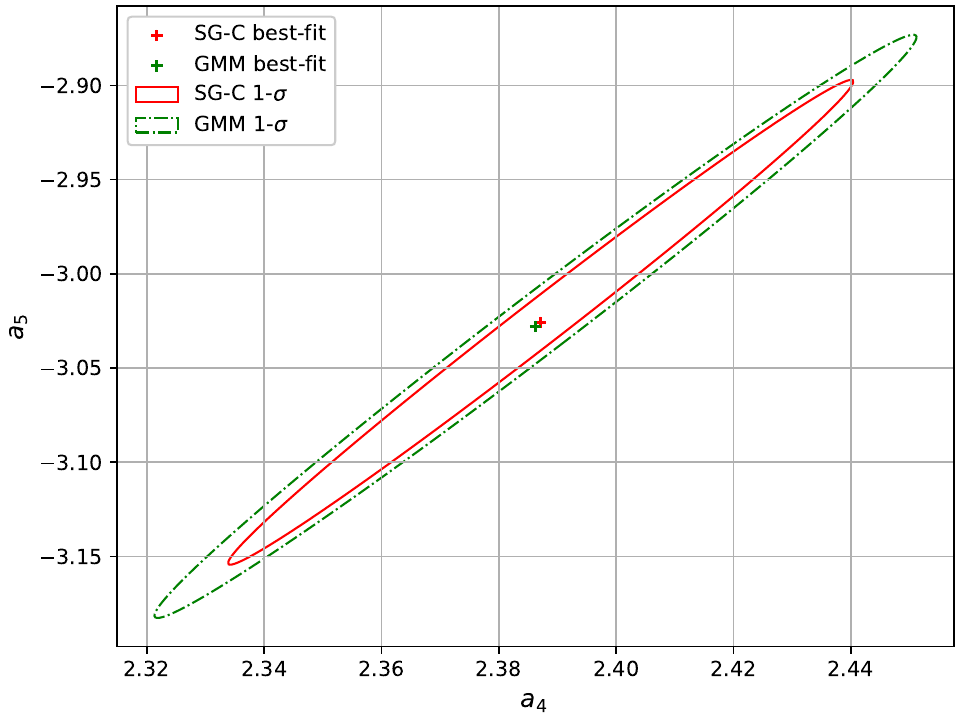}}
\label{fig:consistent_1:b}}
\subfigure[Correlations for Case-4]
{{\includegraphics[width=0.4\columnwidth]{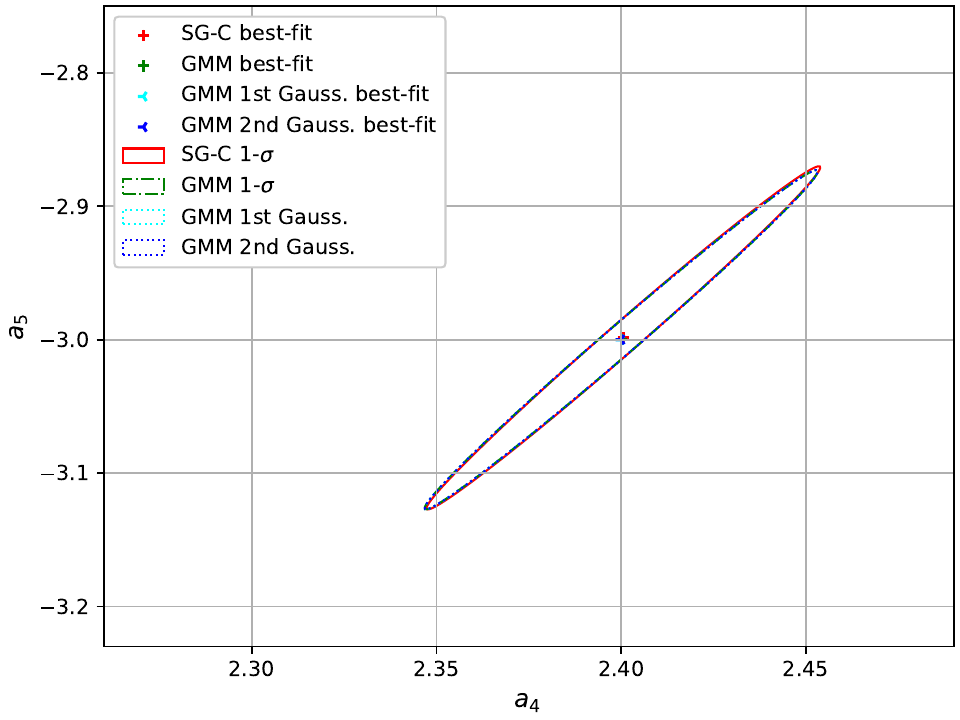}}
\label{fig:consistent_1:c}}
\subfigure[Correlations for Case-5]
{{\includegraphics[width=0.4\columnwidth]{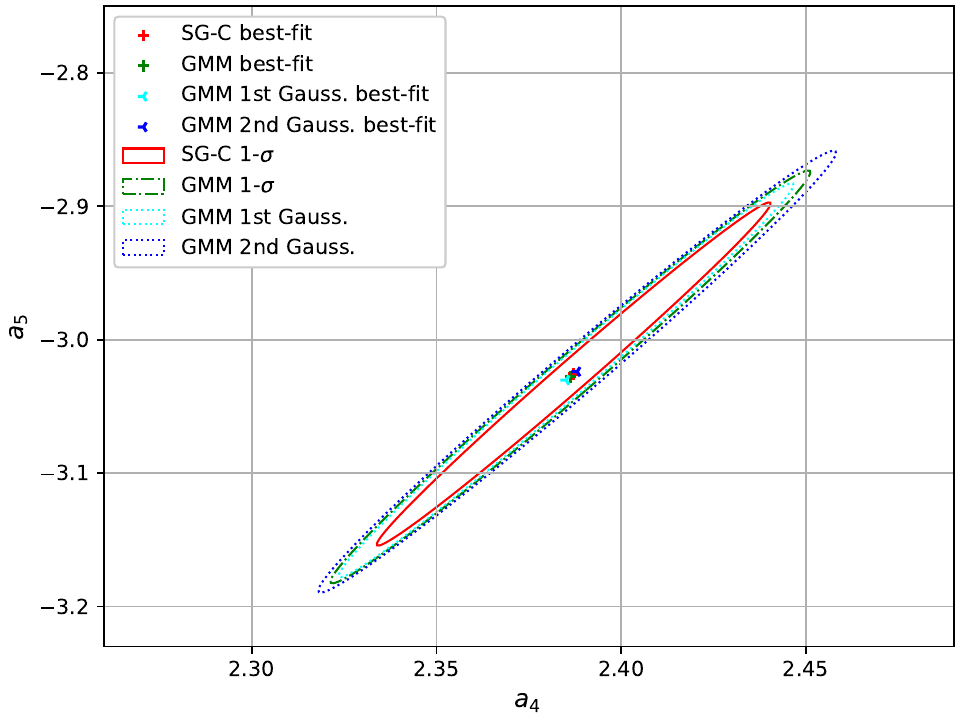}}
\label{fig:consistent_1:d}}
\caption{\label{fig:consistent_1} 
Upper panels: the comparison of covariances obtained by the GMM method and the LS method, for the Case-4 and Case-5.
Lower panels: similar to upper panels, but include the covariances of individual base Gaussian distributions of the GMM.
} 
\end{figure}

\begin{figure}[htbp]  
\centering
\subfigure[Pseudo-PDF ratio for Case-4]
{{\includegraphics[width=0.4\columnwidth]{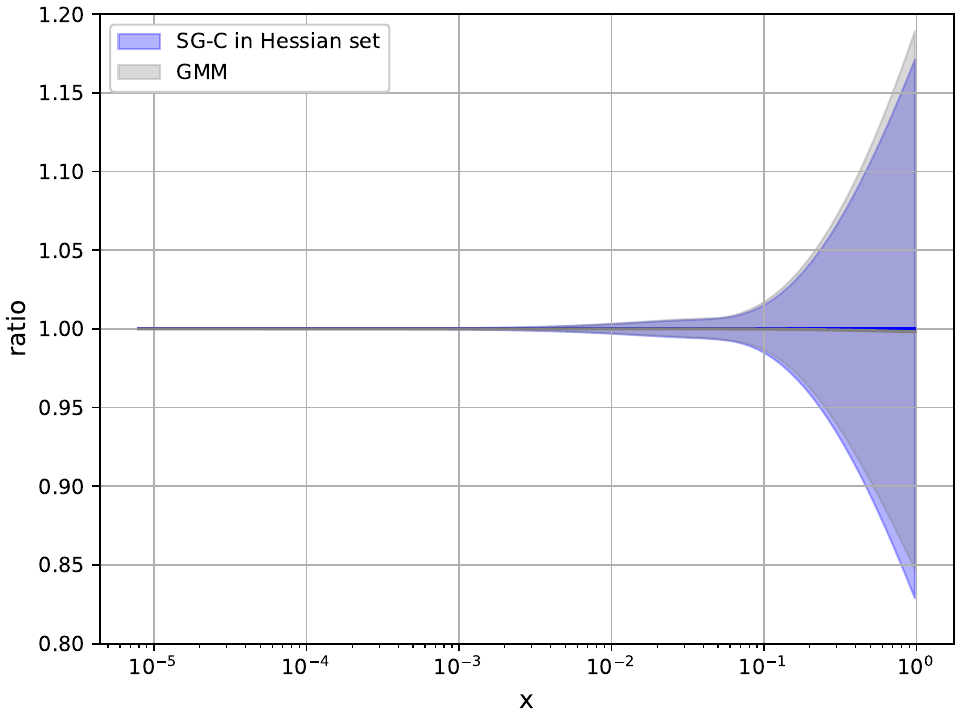}}
\label{fig:consistent_2:a}}
\subfigure[Pseudo-PDF ratio for Case-5]
{{\includegraphics[width=0.4\columnwidth]{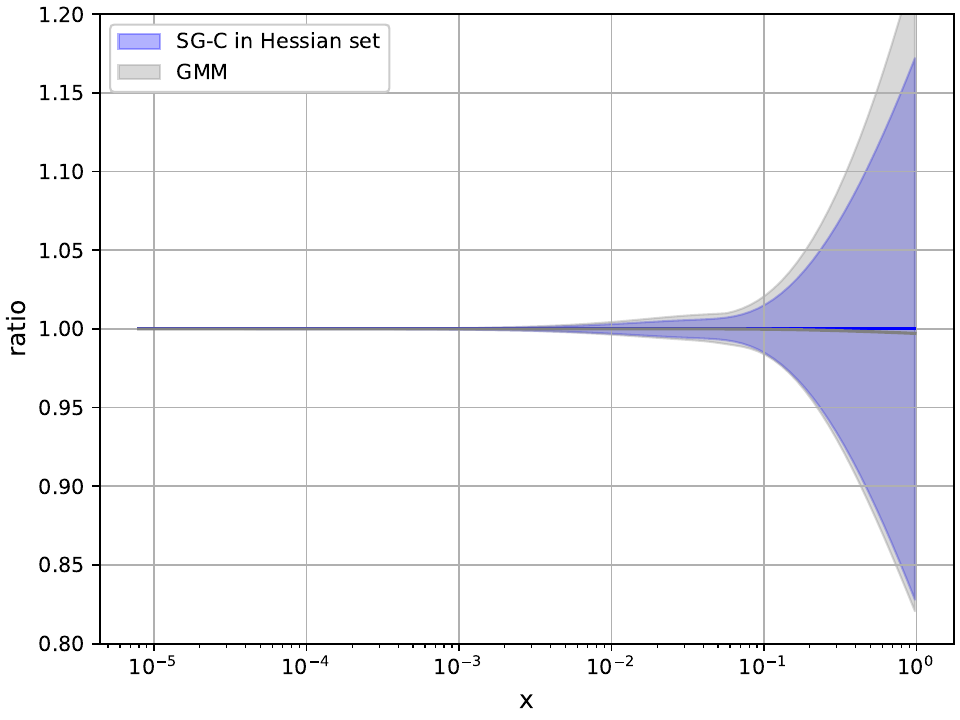}}
\label{fig:consistent_2:b}}
\subfigure[Size of pseudo-PDF uncertainty for Case-4]
{{\includegraphics[width=0.4\columnwidth]{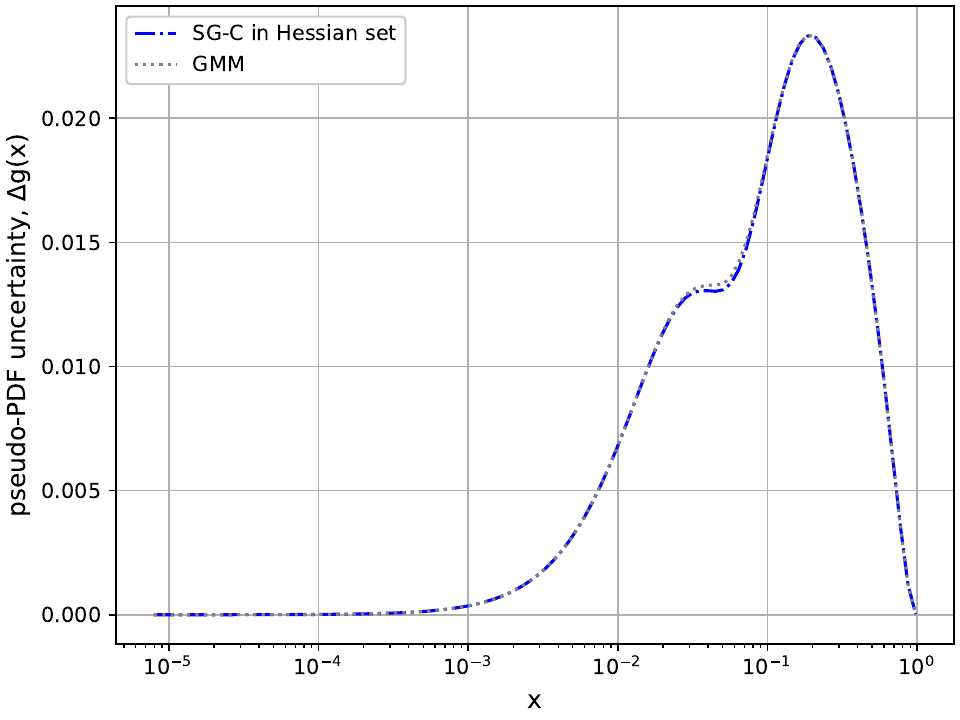}}
\label{fig:consistent_2:c}}
\subfigure[Size of pseudo-PDF uncertainty for Case-5]
{{\includegraphics[width=0.4\columnwidth]{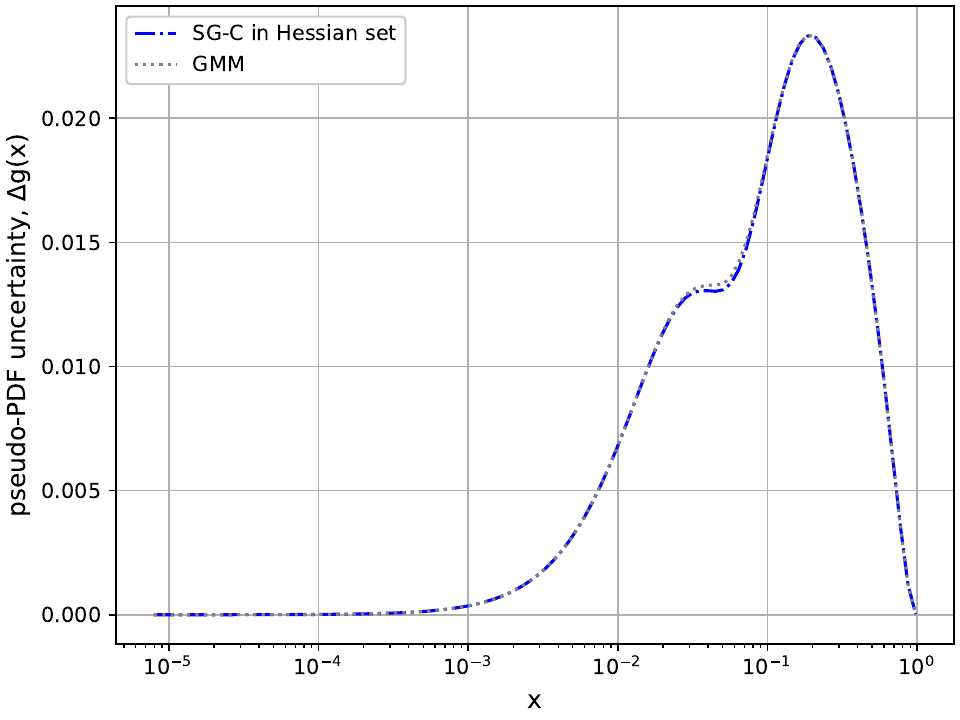}}
\label{fig:consistent_2:d}}
\caption{\label{fig:consistent_2} 
Upper panels: the comparison of pseudo-PDF ratio obtained by both methods for Case-4 and Case-5, given the reference being the central value obtained by the LS method.
Lower panels: the comparison of the sizes of pseudo-PDF uncertainty obtained by the GMM method and the LS method, for  Case-4 and Case-5.
}
\end{figure}

We begin by looking at results from the fit in Case-4. 
The uncertainties predicted by both methods are shown in Figs.~\ref{fig:consistent_1:a},~\ref{fig:consistent_1:c},~\ref{fig:consistent_2:a}, and~\ref{fig:consistent_2:c}. The figures indicate, as expected, that the central value of the fit for both the GMM and the LS methods are in good agreement. In Fig.~\ref{fig:consistent_1:b} and Fig.~\ref{fig:consistent_1:c}, we see that the ellipses from the covariance matrix estimated using both the GMM and the LS methods overlap and are identical. On the other hand, for Case-5, we notice in Fig.~\ref{fig:consistent_1:b} and  Fig.~\ref{fig:consistent_1:d}, that the covariance predicted by the GMM method $\text{cov}_{\text{GMM}}$ is marginally larger than the covariance given by the LS method $\text{cov}_{\chi^2}$. 
Fig.~\ref{fig:consistent_1:d} shows that the covariance of individual base Gaussian distributions of the GMM $\text{cov}_{\text{GMM},i}$, which is larger than that obtained by the LS method $\text{cov}_{\chi^2}$, and that the optimal values of parameters $\hat{\theta}_i$ for both base Gaussian distributions of the GMM differ by a small amount. The overestimation of uncertainty is a result of overfitting in GMM, where the GMM ends up learning statistical fluctuations rather than real tension between data sets.

The GMM method looks for tension and tries to categorize data into two or more different Gaussians. In general, for the GMM, overfitting gives rise to larger uncertainties and in that sense the GMM always provides a conservative estimate of the uncertainty.
However, one would like to provide a precise PDF which is essential for the physics goals of the community. In order to determine if there is overfitting in the GMM, we describe certain information criteria in the next section that can be used to determine the optimal number of Gaussians, {\it i.e.,} the value of $K$.
The result of Case-5 suggest that applying the GMM to a consistent data set will lead to larger uncertainties, which although conservative, should also be avoided.

\subsection{Determining the Number of Gaussian Modes Using Information Criteria}
\label{subsec:GMM_overfit}
The results shown in Sec.~\ref{subsec:consistent_GMM_chi2} indicate that the GMM method overestimates the uncertainty for consistent data sets, especially if one includes more base Gaussian distributions than needed. 

Although the GMM automatically clusters data into sets with tension and those that are consistent, a method to quantify the level of consistency and to determine the optimal number of base Gaussian distributions ($K$) required to avoid overfitting and therefore overestimation of uncertainties.

Here, we propose the use of the Akaike information criterion (AIC)~\cite{Akaike} and Bayesian information criterion (BIC)~\cite{BIC} as indicators that will help determine optimal values for $K$, {\it i.e.,} the number of Gaussians, and as a test of the consistency of fits. 

The values of AIC and BIC, defined in Eq.~\eqref{eq:AIC_BIC}, increase as more parameters are introduced. They also take into account the change in goodness of fit and reduce as the fit improves. Both AIC and BIC are thus able to account for the trade-off between introducing additional parameters and an improvement to the fit. 
The smallest values of the AIC and BIC indicate to us the best choice of parameters, in this case, the best choice for the value of $K$. The AIC and BIC are defined as 
\begin{eqnarray}
 \text{AIC} &=& N_{p} \ \ln N_{\text{pt}} - 2 \ln L \big{|}_{\theta=\hat{\theta}}, \nonumber \\
 \text{BIC} &=& 2 N_{p} - 2 \ln L \big{|}_{\theta=\hat{\theta}}.
 \label{eq:AIC_BIC}
\end{eqnarray}
We count the number of independent parameters $N_{p}$ for the GMM as following,
\begin{equation}
 N_{p} = N_{\text{parm}}^{\text{GMM}} = K \times N_{\text{parm}
 } + ( K - 1 ),
 \label{eq:N_parm}
\end{equation}
where $N_\text{parm}$ is the number of independent parameters for each  base Gaussian distribution. The first term in Eq.~\eqref{eq:N_parm} is the total number of fitting parameters over $K$ copies of base distributions. The second term corresponds to the number of independent weights.
In our pseudo-data study, there are always two free variables in the pseudo-PDF parametrization. Hence the total number of independent parameters for the GMM method in our current fits is
\begin{equation}
 N_{\text{parm}}^{\text{GMM}} \bigg|_{N_{\text{parm}}=2} = 2 K + ( K - 1 ).
 \label{eq:N_parm_used}
\end{equation}

We scan the AIC and BIC of the GMM up to $K=4$ for all 5 cases considered in this work, as summarized in Table~\ref{tab:AIC_BIC}.
For  Case-1 and Case-3, both AIC and BIC suggest two base Gaussian distributions are optimal. This confirms the results shown in Sec.~\ref{sec:result_GMM}, 
particularly in Figs.~\ref{fig:GMM_case1:e} 
and~\ref{fig:GMM_case3:e}, where two base Gaussian distributions can estimate a correlation that covers the two pseudo-data sets in tension.
The AIC and BIC suggest a single Gaussian distribution suffices for Case-2. In Fig.~\ref{fig:GMM_case2:e}, the LS method predicts a correlation, whose slope is close to the correlation estimated with two base Gaussian distributions. This is consistent with the fact that Case-2 has minimal tension due to the large values of data errors. At this point we would like to point out that the GMM in conjunction with AIC and BIC, can be used to determine how much the errors in the data should be inflated in order to make them consistent so that we can use the LS method, as suggested in Ref.~\cite{Kovarik:2019xvh, Erler:2020bif, ParticleDataGroup:2022pth, Cowan:2018lhq, DAgostini:1999niu}.\footnote{ Specifically, one possible measure is to increase the errors in the data sets until the AIC and BIC indicate that $K=1$ is the optimal value of $K$. This provides an alternate way to determine the scale factor used in Ref.~\cite{ParticleDataGroup:2022pth}, c.f. Appendix~\ref{sec:app:W_boson}. }

For Case-4 and Case-5, the AIC and BIC, as expected, suggest that there is no need of using $K >1 $, since by doing so one is over-fitting the data. We reiterate that the consequence of over-fitting is to increase our uncertainty and therefore leads to a conservative estimate of uncertainty, {\it i.e.,} using the optimum value of $K$
provides the most precise PDFs, whereas larger values of $K$ will lead to PDFs with larger uncertainty. We also notice in Table~\ref{tab:AIC_BIC} that a higher number of base distributions $K$ sometimes leads to equal or slightly lower values of $-\ln L$.

    Thus, if we were to only use $-\ln L$ as the measure of the goodness of fit, and not use the AIC and BIC, we would end up overfitting and choosing larger values of $K$.

To conclude, although the GMM method is expected to reduce to the LS method for a consistent data set, one should avoid using it without detailed assessment of the level of tension. The combination of AIC and BIC indicators with the generalized statistical model of the GMM method could effectively reproduce the correct correlation in the presence of inconsistency in data. At the same time we can determine optimal values of $K$ and provide a formalism to determine more precise PDFs.

\begin{table}[htpb]
\begin{center}
\begin{tabular}{cc|cccc}
  & & $K=1$ & $K=2$ & $K=3$ & $K=4$ \\ \hline \hline
 Case-1 & AIC & -102.2 & \textbf{-203.6} & -194.9 & -187.9 \\
   & BIC & -106.1 & \textbf{-211.2} & -206.4 & -203.2 \\
 $N_{\text{pt}}$=100  & $-\ln L$ & -55.0 & \textbf{-109.6} & -109.2 & \textbf{-109.6} \\ \hline
 Case-2 & AIC & \textbf{-21.2} & -15.4 & -7.9 & -0.2 \\
   & BIC & \textbf{-25.0} & -23.0 & -19.3 & -15.5 \\
 $N_{\text{pt}}$=100  & $-\ln L$ & -14.5 & -15.5 & \textbf{-15.7} & \textbf{-15.7} \\ \hline
 Case-3 & AIC & -219.3 & \textbf{-220.2} & -212.8 & -205.0 \\
   & BIC & -223.2 & \textbf{-227.8} & -224.3 & -220.3 \\
 $N_{\text{pt}}$=100  & $-\ln L$ & -113.6 & -117.9 & -117.9 & \textbf{-118.1} \\ \hline
 Case-4 & AIC & \textbf{-117.8} &  -109.9 & -102.1 & -94.3 \\
   & BIC & \textbf{-121.6} & -117.6 & -113.6 & -109.6 \\
 $N_{\text{pt}}$=50  & $-\ln L$ &  \textbf{-62.8} &  \textbf{-62.8} &  \textbf{-62.8} &  \textbf{-62.8} \\ \hline
 Case-5 & AIC & \textbf{-169.3} & -161.5 & -153.6 & -145.8 \\
   & BIC & \textbf{-173.1} & -169.1 & -165.1 & -161.1 \\
 $N_{\text{pt}}$=50  & $-\ln L$ & \textbf{-88.6} & \textbf{-88.6} & \textbf{-88.6} & \textbf{-88.6} \\ \hline
\end{tabular}
\end{center}
\caption{
The AIC, BIC and $-\ln L$ for Case-1 to Case-5 obtained by the GMM with various numbers of base distributions.
The lowest values of AIC, BIC and $-\ln L$ for each cases are written in bold font.
}
\label{tab:AIC_BIC}
\end{table}

\section{Conclusion}
\label{sec:conclusion}

Accurate PDF uncertainty estimation is a theoretical necessity for precision measurements as well as new physics searches at the LHC and future colliders. 
Several sources of PDF uncertainty exist, including 
theoretical uncertainty, experimental
systematic errors, and sampling of PDF functional forms and fitting methodologies, see for example~\cite{Courtoy:2022ocu}.
The existence of  tension among different experimental data sets can reduce the precision of PDFs and necessitates the use of alternate statistical models that provide more precise estimates of uncertainty. In the presence of tension, expansion of uncertainty is necessary to make fits consistent, and methods to do so have been proposed and used by various PDF fitting groups. The nature and amount of expansion of uncertainty differs for different methodologies available in literature, many of which introduce ad hoc criteria.

In this work, we have introduced a novel method to estimate uncertainties for inconsistent data sets that avoids the use of ad hoc criteria, checks consistency and produces expanded uncertainties in a statistically meaningful way.
Specifically, for inconsistent data, uncertainties need to be expanded so that the central values of the separate fits to each consistent data set lies close to or within the uncertainty bands of the theory parameters for the combined fit. This ensures a more precise representation of the range of possible theory values, given the inconsistent  data, and avoids biasing the combined fit. The proposed generalized likelihood, modeled via the Gaussian mixture model (GMM), is able to appropriately map uncertainties by including the possibility that the likelihood can be  either multi-modal or uni-modal.

In this work, we demonstrated the shortcomings of the LS model when combining data sets that are in tension, using a toy model of PDFs. 
For LS fitting, the Monte Carlo method and the Hessian method provide a narrow uncertainty band that is not representative of the tension in the data sets. Increasing the uncertainty, either by scaling the uncertainty on each measurement or equivalently, by introducing a static tolerance criteria, does address the problem but is not very precise. For example, using such criteria may increase uncertainty in regions where it is not necessary to do so whereas the uncertainty in regions of parameter space that do have tension is not suitably adjusted. Hence, the tension is not appropriately represented by the uncertainty bands of the LS fit.

The novel GMM pseudo-PDF fit overcomes these shortcomings of the LS method by parametrizing the likelihood as a mixture of Gaussian distributions and by allowing for a multi-modal feature of the likelihood. We have described how this unsupervised machine learning technique can be leveraged to learn about tension in the fits as well as be used to determine precisely the uncertainties on the PDFs. 
We have shown how to use the  Akaike information criterion (AIC) and Bayesian information criterion (BIC) to avoid overfitting, determine the number of Gaussian modes of the GMM, and check for consistency of the fits. Specifically, it is possible to use the GMM for performing the unsupervised machine learning task of clustering the data in order to identify and learn about the tension in the data sets. Moreover, we have demonstrated how the GMM reduces to the usual LS method in the absence of tension. Finally, we have also provided a more intuitive Bayesian interpretation of the GMM method and discussed its connection to Ensemble Averaging and Bayesian Model Averaging.

\section*{Acknowledgment}
MY thanks Yandong Liu  for the discussions. We thank Pavel Nadolsky and Aurore Courtoy for useful discussions and insights. MY has left PKU at the time of this paper being posted.
This work was supported in part by the National Science Foundation of China under grants No. 12075251, and the Natural Science Foundation of Hunan province of China under Grant No. 2023JJ30496. It is also supported in part by the U.S.~National Science Foundation under Grant No.~PHY-2310291 and PHY-2310497.

\newpage
\appendix

\input{Appendix.tex}
\input{Appendix_W_boson.tex}

\bibliography{main}

\end{document}

%% file: Appendix.tex
\section{Generating Pseudo-Data Sets}
\label{app:sec:pseudo_data}
\renewcommand{\thesubsection}{\thesection.\arabic{subsection}}
We utilize the toy model of pseudo-PDF suggested in Ref.~\cite{Paukkunen:2014zia} and choose the functional form given in Eq.~\eqref{eq:func} to generate pseudo-data sets. For easy reference, we reproduce the function given in Eq.~\eqref{eq:func} below 
\begin{equation}
g(x) = a_0 \ x^{a_1} (1-x)^{a_2} e^{x a_3} ( 1 + x e^{a_4} )^{a_5}.
\label{app:eq:func}
\end{equation}
Here $\left\{a_0,a_1,a_2,a_3,a_4,a_5\right\}$ are parameters of the function. 
We generate pseudo-data sets by setting values of the parameters ($a_i$) to certain fixed values. Then, for each fixed value of $x$ we generate a pseudo-data point by assuming the data follows a normal distribution  with uncertainty given by 
\begin{equation}
\Delta g(x) = \frac{\alpha}{ \sqrt{ g(x) } }. 
\label{eq:error_pData}
\end{equation}
Here, the parameter $\alpha$ in Eq.~\eqref{eq:error_pData} is used to scale the size of uncertainty. 
In the region where the magnitude of pseudo-PDF is small, the size of uncertainty is large, mimicking the case of real world data which has larger statistical uncertainty  in regions where event numbers are small. We generate pseudo-data sets using
\begin{equation}
g_D(x) = \Big{(} 1 + r \times \Delta g(x) \Big{)} g(x),
\label{eq:cv_pData}
\end{equation}
where the random variable $r$ follows a Gaussian distribution with unit standard deviation. 
Although, it is possible to include correlated systematic uncertainties or even full likelihoods from experimental data, for simplicity, we treat the pseudo-PDF uncertainty to be uncorrelated and purely statistical. 

In this work, we create six pseudo-data sets that are summarized  in Table~\ref{tab:GMM_best_fits}. The pseudo-data sets are paired into three cases where we fit the combination of these data sets. 
The choice of parameter values used to generate each of the pseudo-data sets  are presented in Table~\ref{tab:pData_truth_2}. The figures with plots of  the different psuedo-data sets are also referenced in the table.

\begin{table}[htpb]
\begin{center}
\begin{tabular}{cc|cc|cccccc}
 Case & Pseudo-Data & $\alpha$ & $N_{\text{pt}}$ & $a_0$ & $a_1$ & $a_2$ & $a_3$ & $a_4$ & $a_5$ \\ \hline \hline
 Case-1 & \#1 & 0.05 & 50 & 30 & 0.5 & 2.4 & 4.3 & 2.4 & -3.0 \\
 Fig.~\ref{fig:pData} & \#2 & 0.05 & 50 & 30 & 0.5 & 2.4 & 4.3 & 2.6 & -2.8 \\ \hline
 Case-2 & \#3 & 0.15 & 50 & 30 & 0.5 & 2.4 & 4.3 & 2.4 & -3.0 \\
  Fig.~\ref{fig:pData_case23:a} & \#4 & 0.15 & 50 & 30 & 0.5 & 2.4 & 4.3 & 2.6 & -2.8 \\ \hline
 Case-3 & \#5 & 0.05 & 50 & 30 & 0.5 & 2.4 & 4.3 & 2.4 & -3.0 \\
 Fig.~\ref{fig:pData_case23:c} & \#6 & 0.05 & 50 & 30 & 0.5 & 1.4 & 4.3 & 2.4 & -3.0
\\ \hline
\end{tabular}
\end{center}
\caption{
The choice of parameters ($a_i$) of the function $g(x)$ defined in Eq.\eqref{app:eq:func}  used to generate the six different pseudo-data sets \#1 ... \#6. $N_{pt}$ refers to the number of data points in the set and $\alpha$ determines the error via Eq.~\eqref{eq:error_pData}. The pseudo-data sets are paired and combined into a single fit. These fits are referred to as, Case-1, Case-2 and Case-3. We also refer to the figures where a plot of each of these cases can be found.
}
\label{tab:pData_truth_2}
\end{table}

\section{Bayesian view of GMM fits}
\label{sec:appendix:Bayesian}
\renewcommand{\thesubsection}{\thesection.\arabic{subsection}}

Here we describe the Bayesian view of the Gaussian Mixture Model (GMM) as applied to  fitting parton distribution functions.  We begin by first describing the Bayesian view of the usual $\chi^2$ method before describing the same for the GMM. We follow closely the presentation given in Ref.~\cite{Kovarik:2019xvh} and Ref.~\cite{Erler:2020bif}.
\subsection{ Bayesian view of the \texorpdfstring{$\chi^2$}{chi2} likelihood }
Our objective is to be able to fit  theory parameters $a_\alpha$, $\alpha=\{1,...,N_P\}$ given $N_D$ data points $D_k$, $k = {1,..., N_D}$. For data that is Gaussian distributed, we may represent it as 
\begin{align}
    D_i = \braket{D_i} + \sigma_i \Delta_i\ .
\end{align}
Here $\braket{D_i}$ can be thought of as the central value of the $i$-th data point, $\sigma_i$ the uncertainty on the data and $\Delta_i$ is a random variable with a Gaussian distribution with mean $0$ and standard deviation of $1$. $\Delta_i$ represents statistical fluctuations. Note, we have left out the correlated errors to keep the discussion simple and that it is straightforward to reintroduce such errors. 

By definition, our data is Gaussian distributed and if $f$ is some function of $\Delta_i$, then its expectation values is
\begin{align}
    \braket{f}=(2\pi)^{N_D/2} \int f(\Delta) \prod_{i = 1}^{N_D}  d \Delta_i \exp\left(-\frac{1}{2}\Delta_i^2\right) \ .
\end{align}
From this we see immediately that the expectation value of the data is $\braket{D_i}$ and it is also straightforward to show by calculating the second moment $\braket{D_i D_j}$, that the covariance is  
\begin{align}
    C_{ij} = \braket{(D_i - \braket{D_i})(D_j - \braket{D_j})} = \delta_{ij}\sigma_i\sigma_j\ . 
\end{align}
In this case the covariance is diagonal, because we have not considered correlated uncertainties.  If $g$ is some function of the data $D_i$, then it is straightforward to show that the expectation value of $g$ is
\begin{align}
    \braket{g} = \frac{1}{\sqrt{(2 \pi)^{N_D} \det{C}}} \int g(D)\prod_{i,j=1}^{N_D} d D_i \exp{ \left(-\frac{1}{2} (D_i - \braket{D_i})(D_j - \braket{D_j}) C_{ij}^{-1}\right)}\ .
\end{align}
If we now make the identification that $\braket{D_i} = T_i(a)$, where $T_i(a)$ is the theoretical prediction for the datum $D_i$, then we get the conditional probability to obtain the experimental results $D$ if the theory is represented by $T(a)$ is
\begin{align}
    P(D|T(a)) = \frac{1}{\sqrt{(2 \pi)^{N_D} \det{C}}} \prod_{i,j=1}^{N_D} d D_i \exp{ \left(-\frac{1}{2} (D_i - T_i(a))(D_j - T_j(a)) C_{ij}^{-1}\right)} \ .
\end{align}
This can be recast into the more familiar form
\begin{align}
    P(D|T(a)) = \frac{1}{\sqrt{(2 \pi)^{N_D} \det{C}}}  d D\exp{  \left(-\frac{1}{2} \sum_{i,j=1}^{N_D} (D_i - T_i(a))(D_j - T_j(a)) C_{ij}^{-1}\right)} \ ,
\end{align}
where the sum represents the usual $\chi^2$ function that is minimized and $dD = \prod_i^{N_D} d D_i$. 
$P(D|T(a))$ is the likelihood function. Applying Bayes' theorem, it is possible to determine the conditional probability that the theory $T(a)$ is correct given the data $D$
\begin{align}
    P(T(a)|D) = \frac{P(D|T(a))P(T(a))}{P(D)} \ .
\end{align}
Knowledge of $P(D)$ and $P(T(a))$ is limited, making determination of this probability difficult. On the other hand,  the typical fitting problem requires that we make an optimal choice of parameters that maximizes $P(T(a)|D)$. This is easily achieved through the ratio of likelihoods. For example, consider the ratio of the conditional probabilities for two different choices of the theory parameters $a_1$ and $a_2$
\begin{align}
    \frac{P(T(a_1)|D)}{P(T(a_2)|D)} = \frac{P(D|T(a_1)) P(T(a_1))}{P(D|T(a_2)) P(T(a_2))} = \exp\left(-\frac{\chi^2(D,a_1) -\chi^2(D,a_2) }{2}\right)\times \left(\frac{P(T(a_1))}{P(T(a_2))}\right)\ .
\end{align}
Here, $P(T(a_1))/P(T(a_2))$ is the ratio of our prior probabilities that we begin with and which gets updated as we collect more data.\footnote{ See Ref.~\cite{Kovarik:2019xvh} for a more detailed discussion on updating prior and posterior probabilities.}
\subsection{Hierarchical Models and  Bayesian view of the GMM}
\label{app:sec:Bayesian view of the GMM likelihood}
Consider the scenario where we collect data from $K$ different experiments. The data from each of these experiments are represented as 
\begin{align}
    D_i^{(k)} = \braket{D_i^{(k)}} + \sigma_i^{(k)} \Delta_i^{(k)} = T_i(a^{(k)})  + \sigma_i^{(k)} \Delta_i^{(k)}\ .
\end{align}
Here $k=\{1,2,...,K\}$. Note we have assumed, for simplicity, that the data between experiments is uncorrelated. This is not a necessary assumption and can be relaxed in general cases. 
Suppose each data set from each experiment produces a different central value $\braket{D_i^{(k)}}$, indicating a different preferred theory $T_i(a^{(k)})$ for each experiment.
In reality, such differences may be due to an underestimation of experimental uncertainties or the use of an inaccurate theoretical model. Therefore, one should investigate the theory and data more rigorously to identify the source of these discrepancies. However, this is often difficult or even impossible to do with particle physics data. Instead, we can use the Gaussian Mixture Model to estimate uncertainty, given the gaps in our understanding of the source of discrepancy between experiments. For example, factorization in QCD requires there to be only one set of theory parameters of parton distribution functions. The GMM allows for the possibility that several values of our theory parameters $a^{(k)}$ can exist simultaneously. We are not abandoning QCD factorization; rather, we introduce multiple possibilities of the same theory fit parameters in order to capture the discrepancies present in the data. As mentioned earlier, such discrepancies could arise due to imprecise theoretical calculations and/or unaccounted-for experimental biases. The GMM is able to provide an estimate of the probability distribution of the theory, given these unresolved discrepancies in the data.

\begin{figure}
    \centering
    \includegraphics[width=0.45\linewidth]{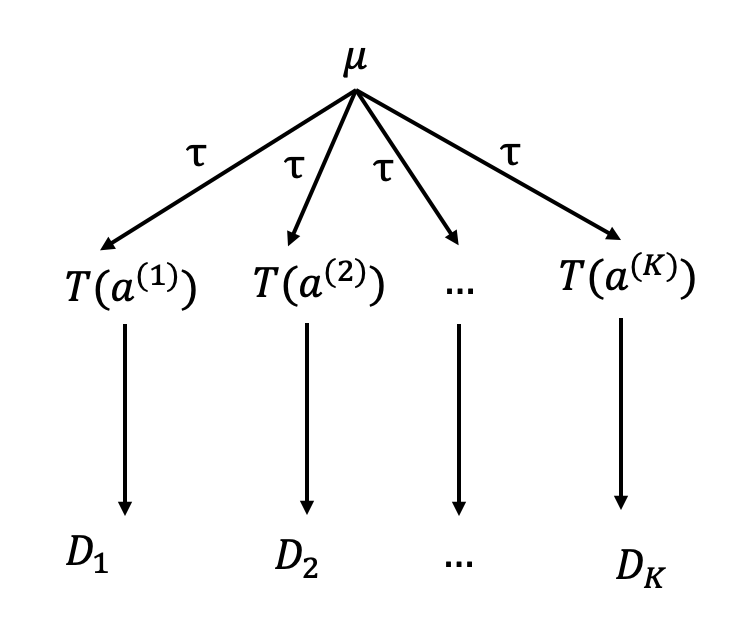}
    \includegraphics[width=0.45\linewidth]{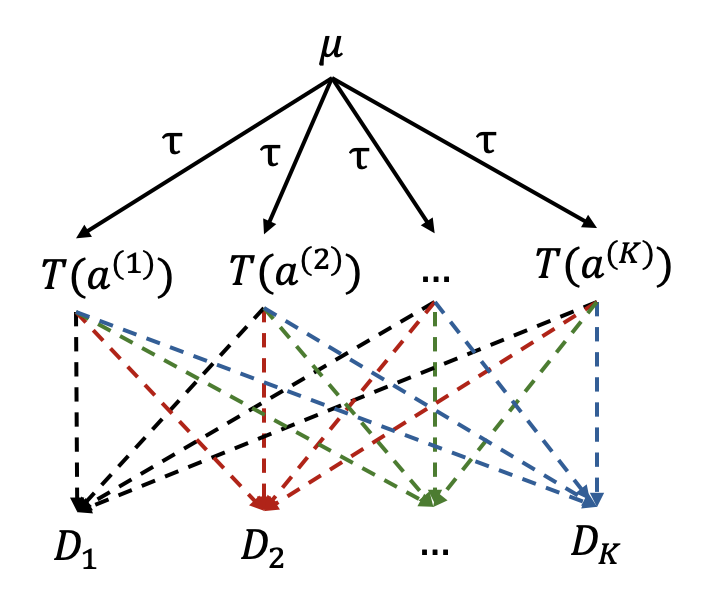}
    \caption{Left:The hierarchical model, as proposed in Ref.~\cite{Erler:2020bif}. Right: the GMM model. Here $\mu$ and $\tau$ correspond to hyper-paramaters from which the theory models ($T(a^{(k)})$) are drawn.  }
    \label{fig:Hierarchical_Model_Schematic}
\end{figure}

A similar approach to the GMM proposed in this work has been used in Ref.~\cite{Erler:2020bif}.  As shown in the diagram on the left in Fig.~\ref{fig:Hierarchical_Model_Schematic}, here each datum ($D_k$) is viewed as measuring different theory model parameters $T(a^{(k)})$, and each $T(a^{(k)})$ originates from some parent distribution governed by hyper-parameters $\mu$ and $\tau$.
The probability $P(T(a^{(k)}))$ is then given by
\begin{align}
  P(T(a^{(k)})) = \int d\mu d\tau   P(T(a^{(k)})|\mu,\tau) p(\mu,\tau) \equiv w_k \ .
  \label{eq:app:hyper_prior}
\end{align}
Here we introduce $w_k = P(T(a^{(k)}))$ so that $\sum_{k=1}^{K} w_k = 1$.
The diagram on the right in Fig.~\ref{fig:Hierarchical_Model_Schematic}
shows the approach used in the GMM. We assume that each datum $D^{(i)}$ could arise due to a random draw from each of the possible theories $T(a^{(k)})$, and so the probability of measuring datum $D^{(i)}$ given the theory $T(a^{(k)})$ is 
\begin{align}
     P(D_i|T(a^{(k)})) P(T(a^{(k)})) =  w_k P(D_i|T(a^{(k)})) = P(T(a^{(k)})|D_i) P(D_i) \ .
\end{align}
Here we have used Bayes' theorem in the last equality. Marginalizing over the different possible theories $T(a^{(k)})$, we get
\begin{align}
    \left( \sum_{k=1}^{K} P(T(a^{(k)})|D_i)\right) P(D_i) = \sum_{k=1}^{K} w_k P(D_i|T(a^{(k)}))\ . 
\end{align}
The term on the left hand side tells us the probability of getting all $T(a^{(k)})$ given the datum $D_i$, whereas the right hand side is the probability of drawing the datum $D_i$ given the set of $T(a^{(k)})$. The objective for a fitting problem using the GMM is to determine the choices of $T(a^{(k)})$ and $w_k$ that maximize the probability on the left hand side. Thus, the right hand side of the equation corresponds to the modified likelihood of the GMM. 
Specifically, assuming Gaussian probability distributions, we have 
\begin{align}
   \prod_{i=1}^{N_D} \left( \sum_{k=1}^{K} P(T(a^{(k)})|D_i)\right) \propto \prod_{i=1}^{N_D}\left(\sum_{k=1}^{K} w_k \mathcal{N}(D_i|T(a^{(k)}), \sigma_i)\right)\ ,
   \label{eq:eq_GMM_Bayesian_likelihood}
\end{align}
which is the likelihood function used in the Gaussian Mixture Model, c.f. Eq.~\eqref{eq:GMM}.
Here $\mathcal{N}(D_i|T(a^{(k)}), \sigma_i)$ is the Normal distribution with mean $T(a^{(k)})$ and standard deviation $\sigma_i$.
Note that in Fig~\ref{fig:Hierarchical_Model_Schematic}, it appears as though for $N_D$ data points one must choose $K=N_D$. This is not required in the GMM and in fact, $K$ is a hyper-parameter that can be determined with the help of information criteria, c.f. Sec.~\ref{subsec:GMM_overfit}. 
Typically, if all the data is consistent, we find $K=1$ which reduces to the usual $\chi^2$ likelihood function. Hence, in Eq.~\eqref{eq:eq_GMM_Bayesian_likelihood}, $K$ and $N_D$ are different with $K \ll N_D$ in typical applications of the GMM. 

One major advantage of using the GMM to construct the Bayesian Hierarchical model is in overcoming a combinatorial problem of partitioning the data into different sets. The GMM partitions the data into consistent data sets, and in the limit of large tension, the partitioned data sets approximate the usual hierarchical model shown in the left of Fig.~\ref{fig:Hierarchical_Model_Schematic}, since the probability of generating the data set $D_i$ from ($T(a^{(j)})$)is large when $i=j$ and negligibly small when $i\neq j$. A further advantage is that it is not necessary to assume a hyper-prior distribution $p(\mu,\tau)$ since the weights $w_k$ are determined through likelihood extremization. Starting with the model in the left of Fig.~\ref{fig:Hierarchical_Model_Schematic} would necessarily require us to specify a hyper-prior distribution or determine a way to partition the data sets. Both of these concerns are directly addressed by the use of the GMM. Finally, we note in passing the interesting possibility of introducing and modifying the hyper-prior distribution in the GMM, which can be achieved by  modifying the weights which could lead to different interpretation of the weights and the statistical model in use.

\subsection{A Note on Bayesian Model Averaging and Ensemble Learning}
\label{sec:app:BMA_and_BMC}
In Appendix~\ref{app:sec:Bayesian view of the GMM likelihood}, we described our approach as a hierarchical model assuming the weights $w_k$ are determined through some parent distribution $p(\mu,\tau)$. However, in practice we do not need to know the parent distribution and, in fact, the weights $w_k$ are determined through a Maximum Likelihood Estimate of the likelihood function in Eq.~\eqref{eq:eq_GMM_Bayesian_likelihood}. The use of the GMM makes the the approach taken here similar to method of Ensemble Learning used frequently in predictive machine learning tasks such as  weather forecasting~\cite{raftery2005using}. The main idea in Ensemble Learning is to define a combined probability distribution as a convex combination of a basis of models
\begin{equation}
    P(y|D, w_1 \dots w_K,T(a^{(1)})\cdots T(a^{(k)}) ) = \sum_{k=1}^{K} w_k P(y|D,T(a^{(k)})) \ ,
\end{equation}
where $y$ is the quantity to be predicted. While several algorithms, such as Stacking~\cite{wolpert1992stacked,yao2018using}, have been proposed to estimate the weights $w_k$, for our purpose, since each of our models $T(a^{(k)})$ are equally complex, a straight forward maximization of the likelihood suffices.

This method is sometimes referred to as the committee method~\cite{Murphy:ML} or as a method of Model Combination~\cite{bishop2007}. In the statistics literature,  a distinction is made between Bayesian Model Combination and another related and popular method known as Bayesian Model Averaging~\cite{leamer1978specification,kass1995bayes,wasserman2000bayesian,hoeting1999}. The former is a convex combination of a basis of models that produces a single combined model that describes the probability distribution of data, such that data could arise from any  of the theoretical model distributions. On the other hand, Bayesian Model Averaging assumes that there is just one model $T(a^{(k)})$ responsible for generating the data and $w_k$ represents the probability  for it to be  the best model. 
For a more in depth discussion of these differences, see for example Refs.~\cite{minka2000bayesian,Monteith:2011}.\footnote{ For more discussion on the use of mixture models as a way to implement Bayesian Model Averaging and Combination, see for example~\cite{kamary2014testing,keller2017bayesian}}
If our goal is to find the best model as is done in Bayesian Model Averaging, then we may interpret the $w_k$ as the posterior probability that $T(a^{(k)})$ is the best model to describe the data. Then what we compute through the use of Eq.~\eqref{eq:eq_GMM_Bayesian_likelihood} is in fact an implementation of Bayesian Model Averaging, see for example Ref.~\cite{raftery2005using}. The formulae we provide for the estimate of the mean and variance in Eq.~\eqref{eq:GMM_mean} and Eq.~\eqref{eq:cov_GMM_2}, respectively, coincide with the formulae used for these quantities in Bayesian Model Averaging~\cite{kass1995bayes,raftery2005using}. Further, one could use values of $w_k$ to determine Bayes factors~\cite{kass1995bayes} and even perform model selection to select the best model out of the mixture of models. Here, we are more interested in a determination of the uncertainty and do not undertake such a program. Importantly, Model Averaging and Combination have been shown to not only provide better predictive power of machine learning tasks, but also provide better estimates of uncertainty~\cite{Murphy:ML}.


%% file: Appendix_W_boson.tex
\section{A Simple 1-D Example of \texorpdfstring{$W$-boson}{W-boson} Mass Combination}
\label{sec:app:W_boson}
\renewcommand{\thesubsection}{\thesection.\arabic{subsection}}
The purpose of this appendix is to demonstrate the use and characteristics of the GMM method,  with the help of a simple one dimensional example. 
Note that a one dimensional fit does not allow us to demonstrate one major advantage that the GMM provides in estimating uncertainties through appropriate modification of eigen-vector directions of the Hessian. Nevertheless, the example provided here is simple and uses real world data. 

Consider the measurement of the $W$-boson mass by various experiments around the world. The measured values of the $W$-boson mass are shown in 
Table~\ref{tab:mw_measurements}. 
\begin{table}
\centering
\begin{tabular}{ |c|c|c| } 
 \hline
 Experiment & $W$-boson mass (GeV)  & $\pm$ Error (GeV) \\ \hline \hline
D0-I~\cite{D0:2002fhu} & 80.483 & 0.084\\ \hline 
CDF-I~\cite{CDF:2000gwd} & 80.433 & 0.079\\ \hline
LEP~\cite{ALEPH:2013dgf} & 80.376 & 0.033 \\ \hline
D0-II~\cite{D0:2012kms} & 80.375 & 0.023\\ \hline
LHCB~\cite{LHCb:2021bjt} & 80.354 & 0.032\\ \hline
CDF-II~\cite{CDF:2022hxs} & 80.4335 & 0.0094\\ \hline
ATLAS23~\cite{ATLAS:2024erm} & 80.36 & 0.016\\ \hline 
\end{tabular}
\caption{Measured values of the $W$-boson mass. 
\label{tab:mw_measurements}
}
\end{table}
CDF and ATLAS recently provided updated measurements of the $W$-boson mass. The reported value from CDF is in tension with the ATLAS measurement. 
In this situation, one might argue that one or both of the analyses are inaccurate. Alternately, either or both experiments could be imprecise, and have underestimated uncertainties.\footnote{ Understanding the source of this discrepancy is a complicated task and studies in this regard are still being undertaken~\cite{Kotwal:2025xsy}.} Here we take an agnostic view on the source of the tension between experiments, which could likely be a result of theoretical inadequacies as  well. 
In this scenario, when we are uncertain about both the source of uncertainty as well as its estimate, how should we combine these disparate measurements in a meaningful way. In other words, in the presence of such tension in the data, how should the measurements be combined to report a mean and uncertainty (point estimates) of the $W$-boson mass? In the following we discuss various possibilities to perform such a combination. Although it is possible to incorporate any correlated uncertainties between the measurements, to keep our presentation simple we treat each measurement as independent and ignore any correlated uncertainties.
\subsection{Least Squares Fit}
\label{sec:app:W boson LS fit}
    In the standard  $\chi^2$ approach, a Gaussian likelihood function from each experiment is multiplied to provide a combined likelihood. Taking the logarithm of the likelihood yields the usual $\chi^2$ log-likelihood up to constant terms as follows
    \begin{equation}
        \chi^2 = \sum_j\frac{ (y_j - m_W)^2}{\Delta y _j ^2 } \ .
    \end{equation}
    Here $y_j$ corresponds to the measured values of the $W$-boson mass and $\Delta y_j $ the uncertainty on the measurement, which can be found in Table~\ref{tab:mw_measurements}. By using the $\chi^2$ we have  implicitly assumed that the likelihood from each measurement is Gaussian with a standard deviation given by the uncertainty.\footnote{Alternately, it is possible, to take a product of the full experimental likelihoods. However, the assumption that each experimental likelihood is Gaussian, usually works well. } Our goal is to learn the distribution of the random variable $m_W$. Although full information about the distribution is available in the log-likelihood, it is usually useful to report the characteristics of the distribution through point estimates such as moments of the distribution. Typically that means reporting the mean or expectation value ($\overline{m}_{W}$) as well as the variance ($\sigma_{m_{W}}$), which is sufficient to fully specify a Gaussian distribution
\begin{align}
    \pi(x ; \overline{m}_W ,\sigma_{m_W}) = \frac{1}{\sigma_{m_W}\sqrt{2\pi}} e^{-\frac{1}{2}\left(\frac{x - \overline{m}_{W}}{ \sigma_{m_{W}}}\right)^2}\ .
\end{align}
The mean, $\overline{m}_{W}$, is determined by minimizing the $\chi^2$ function and the variance can be found from the Hessian. Alternately, since the $m_W$ is a random variable, it follows that the log-likelihood follows a $\chi^2$ distribution. Thus, the variance, $\sigma_{m_W}$, can be obtained by varying the $\chi^2$ away from the minimum up to $\Delta\chi^2 =1$.

Minimizing the $\chi^2$ function, we find the best fit value of the $W$-boson to be
\begin{align}
    \overline{m}_W|_{\chi^2} = 80.4065 \pm 0.0072 \ \text{GeV},
\end{align}
The uncertainty of $7.2$~MeV is determined with the condition $\Delta \chi^2 = 1$, which, in this case is identical to estimate of the standard error using the inverse of the Information matrix or the Hessian matrix, which is related to the covariance of the combined likelihood of $m_W$.
In this simple example, rather than minimizing numerically, it is possible to derive exact formulae, where the mean 
 and uncertainty of the $m_W$ combination is given by
 \begin{align}
     \overline{m}_W|_{\chi^2} = \frac{\sum_j \frac{y_j}{\Delta y_j^2}}{\frac{1}{ \sigma_{m_W}^2}}\ , \quad \text{with} \quad 
     \frac{1}{\sigma_{m_W}^2 }= \sum_j\frac{1}{\Delta y_j^2} 
     \label{eq:app:chi2 formulae}
 \end{align}

As a result of the tension between the ATLAS and CDF measurements, the fit turns out to be quite poor with $\chi^2/\text{d.o.f} \simeq 3.8$. 

One way to proceed with an alternative combination is to inflate the experimental uncertainties by a scale factor ($S_{PDG}$), as proposed by the Particle Data Group~\cite{ParticleDataGroup:2024cfk}, using the conditions described below.
\begin{itemize}
    \item If the $\chi^2/\text{d.o.f} < 1$, the results of the fit are accepted and there is no scaling.
    \item If the $\chi^2/\text{d.o.f} \gg 1$, either a combined mean is not produced or an educated guess of the uncertainty is made.
    \item If the $\chi^2/\text{d.o.f} > 1$, then: 
    \begin{itemize}
        \item If all errors are comparable, then all errors are rescaled by a common scale factor $S_{PDG}= \sqrt{\chi^2/\text{d.o.f}}$. This, by definition, forces the $\chi^2/\text{d.o.f}$ to approach $1$. Note, for the simple examples used here, this is equivalent to introducing a global tolerance criteria with $T=S_{PDG}$.
        \item If some of the individual experimental uncertainties are significantly smaller than others, then $S_{PDG}$ is calculated using only the most precise experiments. The  choice of cutoff to include or exclude experiments by their precision is ad hoc. PDG uses an arbitrary choice of only scaling those experimental uncertainties if the uncertainty for a measurement is less than $3N^{1/2}\bar{\sigma}$, where $N$ is the number of measurements, $\bar{\sigma}$ is the unscaled uncertainty computed from Eq.~\eqref{eq:app:chi2 formulae}.
    \end{itemize}
\end{itemize}

 This idea has been further explored and elucidated in both the Bayesian and Frequentist approach~\cite{Erler:2020bif,Cowan:2018lhq,DAgostini:1999niu}. Using the above recommendations, we find for the data set in Table~\ref{tab:mw_measurements}, all uncertainties are rescaled by a factor of $S_{PDG}\simeq \sqrt{3.8} \simeq 1.95$. This is equivalent to introducing a tolerance of $T\simeq1.95$, without rescaling uncertainties. Applying the second condition to the fit involving only the CDF-II and ATLAS data, we find 
\begin{align}
    \overline{m}_W|_{\chi^2} = 80.4065 \pm 0.0141 \ \text{GeV},
\end{align}
with $\chi^2/\text{d.o.f} = 1 $. The uncertainty, by definition, also increases by a factor of $S_{PDG}$.

A more sophisticated combination of the global measurement of the $W$-boson mass~\cite{LHC-TeVMWWorkingGroup:2023zkn} reveals that the CDF measurement is incompatible with the rest of the measurements and dropping this measurement in the combination improves the probability of compatibility by about $\sim 90 \%$.\footnote{ The probability of compatibility is estimated using the $\chi^2$ distribution. See Ref.~\cite{LHC-TeVMWWorkingGroup:2023zkn} for more details.} A refit after leaving out the CDF-II measurement yields 
\begin{align}
    \overline{m}_W\big|_{\chi^2}^{\text{no CDF-II}} = 80.3683 \pm 0.0112 \ ,
    \label{eq:app:mw_ls_fit_noCDF}
\end{align}
with $\chi^2/\text{d.o.f.} \simeq 0.62$.

In this situation of combining the $W$-boson mass, we not only have ample data to compare, but also theoretical prediction of $m_W|_{SM} = 80.355 \pm 0.006$ GeV~\cite{deBlas:2021wap,Haller:2022eyb}. 
On the other hand, PDF fits do not have strong theoretical information and often data can be sparse so that a determination of compatibility is not straightforward and in some cases not even possible.
To draw an analogy with the $W$-boson mass combination, let us consider the hypothetical situation in which we have only the CDF-II and ATLAS measurements of the $W$-boson mass and no theoretical insight on this quantity, as is the case for non-perturbative parton distribution functions.\footnote{Recently, there have been some attempts to determine the PDFs using Lattice QCD, see for example Ref.~\cite{Lin:2017snn}.}
If we redo the fits with only the CDF-II and ATLAS data, we get a very poor fit with $\chi^2/\text{d.o.f} \simeq 15.7$ and 
\begin{align}
    \overline{m}_W\big|^{\text{ATLAS23+CDF-II}}_{\chi^2} = 80.4146 \pm 0.0081 \ \text{GeV}.
\end{align}
For such a poor fit, PDG prescription suggests either not doing a combination at all or making an educated guess for the uncertainty. If we were to rescale the uncertainties, the combined uncertainty for this fit would increase approximately by a factor of 4 to about $\pm 0.0321$. 
Note that, in spite of the large four fold increase in uncertainty, the ATLAS measurements central value does not fall within the 1-$\sigma$ uncertainty of this fit, neither does the theoretical prediction of $m_W|_{SM} = 80.355 \pm 0.006$ GeV~\cite{deBlas:2021wap,Haller:2022eyb}. Hence, as suggested by PDG, an educated guess for the uncertainty is necessary to account for the bias introduced by the combination. Rather than making an educated guess, we can use the GMM method proposed in this work to determine the uncertainty.

An important question to ask is why  the uncertainty should increase as more data is collected? This appears to be contrary to what we learn in statistics where the statistical uncertainty reduces with increasing sample size. One way to try and understand this is to recall that experiments usually need to unbias their measurements through the use of some procedure. For example, proper calibration is necessary to create a map ($f: \theta_{\text{M}} \to \theta_{\text{Th}}$) that translates the numbers read out from the experimental device ($\theta_M$), such as a data acquisition system, to the quantity of interest $\theta_{\text{Th}}$. The procedure to determine this map is usually statistical in nature and introduces systematic uncertainties. When fitting to a consistent data set, we assume that the map ($f: \theta_{\text{M}} \to \theta_{\text{Th}}$) has been accurately determined and any uncertainty is accounted for by the systematic uncertainties provided by the experiment. 
If a second experiment produces data that is in tension with the first one, then this provides evidence that the map $f: \theta_{\text{M}} \to \theta_{\text{Th}}$ has not been determined accurately for either or both experiments. 
In this situation, 
our initial assumption that the maps have been determined accurately is incorrect, and we must incorporate a new uncertainty that accounts for the presence of a systematic bias. One way to do this is to marginalize over the space of possible maps ($f: \theta_{\text{M}} \to \theta_{\text{Th}}$), or in other words, 
we must marginalize over the possible distributions of these maps given the experimental data. The hyper-prior distribution of the hierarchical model ($p(\mu,\tau)$) in Eq.~\eqref{eq:app:hyper_prior} approximates the possible distribution of such maps. When all experiments are consistent, then it is consistent to use a Dirac Delta distribution for the hyper-prior. On the other hand, information about the possible inaccuracy of such maps changes the hyper-prior away from a Dirac Delta distribution, which in turn necessarily increases the uncertainty. This can be avoided in the ideal situation if there is a way to choose between experiments, or there is a way to reanalyze and determine the maps more accurately.

We reiterate that, in this simple example, we are able to compare with several other $W$-boson mass measurements as well as one from electro-weak fits, to understand the compatibility of various experimental measurements. On the other hand, parton distribution functions do not have a theoretical counterpart to be compared to and several portions of the fit might also have very sparse amounts of data. 
Hence, uncertainties on the PDFs should be more representative of the data that they are extracted from. In the next subsection, we describe how to improve the probability distribution on measured quantities and hence provide more precise estimates of uncertainty by using a mixture model and  allowing for the likelihood to be multi-modal.

\subsection{GMM Fit}
Here we apply the GMM method to the simple case of $W$-boson mass combination. We begin by defining the likelihood as follows:  
\begin{eqnarray}
 - \text{log} L &=& - \text{log} \bigg{(} \prod_{j=1}^{N_{\text{pt}}} L(y_j, \Delta y_j|\vec{\theta}) \bigg{)} = - \sum_{j=1}^{N_{\text{pt}}} \text{log} \bigg{(} \sum_{i=1}^K \omega_i \mathcal{N}(y_j, \Delta y_j|\theta_i) \bigg{)}\nonumber \\
 &=& - \sum_{j=1}^{N_{\text{pt}}} \text{log} \bigg{(} \sum_{i=1}^K \frac{\omega_i}{\sqrt{2\pi} \Delta y_j} \text{exp} \Big{[} -\frac{1}{2} \Big{(} \frac{y_j - y_j(\theta_i)}{\Delta y_j} \Big{)}^2 \Big{]} \bigg{)}.
 \label{eq:likelihood_GMM_W_boson}
\end{eqnarray}
By doing so we have introduced $K$ different random variables $\theta_i$, and we would like to learn about the probability distribution of these $K$ random variables. 
While information about the probability distribution of the $K$ variables is available in the likelihood function above, we would like to report the point estimates of the distribution, namely the mean and variance of the distribution.\footnote{Other measures of uncertainty can be used, however, using the mean and variance makes it straightforward to compare with $\chi^2$ fits.} 
The best fit is determined through Maximum Likelihood Estimation by varying and finding the $K$ different values of $m_W$  as well as the weights $\omega_i$ of each Gaussian that maximize the full log-likelihood. 
Once we have minimized the loss function, determination of an estimate of the mean is straightforward and simply given by
\begin{equation}
 \overline{m}_W|_{GMM}=\mathbb{E}[\theta] = \sum_{i=1}^K \omega_i \hat{\theta}_i.
 \label{eq:GMM_expectation}
\end{equation}
Here $\hat{\theta}_i$ are the $K$ different values of $m_W$ that minimize the loss function defined in Eq.~\eqref{eq:likelihood_GMM_W_boson}.
Determination of the variance requires more work. We begin with the ansatz that the probability distribution of $m_W$ is, to a good approximation, given by a Gaussian mixture model of the form
\begin{equation}
 \pi(x | \hat{\theta}_i ,\sigma_{i}) = \sum_{i}^{K} \frac{\omega_i}{\sigma_{i}\sqrt{2\pi}} e^{-\frac{1}{2}\left(\frac{x - \hat{\theta}_i}{ \sigma_{i}}\right)^2}\ .
 \label{eq:gmm_ansatz}
\end{equation}
The variance for the the probability distribution function defined above is 
\begin{equation}
 \sigma^2 = \text{cov}_{\text{GMM}} = \sum_{i=1}^K \omega_i \sigma_i^2 + 
 \sum_{i=1}^K \omega_i (\mathbb{E}[\theta] - \hat{\theta}_i )^2 \ .
\end{equation}
While the second term in the equation above is already determined with the help of Eq.~\eqref{eq:GMM_expectation}, the first terms requires knowledge of the variance ($\sigma_i^2$) for each Gaussian. When using the Expectation Maximization algorithm, the individual variances are calculated automatically. Here, we approximate the variance by calculating the observed Fisher Information Matrix.\footnote{It is also possible to sample the likelihood function in Eq.~\eqref{eq:likelihood_GMM_W_boson} to produce a Monte Carlo set on which we can run the Expectation Maximization Algorithm.} 
The results  of minimizing the likelihood defined in Eq.~\eqref{eq:likelihood_GMM_W_boson} for different values of $K$ are shown in Table~\ref{tab:mW_GMM}. The mean ($\overline{m}_W$), uncertainty ($\sigma$), as well as the values of the two information criteria (AIC, BIC) are presented. We see that the mean and variances for $K=2$ through $K=5$ are similar and this is a result of the values of $m_W$ for each Gaussian settling either around the latest measured values of CDF-II or ATLAS and thus creating a degeneracy for values $K>2$. The fact that adding additional Gaussians and going beyond $K=2$ is unnecessary can also be seen by looking at the values of AIC and BIC in Table~\ref{tab:mW_GMM}. Importantly, we see that the AIC and BIC are minimized for $K=2$ and thus this model provides the best fit to the data. 
As expected, the uncertainty suggested by this model is $30.8$~MeV and is significantly larger than the $7.2$~MeV suggested by the $K=1$ model, which is identical to the LS fit in Appendix~\ref{sec:app:W boson LS fit}, which provided a poor fit to the data. 
\begin{table}[htb]
    \centering
    \begin{tabular}{|c|c|c|c|c|}
    \hline
        $K$ & $\bar{m}_W$ & $\sigma$ & AIC & BIC  \\ \hline \hline
        1  &  80.4065  &  0.0072  &  -11.01 & -10.96 \\  \hline
2  &  80.3850  &  0.03081 &  \textbf{-22.89 } &  \textbf{-22.73} \\  \hline 
3  &  80.3850 &  0.03091 &  -19.00 & -18.73  \\ \hline
4  &  80.3850  &  0.03091 & -15.10 & -14.73\\ \hline 
5  &  80.3850 &  0.03095  &  -11.21 & -10.73\\ \hline 
    \end{tabular}
    \caption{The Mean ($\bar{m}_W$), uncertainty ($\sigma$), AIC and BIC values for different values of $K$ are listed. The numbers in bold correspond to the smallest values of AIC and BIC.}
    \label{tab:mW_GMM}
\end{table}

One could also consider the possibility of performing the fits without the latest CDF-II~\cite{CDF:2022hxs} data that is in tension with the ATLAS~\cite{ATLAS:2024erm} data. When excluding the CDF-II data, we find that the AIC and BIC are minimized for $K=1$, suggesting that the data is consistent and that the usual $\chi^2$ approach should work. The result of the combination for $K=1$, which is identical to the LS methods, is given in Eq.~\eqref{eq:app:mw_ls_fit_noCDF}.
On the other hand, excluding only the ATLAS23 result and including CDF-II in the data set does not lead to a consistent fit for $K=1$. Moreover, the AIC and BIC are minimized for $K=2$, suggesting that there is lingering tension with the rest of the data, especially since LEP, D0-II and LHCb are in tension with CDF-II. The result of the fit is 
\begin{align}
    \overline{m}_W\bigg|_{\text{no ATLAS23}}^{K=2} = 80.393 \pm  0.031 \ \text{GeV}\ ,
\end{align}
The uncertainty suggested by $K=2$ model when excluding ATLAS23 is about $31$~MeV which is similar to the uncertainty found with the full data set. 
\begin{figure}
    \centering
    \includegraphics[width=0.45\linewidth]{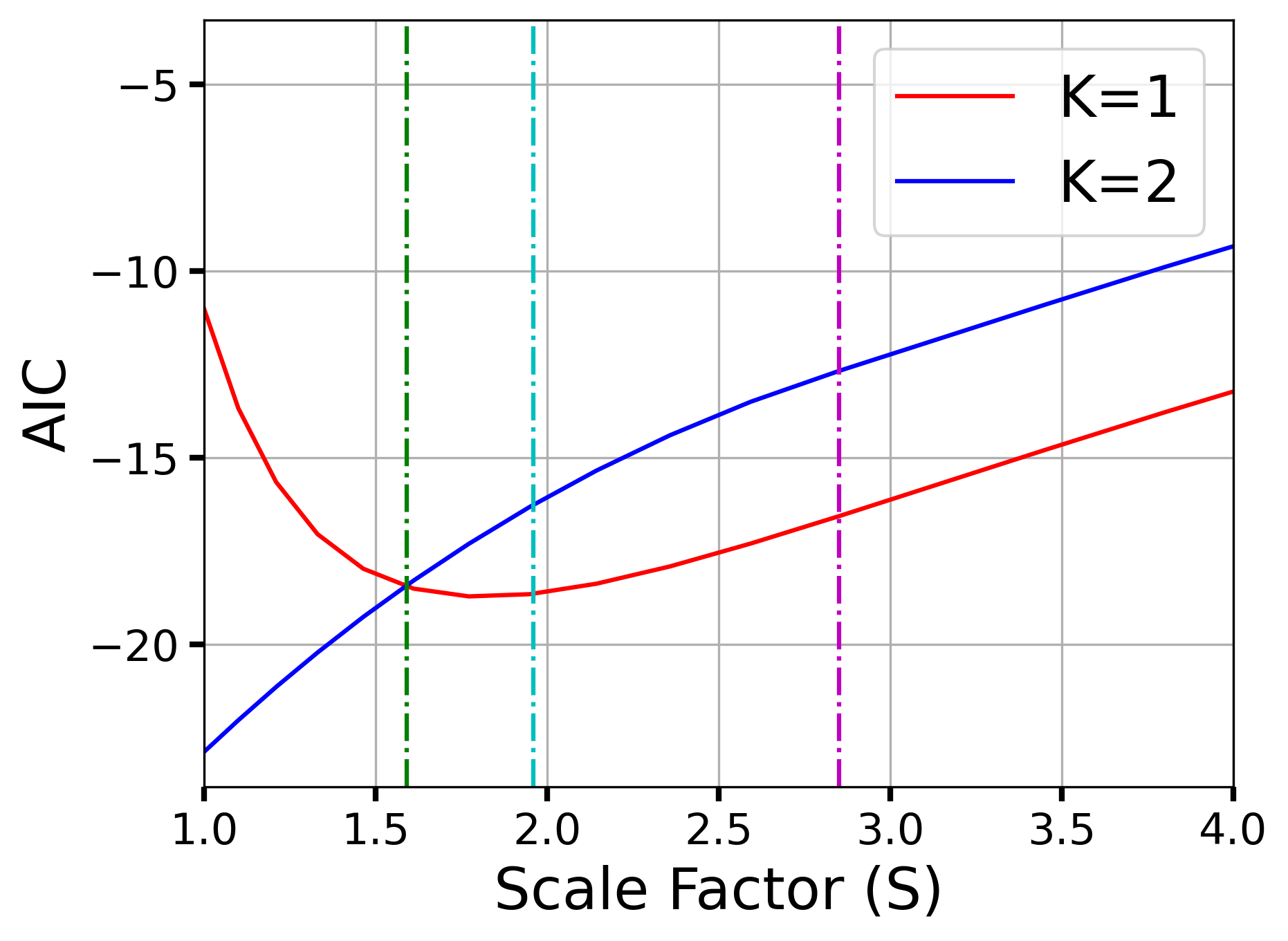}
    \includegraphics[width=0.45\linewidth]{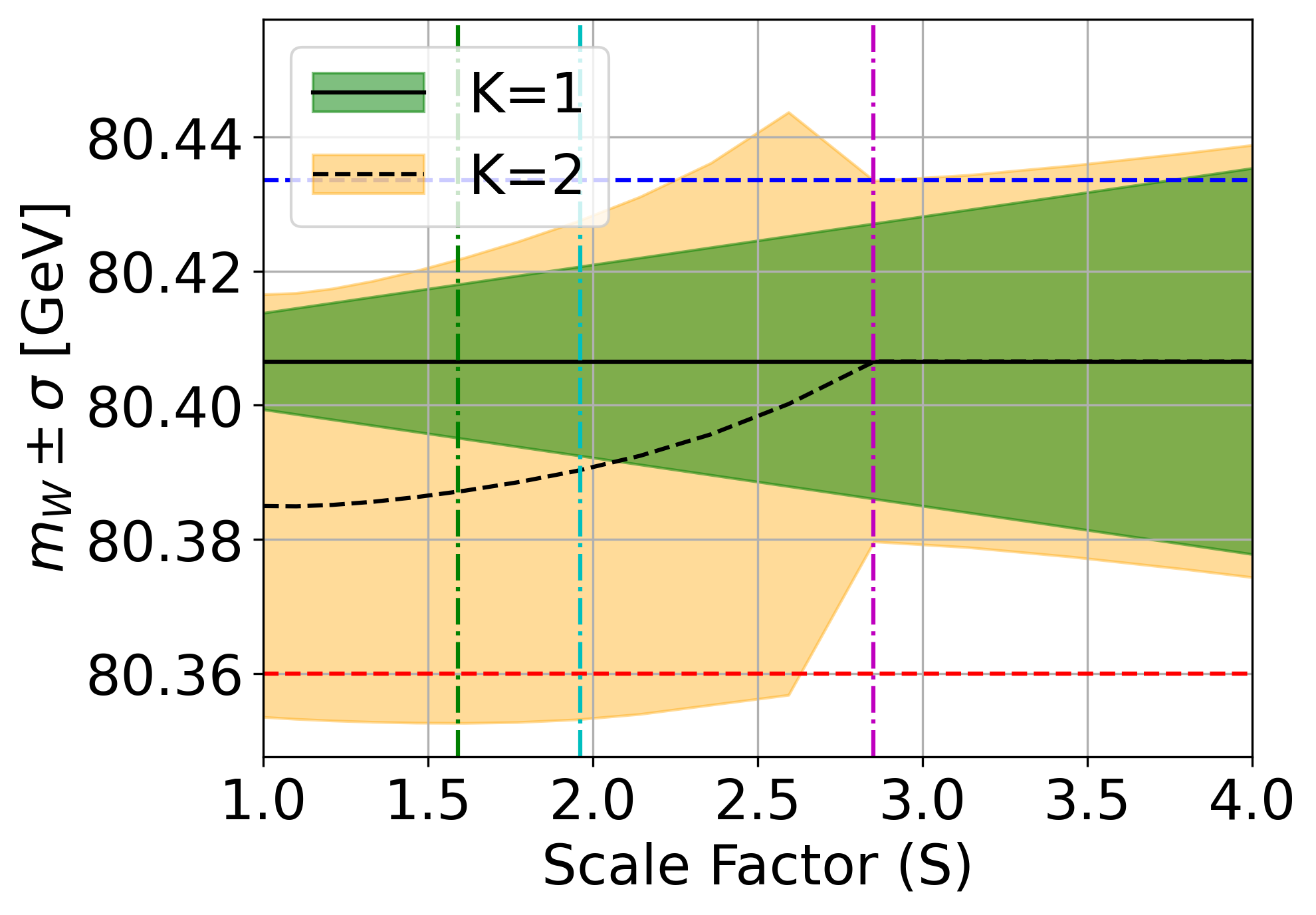}
    \caption{\textbf{Left:} Values of AIC as a function of the Scale factor (S) by which the uncertainties of experimental  measurements of the $W$-boson mass in Table~\ref{tab:mw_measurements} are multiplied. The red and blue lines represent the AIC values for fits with $K=1$ and $K=2$, respectively. The three vertical dot-dashed lines represent the values of S when the AIC for $K=1$ and $K=2$ are equal (green), the value S calculated according the the PDG recommendation (cyan) and the value of S when the mean for $K=1$ and $K=2$ coincide (magenta).
    \textbf{Right:} The mean and uncertainty of the combination of the data in Table~\ref{tab:mw_measurements} as a function of the Scale factor (S) that multiplies the uncertainties of experimental  measurements of the $W$-boson mass. The horizontal red dashed line represents the central value of the ATLAS23 measurement and the horizontal blue dashed line represents the central value of the CDF-II measurement. The three vertical lines are identical to the plot on the left and represent the same thing.}
    \label{fig:mW_scaling_uncertainty}
\end{figure}

We now consider the effect of artificially increasing the measured uncertainties on the GMM fits, which, for this simple example, is equivalent to introducing a global tolerance criteria. The GMM approach described here provides alternate measures for determining the amount by which the uncertainties should be rescaled to make the fit consistent. There are many ways to increase the uncertainty. For the purpose of simple demonstration, we choose to increase the uncertainty on all measurements given in Table~\ref{tab:mw_measurements}. We track the value of the AIC and BIC, for $K=1$ and $K=2$, as the uncertainty on each measurement is increased by scale factor ($S$). The result of doing so is shown in Fig.~\ref{fig:mW_scaling_uncertainty}.

The plot on the left in Fig.~\ref{fig:mW_scaling_uncertainty} shows the variation of the AIC for $K=1$ and $K=2$ as a function of the scaling factor ($S$). Initially, when $S=1$, the AIC is smaller for $K=2$, indicating that this is a preferred model. As $S$ increases, the AIC increases for $K=2$ and decreases for $K=1$ until they intersect at around $S \simeq 1.6$, suggesting that increasing the uncertainties of all data sets by at least this factor makes the data consistent. Also shown on the figure is $S=S_{PDG} = 1.95$. 

The plot on the right shows the mean and estimates of uncertainty of the fits for $K=1$ and $K=2$, for various values of $S$. The mean value for $K=2$ below $S\simeq 2.8$ is lower than the mean for $K=1$, since most of the data is more compatible with the ATLAS23 measurement. 
Without knowledge of what the true value of the $W$-boson mass should be, the larger uncertainty and lower value of the mean for $K=2$ and $S=1$ is more representative of the measured values across all experiments, whereas the result for $K=1$ underestimates the uncertainty and also prefers a higher value of the mean, indicating the presence of a bias. 
Importantly, the bias introduced by combining all the data shifts the $W$-boson mass to larger values and this shift is larger than the (green) uncertainty band even after scaling the uncertainties to make the data consistent. If it is not possible to unbias the fit, by omitting inconsistent data sets, then the uncertainty should be expanded to account for this bias.

Another way to say this is that if we were to predict the outcome of future experiments, based on the combined probability distribution for $K=1$, then we should expect the central value to be closer to $80.4$ with a high probability even after scaling the uncertainty. Furthermore, we  should not expect future measurements to be close to the SM theory prediction ($80.358 \pm 
 0.008$~GeV)~\cite{Baak:2014ora}. On the other hand, the result of  using $K = 2$ correctly indicates that there is a much higher probability for predicting lower values of the $W$-boson mass, since most past experiments are compatible with a lower value of the $W$-boson mass. Further, the larger uncertainty accommodates  the discrepancy between experimental measurements and allows us to predict larger values for future measurements of the $W$-boson mass as well.  

We end with a reminder that in the ideal situation, we would be able to go back to the experimental measurements and identify possible reasons for the discrepancy before combining the fits. While this may be possible for the $W$-boson mass, it is not typically the case for fits to parton distribution functions. We must therefore rely almost entirely on the existing data to provide a better estimate of the uncertainty of PDF fits.

%% file: main.bbl
\begin{thebibliography}{74}%
\makeatletter
\providecommand \@ifxundefined [1]{%
 \@ifx{#1\undefined}
}%
\providecommand \@ifnum [1]{%
 \ifnum #1\expandafter \@firstoftwo
 \else \expandafter \@secondoftwo
 \fi
}%
\providecommand \@ifx [1]{%
 \ifx #1\expandafter \@firstoftwo
 \else \expandafter \@secondoftwo
 \fi
}%
\providecommand \natexlab [1]{#1}%
\providecommand \enquote  [1]{``#1''}%
\providecommand \bibnamefont  [1]{#1}%
\providecommand \bibfnamefont [1]{#1}%
\providecommand \citenamefont [1]{#1}%
\providecommand \href@noop [0]{\@secondoftwo}%
\providecommand \href [0]{\begingroup \@sanitize@url \@href}%
\providecommand \@href[1]{\@@startlink{#1}\@@href}%
\providecommand \@@href[1]{\endgroup#1\@@endlink}%
\providecommand \@sanitize@url [0]{\catcode `\\12\catcode `\$12\catcode
  `\&12\catcode `\#12\catcode `\^12\catcode `\_12\catcode `\%12\relax}%
\providecommand \@@startlink[1]{}%
\providecommand \@@endlink[0]{}%
\providecommand \url  [0]{\begingroup\@sanitize@url \@url }%
\providecommand \@url [1]{\endgroup\@href {#1}{\urlprefix }}%
\providecommand \urlprefix  [0]{URL }%
\providecommand \Eprint [0]{\href }%
\providecommand \doibase [0]{https://doi.org/}%
\providecommand \selectlanguage [0]{\@gobble}%
\providecommand \bibinfo  [0]{\@secondoftwo}%
\providecommand \bibfield  [0]{\@secondoftwo}%
\providecommand \translation [1]{[#1]}%
\providecommand \BibitemOpen [0]{}%
\providecommand \bibitemStop [0]{}%
\providecommand \bibitemNoStop [0]{.\EOS\space}%
\providecommand \EOS [0]{\spacefactor3000\relax}%
\providecommand \BibitemShut  [1]{\csname bibitem#1\endcsname}%
\let\auto@bib@innerbib\@empty
\bibitem [{\citenamefont {Hou}\ \emph {et~al.}(2021)\citenamefont {Hou} \emph
  {et~al.}}]{Hou:2019efy}%
  \BibitemOpen
  \bibfield  {author} {\bibinfo {author} {\bibfnamefont {T.-J.}\ \bibnamefont
  {Hou}} \emph {et~al.},\ }\bibfield  {title} {\bibinfo {title} {{New CTEQ
  global analysis of quantum chromodynamics with high-precision data from the
  LHC}},\ }\href {https://doi.org/10.1103/PhysRevD.103.014013} {\bibfield
  {journal} {\bibinfo  {journal} {Phys. Rev. D}\ }\textbf {\bibinfo {volume}
  {103}},\ \bibinfo {pages} {014013} (\bibinfo {year} {2021})},\ \Eprint
  {https://arxiv.org/abs/1912.10053} {arXiv:1912.10053 [hep-ph]} \BibitemShut
  {NoStop}%
\bibitem [{\citenamefont {Bailey}\ \emph {et~al.}(2021)\citenamefont {Bailey},
  \citenamefont {Cridge}, \citenamefont {Harland-Lang}, \citenamefont
  {Martin},\ and\ \citenamefont {Thorne}}]{Bailey:2020ooq}%
  \BibitemOpen
  \bibfield  {author} {\bibinfo {author} {\bibfnamefont {S.}~\bibnamefont
  {Bailey}}, \bibinfo {author} {\bibfnamefont {T.}~\bibnamefont {Cridge}},
  \bibinfo {author} {\bibfnamefont {L.~A.}\ \bibnamefont {Harland-Lang}},
  \bibinfo {author} {\bibfnamefont {A.~D.}\ \bibnamefont {Martin}},\ and\
  \bibinfo {author} {\bibfnamefont {R.~S.}\ \bibnamefont {Thorne}},\ }\bibfield
   {title} {\bibinfo {title} {{Parton distributions from LHC, HERA, Tevatron
  and fixed target data: MSHT20 PDFs}},\ }\href
  {https://doi.org/10.1140/epjc/s10052-021-09057-0} {\bibfield  {journal}
  {\bibinfo  {journal} {Eur. Phys. J. C}\ }\textbf {\bibinfo {volume} {81}},\
  \bibinfo {pages} {341} (\bibinfo {year} {2021})},\ \Eprint
  {https://arxiv.org/abs/2012.04684} {arXiv:2012.04684 [hep-ph]} \BibitemShut
  {NoStop}%
\bibitem [{\citenamefont {Ball}\ \emph
  {et~al.}(2022{\natexlab{a}})\citenamefont {Ball} \emph
  {et~al.}}]{NNPDF:2021njg}%
  \BibitemOpen
  \bibfield  {author} {\bibinfo {author} {\bibfnamefont {R.~D.}\ \bibnamefont
  {Ball}} \emph {et~al.} (\bibinfo {collaboration} {NNPDF}),\ }\bibfield
  {title} {\bibinfo {title} {{The path to proton structure at 1\% accuracy}},\
  }\href {https://doi.org/10.1140/epjc/s10052-022-10328-7} {\bibfield
  {journal} {\bibinfo  {journal} {Eur. Phys. J. C}\ }\textbf {\bibinfo {volume}
  {82}},\ \bibinfo {pages} {428} (\bibinfo {year} {2022}{\natexlab{a}})},\
  \Eprint {https://arxiv.org/abs/2109.02653} {arXiv:2109.02653 [hep-ph]}
  \BibitemShut {NoStop}%
\bibitem [{\citenamefont {Alekhin}\ \emph {et~al.}(2017)\citenamefont
  {Alekhin}, \citenamefont {Bl\"umlein}, \citenamefont {Moch},\ and\
  \citenamefont {Placakyte}}]{Alekhin:2017kpj}%
  \BibitemOpen
  \bibfield  {author} {\bibinfo {author} {\bibfnamefont {S.}~\bibnamefont
  {Alekhin}}, \bibinfo {author} {\bibfnamefont {J.}~\bibnamefont {Bl\"umlein}},
  \bibinfo {author} {\bibfnamefont {S.}~\bibnamefont {Moch}},\ and\ \bibinfo
  {author} {\bibfnamefont {R.}~\bibnamefont {Placakyte}},\ }\bibfield  {title}
  {\bibinfo {title} {{Parton distribution functions, $\alpha_s$, and
  heavy-quark masses for LHC Run II}},\ }\href
  {https://doi.org/10.1103/PhysRevD.96.014011} {\bibfield  {journal} {\bibinfo
  {journal} {Phys. Rev. D}\ }\textbf {\bibinfo {volume} {96}},\ \bibinfo
  {pages} {014011} (\bibinfo {year} {2017})},\ \Eprint
  {https://arxiv.org/abs/1701.05838} {arXiv:1701.05838 [hep-ph]} \BibitemShut
  {NoStop}%
\bibitem [{\citenamefont {Aad}\ \emph {et~al.}(2022)\citenamefont {Aad} \emph
  {et~al.}}]{ATLAS:2021vod}%
  \BibitemOpen
  \bibfield  {author} {\bibinfo {author} {\bibfnamefont {G.}~\bibnamefont
  {Aad}} \emph {et~al.} (\bibinfo {collaboration} {ATLAS}),\ }\bibfield
  {title} {\bibinfo {title} {{Determination of the parton distribution
  functions of the proton using diverse ATLAS data from $pp$ collisions at
  $\sqrt{s} = 7$, 8 and 13~TeV}},\ }\href
  {https://doi.org/10.1140/epjc/s10052-022-10217-z} {\bibfield  {journal}
  {\bibinfo  {journal} {Eur. Phys. J. C}\ }\textbf {\bibinfo {volume} {82}},\
  \bibinfo {pages} {438} (\bibinfo {year} {2022})},\ \Eprint
  {https://arxiv.org/abs/2112.11266} {arXiv:2112.11266 [hep-ex]} \BibitemShut
  {NoStop}%
\bibitem [{\citenamefont {Ball}\ \emph {et~al.}(2021)\citenamefont {Ball} \emph
  {et~al.}}]{NNPDF:2021uiq}%
  \BibitemOpen
  \bibfield  {author} {\bibinfo {author} {\bibfnamefont {R.~D.}\ \bibnamefont
  {Ball}} \emph {et~al.} (\bibinfo {collaboration} {NNPDF}),\ }\bibfield
  {title} {\bibinfo {title} {{An open-source machine learning framework for
  global analyses of parton distributions}},\ }\href
  {https://doi.org/10.1140/epjc/s10052-021-09747-9} {\bibfield  {journal}
  {\bibinfo  {journal} {Eur. Phys. J. C}\ }\textbf {\bibinfo {volume} {81}},\
  \bibinfo {pages} {958} (\bibinfo {year} {2021})},\ \Eprint
  {https://arxiv.org/abs/2109.02671} {arXiv:2109.02671 [hep-ph]} \BibitemShut
  {NoStop}%
\bibitem [{\citenamefont {Alekhin}\ \emph {et~al.}(2015)\citenamefont {Alekhin}
  \emph {et~al.}}]{Alekhin:2014irh}%
  \BibitemOpen
  \bibfield  {author} {\bibinfo {author} {\bibfnamefont {S.}~\bibnamefont
  {Alekhin}} \emph {et~al.},\ }\bibfield  {title} {\bibinfo {title}
  {{HERAFitter}},\ }\href {https://doi.org/10.1140/epjc/s10052-015-3480-z}
  {\bibfield  {journal} {\bibinfo  {journal} {Eur. Phys. J. C}\ }\textbf
  {\bibinfo {volume} {75}},\ \bibinfo {pages} {304} (\bibinfo {year} {2015})},\
  \Eprint {https://arxiv.org/abs/1410.4412} {arXiv:1410.4412 [hep-ph]}
  \BibitemShut {NoStop}%
\bibitem [{\citenamefont {Amoroso}\ \emph {et~al.}(2022)\citenamefont {Amoroso}
  \emph {et~al.}}]{Amoroso:2022eow}%
  \BibitemOpen
  \bibfield  {author} {\bibinfo {author} {\bibfnamefont {S.}~\bibnamefont
  {Amoroso}} \emph {et~al.},\ }\bibfield  {title} {\bibinfo {title} {{Snowmass
  2021 whitepaper: Proton structure at the precision frontier}},\ }\href
  {https://doi.org/10.5506/APhysPolB.53.12-A1} {\bibfield  {journal} {\bibinfo
  {journal} {Acta Phys. Polon. B}\ }\textbf {\bibinfo {volume} {53}},\ \bibinfo
  {pages} {A1} (\bibinfo {year} {2022})},\ \Eprint
  {https://arxiv.org/abs/2203.13923} {arXiv:2203.13923 [hep-ph]} \BibitemShut
  {NoStop}%
\bibitem [{\citenamefont {Aaltonen}\ \emph {et~al.}(2022)\citenamefont
  {Aaltonen} \emph {et~al.}}]{CDF:2022hxs}%
  \BibitemOpen
  \bibfield  {author} {\bibinfo {author} {\bibfnamefont {T.}~\bibnamefont
  {Aaltonen}} \emph {et~al.} (\bibinfo {collaboration} {CDF}),\ }\bibfield
  {title} {\bibinfo {title} {{High-precision measurement of the W boson mass
  with the CDF II detector}},\ }\href {https://doi.org/10.1126/science.abk1781}
  {\bibfield  {journal} {\bibinfo  {journal} {Science}\ }\textbf {\bibinfo
  {volume} {376}},\ \bibinfo {pages} {170} (\bibinfo {year}
  {2022})}\BibitemShut {NoStop}%
\bibitem [{\citenamefont {Cepeda}\ \emph {et~al.}(2019)\citenamefont {Cepeda}
  \emph {et~al.}}]{Cepeda:2019klc}%
  \BibitemOpen
  \bibfield  {author} {\bibinfo {author} {\bibfnamefont {M.}~\bibnamefont
  {Cepeda}} \emph {et~al.},\ }\bibfield  {title} {\bibinfo {title} {{Report
  from Working Group 2}: {Higgs Physics at the HL-LHC and HE-LHC}},\ }\href
  {https://doi.org/10.23731/CYRM-2019-007.221} {\bibfield  {journal} {\bibinfo
  {journal} {CERN Yellow Rep. Monogr.}\ }\textbf {\bibinfo {volume} {7}},\
  \bibinfo {pages} {221} (\bibinfo {year} {2019})},\ \Eprint
  {https://arxiv.org/abs/1902.00134} {arXiv:1902.00134 [hep-ph]} \BibitemShut
  {NoStop}%
\bibitem [{\citenamefont {de~Florian}\ \emph {et~al.}(2016)\citenamefont
  {de~Florian} \emph {et~al.}}]{LHCHiggsCrossSectionWorkingGroup:2016ypw}%
  \BibitemOpen
  \bibfield  {author} {\bibinfo {author} {\bibfnamefont {D.}~\bibnamefont
  {de~Florian}} \emph {et~al.} (\bibinfo {collaboration} {LHC Higgs Cross
  Section Working Group}),\ }\href {https://doi.org/10.23731/CYRM-2017-002}
  {\bibinfo {title} {{Handbook of LHC Higgs Cross Sections: 4. Deciphering the
  Nature of the Higgs Sector}}} (\bibinfo {year} {2016}),\ \Eprint
  {https://arxiv.org/abs/1610.07922} {arXiv:1610.07922 [hep-ph]} \BibitemShut
  {NoStop}%
\bibitem [{\citenamefont {Aaboud}\ \emph {et~al.}(2017)\citenamefont {Aaboud}
  \emph {et~al.}}]{ATLAS:2016nqi}%
  \BibitemOpen
  \bibfield  {author} {\bibinfo {author} {\bibfnamefont {M.}~\bibnamefont
  {Aaboud}} \emph {et~al.} (\bibinfo {collaboration} {ATLAS}),\ }\bibfield
  {title} {\bibinfo {title} {{Precision measurement and interpretation of
  inclusive $W^+$ , $W^-$ and $Z/\gamma ^*$ production cross sections with the
  ATLAS detector}},\ }\href {https://doi.org/10.1140/epjc/s10052-017-4911-9}
  {\bibfield  {journal} {\bibinfo  {journal} {Eur. Phys. J. C}\ }\textbf
  {\bibinfo {volume} {77}},\ \bibinfo {pages} {367} (\bibinfo {year} {2017})},\
  \Eprint {https://arxiv.org/abs/1612.03016} {arXiv:1612.03016 [hep-ex]}
  \BibitemShut {NoStop}%
\bibitem [{\citenamefont {Hou}\ \emph {et~al.}(2022{\natexlab{a}})\citenamefont
  {Hou}, \citenamefont {Lin}, \citenamefont {Yan},\ and\ \citenamefont
  {Yuan}}]{Hou:2022sdf}%
  \BibitemOpen
  \bibfield  {author} {\bibinfo {author} {\bibfnamefont {T.-J.}\ \bibnamefont
  {Hou}}, \bibinfo {author} {\bibfnamefont {H.-W.}\ \bibnamefont {Lin}},
  \bibinfo {author} {\bibfnamefont {M.}~\bibnamefont {Yan}},\ and\ \bibinfo
  {author} {\bibfnamefont {C.~P.}\ \bibnamefont {Yuan}},\ }\href@noop {}
  {\bibinfo {title} {{Impact of lattice $s(x)-\bar{s}(x)$ data in the CTEQ-TEA
  global analysis}}} (\bibinfo {year} {2022}{\natexlab{a}}),\ \Eprint
  {https://arxiv.org/abs/2204.07944} {2204.07944 [hep-ph]} \BibitemShut
  {NoStop}%
\bibitem [{\citenamefont {Guzzi}\ \emph {et~al.}(2022)\citenamefont {Guzzi}
  \emph {et~al.}}]{Guzzi:2021fre}%
  \BibitemOpen
  \bibfield  {author} {\bibinfo {author} {\bibfnamefont {M.}~\bibnamefont
  {Guzzi}} \emph {et~al.},\ }\bibfield  {title} {\bibinfo {title} {{NNLO
  constraints on proton PDFs from the SeaQuest and STAR experiments and other
  developments in the CTEQ-TEA global analysis}},\ }\href
  {https://doi.org/10.21468/SciPostPhysProc.8.005} {\bibfield  {journal}
  {\bibinfo  {journal} {SciPost Phys. Proc.}\ }\textbf {\bibinfo {volume}
  {8}},\ \bibinfo {pages} {005} (\bibinfo {year} {2022})},\ \Eprint
  {https://arxiv.org/abs/2108.06596} {arXiv:2108.06596 [hep-ph]} \BibitemShut
  {NoStop}%
\bibitem [{\citenamefont {Hou}\ \emph {et~al.}(2022{\natexlab{b}})\citenamefont
  {Hou}, \citenamefont {Yan}, \citenamefont {Liang}, \citenamefont {Liu},\ and\
  \citenamefont {Yuan}}]{Hou:2022ajg}%
  \BibitemOpen
  \bibfield  {author} {\bibinfo {author} {\bibfnamefont {T.-J.}\ \bibnamefont
  {Hou}}, \bibinfo {author} {\bibfnamefont {M.}~\bibnamefont {Yan}}, \bibinfo
  {author} {\bibfnamefont {J.}~\bibnamefont {Liang}}, \bibinfo {author}
  {\bibfnamefont {K.-F.}\ \bibnamefont {Liu}},\ and\ \bibinfo {author}
  {\bibfnamefont {C.~P.}\ \bibnamefont {Yuan}},\ }\bibfield  {title} {\bibinfo
  {title} {{Connected and disconnected sea partons from the CT18
  parametrization of PDFs}},\ }\href
  {https://doi.org/10.1103/PhysRevD.106.096008} {\bibfield  {journal} {\bibinfo
   {journal} {Phys. Rev. D}\ }\textbf {\bibinfo {volume} {106}},\ \bibinfo
  {pages} {096008} (\bibinfo {year} {2022}{\natexlab{b}})},\ \Eprint
  {https://arxiv.org/abs/2206.02431} {arXiv:2206.02431 [hep-ph]} \BibitemShut
  {NoStop}%
\bibitem [{\citenamefont {Kova\v{r}\'\i{}k}\ \emph {et~al.}(2020)\citenamefont
  {Kova\v{r}\'\i{}k}, \citenamefont {Nadolsky},\ and\ \citenamefont
  {Soper}}]{Kovarik:2019xvh}%
  \BibitemOpen
  \bibfield  {author} {\bibinfo {author} {\bibfnamefont {K.}~\bibnamefont
  {Kova\v{r}\'\i{}k}}, \bibinfo {author} {\bibfnamefont {P.~M.}\ \bibnamefont
  {Nadolsky}},\ and\ \bibinfo {author} {\bibfnamefont {D.~E.}\ \bibnamefont
  {Soper}},\ }\bibfield  {title} {\bibinfo {title} {{Hadronic structure in
  high-energy collisions}},\ }\href
  {https://doi.org/10.1103/RevModPhys.92.045003} {\bibfield  {journal}
  {\bibinfo  {journal} {Rev. Mod. Phys.}\ }\textbf {\bibinfo {volume} {92}},\
  \bibinfo {pages} {045003} (\bibinfo {year} {2020})},\ \Eprint
  {https://arxiv.org/abs/1905.06957} {arXiv:1905.06957 [hep-ph]} \BibitemShut
  {NoStop}%
\bibitem [{\citenamefont {Ball}\ \emph
  {et~al.}(2022{\natexlab{b}})\citenamefont {Ball}, \citenamefont {Candido},
  \citenamefont {Cruz-Martinez}, \citenamefont {Forte}, \citenamefont {Giani},
  \citenamefont {Hekhorn}, \citenamefont {Kudashkin}, \citenamefont {Magni},\
  and\ \citenamefont {Rojo}}]{Ball:2022qks}%
  \BibitemOpen
  \bibfield  {author} {\bibinfo {author} {\bibfnamefont {R.~D.}\ \bibnamefont
  {Ball}}, \bibinfo {author} {\bibfnamefont {A.}~\bibnamefont {Candido}},
  \bibinfo {author} {\bibfnamefont {J.}~\bibnamefont {Cruz-Martinez}}, \bibinfo
  {author} {\bibfnamefont {S.}~\bibnamefont {Forte}}, \bibinfo {author}
  {\bibfnamefont {T.}~\bibnamefont {Giani}}, \bibinfo {author} {\bibfnamefont
  {F.}~\bibnamefont {Hekhorn}}, \bibinfo {author} {\bibfnamefont
  {K.}~\bibnamefont {Kudashkin}}, \bibinfo {author} {\bibfnamefont
  {G.}~\bibnamefont {Magni}},\ and\ \bibinfo {author} {\bibfnamefont
  {J.}~\bibnamefont {Rojo}} (\bibinfo {collaboration} {NNPDF}),\ }\bibfield
  {title} {\bibinfo {title} {{Evidence for intrinsic charm quarks in the
  proton}},\ }\href {https://doi.org/10.1038/s41586-022-04998-2} {\bibfield
  {journal} {\bibinfo  {journal} {Nature}\ }\textbf {\bibinfo {volume} {608}},\
  \bibinfo {pages} {483} (\bibinfo {year} {2022}{\natexlab{b}})},\ \Eprint
  {https://arxiv.org/abs/2208.08372} {arXiv:2208.08372 [hep-ph]} \BibitemShut
  {NoStop}%
\bibitem [{\citenamefont {Stegeman}(2022)}]{Stegeman:2022wrn}%
  \BibitemOpen
  \bibfield  {author} {\bibinfo {author} {\bibfnamefont {R.}~\bibnamefont
  {Stegeman}},\ }\bibfield  {title} {\bibinfo {title} {{The negligible impact
  of experimental inconsistencies in the NNPDF4.0 global dataset}},\ }\href
  {https://doi.org/10.22323/1.414.0787} {\bibfield  {journal} {\bibinfo
  {journal} {PoS}\ }\textbf {\bibinfo {volume} {ICHEP2022}},\ \bibinfo {pages}
  {787} (\bibinfo {year} {2022})},\ \Eprint {https://arxiv.org/abs/2212.07703}
  {arXiv:2212.07703 [hep-ph]} \BibitemShut {NoStop}%
\bibitem [{\citenamefont {Lambri}(2020)}]{lambri2020thesis}%
  \BibitemOpen
  \bibfield  {author} {\bibinfo {author} {\bibfnamefont {N.}~\bibnamefont
  {Lambri}},\ }\emph {\bibinfo {title} {Optimized Regression Models for Parton
  Distribution Functions Determination Using Deep Learning Models}},\ \href
  {http://nnpdf.mi.infn.it/wp-content/uploads/2021/01/thesis_lambri.pdf}
  {Master's thesis},\ \bibinfo  {school} {Università degli Studi di Milano}
  (\bibinfo {year} {2020}),\ \bibinfo {note} {master's thesis, accessed:
  2024-07-15}\BibitemShut {NoStop}%
\bibitem [{\citenamefont {Talon}(2019)}]{thesis:talon}%
  \BibitemOpen
  \bibfield  {author} {\bibinfo {author} {\bibfnamefont {C.}~\bibnamefont
  {Talon}},\ }\emph {\bibinfo {title} {Machine Learning and Parton
  Distributions}},\ \href
  {http://nnpdf.mi.infn.it/wp-content/uploads/2019/06/Talon_MS_thesis.pdf}
  {\bibinfo {type} {Master's thesis}},\ \bibinfo  {school} {University of
  Milan} (\bibinfo {year} {2019}),\ \bibinfo {note} {accessed:
  2024-07-15}\BibitemShut {NoStop}%
\bibitem [{\citenamefont {Cridge}(2022)}]{Cridge:2021qjj}%
  \BibitemOpen
  \bibfield  {author} {\bibinfo {author} {\bibfnamefont {T.}~\bibnamefont
  {Cridge}} (\bibinfo {collaboration} {PDF4LHC21 combination group}),\
  }\bibfield  {title} {\bibinfo {title} {{PDF4LHC21: Update on the benchmarking
  of the CT, MSHT and NNPDF global PDF fits}},\ }\href
  {https://doi.org/10.21468/SciPostPhysProc.8.101} {\bibfield  {journal}
  {\bibinfo  {journal} {SciPost Phys. Proc.}\ }\textbf {\bibinfo {volume}
  {8}},\ \bibinfo {pages} {101} (\bibinfo {year} {2022})},\ \Eprint
  {https://arxiv.org/abs/2108.09099} {arXiv:2108.09099 [hep-ph]} \BibitemShut
  {NoStop}%
\bibitem [{\citenamefont {Gelman}\ \emph {et~al.}(2013)\citenamefont {Gelman},
  \citenamefont {Carlin}, \citenamefont {Stern}, \citenamefont {Dunson},
  \citenamefont {Vehtari},\ and\ \citenamefont {Rubin}}]{gelman2013bayesian}%
  \BibitemOpen
  \bibfield  {author} {\bibinfo {author} {\bibfnamefont {A.}~\bibnamefont
  {Gelman}}, \bibinfo {author} {\bibfnamefont {J.}~\bibnamefont {Carlin}},
  \bibinfo {author} {\bibfnamefont {H.}~\bibnamefont {Stern}}, \bibinfo
  {author} {\bibfnamefont {D.}~\bibnamefont {Dunson}}, \bibinfo {author}
  {\bibfnamefont {A.}~\bibnamefont {Vehtari}},\ and\ \bibinfo {author}
  {\bibfnamefont {D.}~\bibnamefont {Rubin}},\ }\href
  {https://doi.org/10.1201/b16018} {\emph {\bibinfo {title} {Bayesian Data
  Analysis, Third Edition}}},\ Chapman \& Hall/CRC Texts in Statistical
  Science\ (\bibinfo  {publisher} {Taylor \& Francis},\ \bibinfo {year}
  {2013})\BibitemShut {NoStop}%
\bibitem [{\citenamefont {McGlothlin}\ and\ \citenamefont
  {Viele}(2018)}]{mcglothlin2018bayesian}%
  \BibitemOpen
  \bibfield  {author} {\bibinfo {author} {\bibfnamefont {A.~E.}\ \bibnamefont
  {McGlothlin}}\ and\ \bibinfo {author} {\bibfnamefont {K.}~\bibnamefont
  {Viele}},\ }\bibfield  {title} {\bibinfo {title} {Bayesian hierarchical
  models},\ }\href@noop {} {\bibfield  {journal} {\bibinfo  {journal} {Jama}\
  }\textbf {\bibinfo {volume} {320}},\ \bibinfo {pages} {2365} (\bibinfo {year}
  {2018})}\BibitemShut {NoStop}%
\bibitem [{\citenamefont {Erler}\ and\ \citenamefont
  {Ferro-Hern\'andez}(2020)}]{Erler:2020bif}%
  \BibitemOpen
  \bibfield  {author} {\bibinfo {author} {\bibfnamefont {J.}~\bibnamefont
  {Erler}}\ and\ \bibinfo {author} {\bibfnamefont {R.}~\bibnamefont
  {Ferro-Hern\'andez}},\ }\bibfield  {title} {\bibinfo {title} {{Alternative to
  the application of PDG scale factors}},\ }\href
  {https://doi.org/10.1140/epjc/s10052-020-8115-3} {\bibfield  {journal}
  {\bibinfo  {journal} {Eur. Phys. J. C}\ }\textbf {\bibinfo {volume} {80}},\
  \bibinfo {pages} {541} (\bibinfo {year} {2020})},\ \Eprint
  {https://arxiv.org/abs/2004.01219} {arXiv:2004.01219 [physics.data-an]}
  \BibitemShut {NoStop}%
\bibitem [{\citenamefont {Murphy}(2012)}]{Murphy:ML}%
  \BibitemOpen
  \bibfield  {author} {\bibinfo {author} {\bibfnamefont {K.~P.}\ \bibnamefont
  {Murphy}},\ }\href@noop {} {\emph {\bibinfo {title} {Machine learning: a
  probabilistic perspective}}}\ (\bibinfo  {publisher} {MIT Press},\ \bibinfo
  {address} {Cambridge, MA},\ \bibinfo {year} {2012})\BibitemShut {NoStop}%
\bibitem [{\citenamefont {Bishop}(2007)}]{bishop2007}%
  \BibitemOpen
  \bibfield  {author} {\bibinfo {author} {\bibfnamefont {C.~M.}\ \bibnamefont
  {Bishop}},\ }\href
  {http://www.amazon.com/Pattern-Recognition-Learning-Information-Statistics/dp/0387310738%3FSubscriptionId%3D13CT5CVB80YFWJEPWS02%26tag%3Dws%26linkCode%3Dxm2%26camp%3D2025%26creative%3D165953%26creativeASIN%3D0387310738}
  {\emph {\bibinfo {title} {Pattern Recognition and Machine Learning
  (Information Science and Statistics)}}},\ \bibinfo {edition} {1st}\ ed.\
  (\bibinfo  {publisher} {Springer},\ \bibinfo {year} {2007})\BibitemShut
  {NoStop}%
\bibitem [{\citenamefont {Leamer}\ and\ \citenamefont
  {Leamer}(1978)}]{leamer1978specification}%
  \BibitemOpen
  \bibfield  {author} {\bibinfo {author} {\bibfnamefont {E.~E.}\ \bibnamefont
  {Leamer}}\ and\ \bibinfo {author} {\bibfnamefont {E.~E.}\ \bibnamefont
  {Leamer}},\ }\href@noop {} {\emph {\bibinfo {title} {Specification searches:
  Ad hoc inference with nonexperimental data}}},\ Vol.~\bibinfo {volume} {53}\
  (\bibinfo  {publisher} {Wiley New York},\ \bibinfo {year} {1978})\BibitemShut
  {NoStop}%
\bibitem [{\citenamefont {George}\ and\ \citenamefont
  {McCulloch}(1993)}]{george1993variable}%
  \BibitemOpen
  \bibfield  {author} {\bibinfo {author} {\bibfnamefont {E.~I.}\ \bibnamefont
  {George}}\ and\ \bibinfo {author} {\bibfnamefont {R.~E.}\ \bibnamefont
  {McCulloch}},\ }\bibfield  {title} {\bibinfo {title} {Variable selection via
  gibbs sampling},\ }\href@noop {} {\bibfield  {journal} {\bibinfo  {journal}
  {Journal of the American Statistical Association}\ }\textbf {\bibinfo
  {volume} {88}},\ \bibinfo {pages} {881} (\bibinfo {year} {1993})}\BibitemShut
  {NoStop}%
\bibitem [{\citenamefont {Kass}\ and\ \citenamefont
  {Raftery}(1995)}]{kass1995bayes}%
  \BibitemOpen
  \bibfield  {author} {\bibinfo {author} {\bibfnamefont {R.~E.}\ \bibnamefont
  {Kass}}\ and\ \bibinfo {author} {\bibfnamefont {A.~E.}\ \bibnamefont
  {Raftery}},\ }\bibfield  {title} {\bibinfo {title} {Bayes factors},\
  }\href@noop {} {\bibfield  {journal} {\bibinfo  {journal} {Journal of the
  american statistical association}\ }\textbf {\bibinfo {volume} {90}},\
  \bibinfo {pages} {773} (\bibinfo {year} {1995})}\BibitemShut {NoStop}%
\bibitem [{\citenamefont {Hoeting}\ \emph {et~al.}(1999)\citenamefont
  {Hoeting}, \citenamefont {Madigan}, \citenamefont {Raftery},\ and\
  \citenamefont {Volinsky}}]{hoeting1999}%
  \BibitemOpen
  \bibfield  {author} {\bibinfo {author} {\bibfnamefont {J.~A.}\ \bibnamefont
  {Hoeting}}, \bibinfo {author} {\bibfnamefont {D.}~\bibnamefont {Madigan}},
  \bibinfo {author} {\bibfnamefont {A.~E.}\ \bibnamefont {Raftery}},\ and\
  \bibinfo {author} {\bibfnamefont {C.~T.}\ \bibnamefont {Volinsky}},\
  }\bibfield  {title} {\bibinfo {title} {{Bayesian model averaging: a tutorial
  (with comments by M. Clyde, David Draper and E. I. George, and a rejoinder by
  the authors}},\ }\href {https://doi.org/10.1214/ss/1009212519} {\bibfield
  {journal} {\bibinfo  {journal} {Statistical Science}\ }\textbf {\bibinfo
  {volume} {14}},\ \bibinfo {pages} {382 } (\bibinfo {year}
  {1999})}\BibitemShut {NoStop}%
\bibitem [{\citenamefont {Raftery}\ and\ \citenamefont
  {Zheng}(2003)}]{raftery2003discussion}%
  \BibitemOpen
  \bibfield  {author} {\bibinfo {author} {\bibfnamefont {A.~E.}\ \bibnamefont
  {Raftery}}\ and\ \bibinfo {author} {\bibfnamefont {Y.}~\bibnamefont
  {Zheng}},\ }\bibfield  {title} {\bibinfo {title} {Discussion: Performance of
  bayesian model averaging},\ }\href@noop {} {\bibfield  {journal} {\bibinfo
  {journal} {Journal of the American Statistical Association}\ }\textbf
  {\bibinfo {volume} {98}},\ \bibinfo {pages} {931} (\bibinfo {year}
  {2003})}\BibitemShut {NoStop}%
\bibitem [{\citenamefont {Jay}\ and\ \citenamefont {Neil}(2021)}]{Jay:2020jkz}%
  \BibitemOpen
  \bibfield  {author} {\bibinfo {author} {\bibfnamefont {W.~I.}\ \bibnamefont
  {Jay}}\ and\ \bibinfo {author} {\bibfnamefont {E.~T.}\ \bibnamefont {Neil}},\
  }\bibfield  {title} {\bibinfo {title} {{Bayesian model averaging for analysis
  of lattice field theory results}},\ }\href
  {https://doi.org/10.1103/PhysRevD.103.114502} {\bibfield  {journal} {\bibinfo
   {journal} {Phys. Rev. D}\ }\textbf {\bibinfo {volume} {103}},\ \bibinfo
  {pages} {114502} (\bibinfo {year} {2021})},\ \Eprint
  {https://arxiv.org/abs/2008.01069} {arXiv:2008.01069 [stat.ME]} \BibitemShut
  {NoStop}%
\bibitem [{\citenamefont {Neil}\ and\ \citenamefont
  {Sitison}(2024)}]{Neil:2022joj}%
  \BibitemOpen
  \bibfield  {author} {\bibinfo {author} {\bibfnamefont {E.~T.}\ \bibnamefont
  {Neil}}\ and\ \bibinfo {author} {\bibfnamefont {J.~W.}\ \bibnamefont
  {Sitison}},\ }\bibfield  {title} {\bibinfo {title} {{Improved information
  criteria for Bayesian model averaging in lattice field theory}},\ }\href
  {https://doi.org/10.1103/PhysRevD.109.014510} {\bibfield  {journal} {\bibinfo
   {journal} {Phys. Rev. D}\ }\textbf {\bibinfo {volume} {109}},\ \bibinfo
  {pages} {014510} (\bibinfo {year} {2024})},\ \Eprint
  {https://arxiv.org/abs/2208.14983} {arXiv:2208.14983 [stat.ME]} \BibitemShut
  {NoStop}%
\bibitem [{\citenamefont {Neil}\ and\ \citenamefont
  {Sitison}(2023)}]{Neil:2023pgt}%
  \BibitemOpen
  \bibfield  {author} {\bibinfo {author} {\bibfnamefont {E.~T.}\ \bibnamefont
  {Neil}}\ and\ \bibinfo {author} {\bibfnamefont {J.~W.}\ \bibnamefont
  {Sitison}},\ }\bibfield  {title} {\bibinfo {title} {{Model averaging
  approaches to data subset selection}},\ }\href
  {https://doi.org/10.1103/PhysRevE.108.045308} {\bibfield  {journal} {\bibinfo
   {journal} {Phys. Rev. E}\ }\textbf {\bibinfo {volume} {108}},\ \bibinfo
  {pages} {045308} (\bibinfo {year} {2023})},\ \Eprint
  {https://arxiv.org/abs/2305.19417} {arXiv:2305.19417 [stat.ME]} \BibitemShut
  {NoStop}%
\bibitem [{\citenamefont {Cybenko}(1989)}]{Cybenko:1989}%
  \BibitemOpen
  \bibfield  {author} {\bibinfo {author} {\bibfnamefont {G.}~\bibnamefont
  {Cybenko}},\ }\bibfield  {title} {\bibinfo {title} {Approximation by
  superpositions of a sigmoidal function},\ }\href@noop {} {\bibfield
  {journal} {\bibinfo  {journal} {Mathematics of control, signals, and
  systems}\ }\textbf {\bibinfo {volume} {2}},\ \bibinfo {pages} {303} (\bibinfo
  {year} {1989})}\BibitemShut {NoStop}%
\bibitem [{\citenamefont {Stone}(1937)}]{Stone:1937}%
  \BibitemOpen
  \bibfield  {author} {\bibinfo {author} {\bibfnamefont {M.~H.}\ \bibnamefont
  {Stone}},\ }\bibfield  {title} {\bibinfo {title} {Applications of the theory
  of boolean rings to general topology},\ }\href@noop {} {\bibfield  {journal}
  {\bibinfo  {journal} {Transactions of the American Mathematical Society}\
  }\textbf {\bibinfo {volume} {41}},\ \bibinfo {pages} {375} (\bibinfo {year}
  {1937})}\BibitemShut {NoStop}%
\bibitem [{\citenamefont {Stone}(1948)}]{Stone:1948}%
  \BibitemOpen
  \bibfield  {author} {\bibinfo {author} {\bibfnamefont {M.~H.}\ \bibnamefont
  {Stone}},\ }\href@noop {} {\bibinfo {title} {The generalized weierstrass
  approximation theorem}} (\bibinfo {year} {1948})\BibitemShut {NoStop}%
\bibitem [{\citenamefont {Workman}\ \emph {et~al.}(2022)\citenamefont {Workman}
  \emph {et~al.}}]{ParticleDataGroup:2022pth}%
  \BibitemOpen
  \bibfield  {author} {\bibinfo {author} {\bibfnamefont {R.~L.}\ \bibnamefont
  {Workman}} \emph {et~al.} (\bibinfo {collaboration} {Particle Data Group}),\
  }\bibfield  {title} {\bibinfo {title} {{Review of Particle Physics}},\ }\href
  {https://doi.org/10.1093/ptep/ptac097} {\bibfield  {journal} {\bibinfo
  {journal} {PTEP}\ }\textbf {\bibinfo {volume} {2022}},\ \bibinfo {pages}
  {083C01} (\bibinfo {year} {2022})}\BibitemShut {NoStop}%
\bibitem [{\citenamefont {Pearson}(1894)}]{pearson1894contributions}%
  \BibitemOpen
  \bibfield  {author} {\bibinfo {author} {\bibfnamefont {K.}~\bibnamefont
  {Pearson}},\ }\bibfield  {title} {\bibinfo {title} {Contributions to the
  mathematical theory of evolution},\ }\href@noop {} {\bibfield  {journal}
  {\bibinfo  {journal} {Philosophical Transactions of the Royal Society of
  London. A}\ }\textbf {\bibinfo {volume} {185}},\ \bibinfo {pages} {71}
  (\bibinfo {year} {1894})}\BibitemShut {NoStop}%
\bibitem [{\citenamefont {Abadi}\ \emph {et~al.}(2016)\citenamefont {Abadi},
  \citenamefont {Agarwal}, \citenamefont {Barham}, \citenamefont {Brevdo},
  \citenamefont {Chen}, \citenamefont {Citro}, \citenamefont {Corrado},
  \citenamefont {Davis}, \citenamefont {Dean}, \citenamefont {Devin},
  \citenamefont {Ghemawat}, \citenamefont {Goodfellow}, \citenamefont {Harp},
  \citenamefont {Irving}, \citenamefont {Isard}, \citenamefont {Jia},
  \citenamefont {Jozefowicz}, \citenamefont {Kaiser}, \citenamefont {Kudlur},
  \citenamefont {Levenberg}, \citenamefont {Mane}, \citenamefont {Monga},
  \citenamefont {Moore}, \citenamefont {Murray}, \citenamefont {Olah},
  \citenamefont {Schuster}, \citenamefont {Shlens}, \citenamefont {Steiner},
  \citenamefont {Sutskever}, \citenamefont {Talwar}, \citenamefont {Tucker},
  \citenamefont {Vanhoucke}, \citenamefont {Vasudevan}, \citenamefont {Viegas},
  \citenamefont {Vinyals}, \citenamefont {Warden}, \citenamefont {Wattenberg},
  \citenamefont {Wicke}, \citenamefont {Yu},\ and\ \citenamefont
  {Zheng}}]{TensorFlow:2016}%
  \BibitemOpen
  \bibfield  {author} {\bibinfo {author} {\bibfnamefont {M.}~\bibnamefont
  {Abadi}}, \bibinfo {author} {\bibfnamefont {A.}~\bibnamefont {Agarwal}},
  \bibinfo {author} {\bibfnamefont {P.}~\bibnamefont {Barham}}, \bibinfo
  {author} {\bibfnamefont {E.}~\bibnamefont {Brevdo}}, \bibinfo {author}
  {\bibfnamefont {Z.}~\bibnamefont {Chen}}, \bibinfo {author} {\bibfnamefont
  {C.}~\bibnamefont {Citro}}, \bibinfo {author} {\bibfnamefont {G.~S.}\
  \bibnamefont {Corrado}}, \bibinfo {author} {\bibfnamefont {A.}~\bibnamefont
  {Davis}}, \bibinfo {author} {\bibfnamefont {J.}~\bibnamefont {Dean}},
  \bibinfo {author} {\bibfnamefont {M.}~\bibnamefont {Devin}}, \bibinfo
  {author} {\bibfnamefont {S.}~\bibnamefont {Ghemawat}}, \bibinfo {author}
  {\bibfnamefont {I.}~\bibnamefont {Goodfellow}}, \bibinfo {author}
  {\bibfnamefont {A.}~\bibnamefont {Harp}}, \bibinfo {author} {\bibfnamefont
  {G.}~\bibnamefont {Irving}}, \bibinfo {author} {\bibfnamefont
  {M.}~\bibnamefont {Isard}}, \bibinfo {author} {\bibfnamefont
  {Y.}~\bibnamefont {Jia}}, \bibinfo {author} {\bibfnamefont {R.}~\bibnamefont
  {Jozefowicz}}, \bibinfo {author} {\bibfnamefont {L.}~\bibnamefont {Kaiser}},
  \bibinfo {author} {\bibfnamefont {M.}~\bibnamefont {Kudlur}}, \bibinfo
  {author} {\bibfnamefont {J.}~\bibnamefont {Levenberg}}, \bibinfo {author}
  {\bibfnamefont {D.}~\bibnamefont {Mane}}, \bibinfo {author} {\bibfnamefont
  {R.}~\bibnamefont {Monga}}, \bibinfo {author} {\bibfnamefont
  {S.}~\bibnamefont {Moore}}, \bibinfo {author} {\bibfnamefont
  {D.}~\bibnamefont {Murray}}, \bibinfo {author} {\bibfnamefont
  {C.}~\bibnamefont {Olah}}, \bibinfo {author} {\bibfnamefont {M.}~\bibnamefont
  {Schuster}}, \bibinfo {author} {\bibfnamefont {J.}~\bibnamefont {Shlens}},
  \bibinfo {author} {\bibfnamefont {B.}~\bibnamefont {Steiner}}, \bibinfo
  {author} {\bibfnamefont {I.}~\bibnamefont {Sutskever}}, \bibinfo {author}
  {\bibfnamefont {K.}~\bibnamefont {Talwar}}, \bibinfo {author} {\bibfnamefont
  {P.}~\bibnamefont {Tucker}}, \bibinfo {author} {\bibfnamefont
  {V.}~\bibnamefont {Vanhoucke}}, \bibinfo {author} {\bibfnamefont
  {V.}~\bibnamefont {Vasudevan}}, \bibinfo {author} {\bibfnamefont
  {F.}~\bibnamefont {Viegas}}, \bibinfo {author} {\bibfnamefont
  {O.}~\bibnamefont {Vinyals}}, \bibinfo {author} {\bibfnamefont
  {P.}~\bibnamefont {Warden}}, \bibinfo {author} {\bibfnamefont
  {M.}~\bibnamefont {Wattenberg}}, \bibinfo {author} {\bibfnamefont
  {M.}~\bibnamefont {Wicke}}, \bibinfo {author} {\bibfnamefont
  {Y.}~\bibnamefont {Yu}},\ and\ \bibinfo {author} {\bibfnamefont
  {X.}~\bibnamefont {Zheng}},\ }\href@noop {} {\bibinfo {title} {{TensorFlow:
  Large-Scale Machine Learning on Heterogeneous Distributed Systems}}}
  (\bibinfo {year} {2016}),\ \Eprint {https://arxiv.org/abs/1603.04467}
  {arXiv:1603.04467} \BibitemShut {NoStop}%
\bibitem [{\citenamefont {McLachlan}\ and\ \citenamefont
  {Peel}(2000)}]{mclachlan2000finite}%
  \BibitemOpen
  \bibfield  {author} {\bibinfo {author} {\bibfnamefont {G.}~\bibnamefont
  {McLachlan}}\ and\ \bibinfo {author} {\bibfnamefont {D.}~\bibnamefont
  {Peel}},\ }\href {https://doi.org/10.1002/0471721182} {\emph {\bibinfo
  {title} {Finite Mixture Models}}},\ Wiley Series in Probability and
  Statistics\ (\bibinfo  {publisher} {John Wiley \& Sons},\ \bibinfo {year}
  {2000})\BibitemShut {NoStop}%
\bibitem [{\citenamefont {Raftery}\ \emph {et~al.}(2005)\citenamefont
  {Raftery}, \citenamefont {Gneiting}, \citenamefont {Balabdaoui},\ and\
  \citenamefont {Polakowski}}]{raftery2005using}%
  \BibitemOpen
  \bibfield  {author} {\bibinfo {author} {\bibfnamefont {A.~E.}\ \bibnamefont
  {Raftery}}, \bibinfo {author} {\bibfnamefont {T.}~\bibnamefont {Gneiting}},
  \bibinfo {author} {\bibfnamefont {F.}~\bibnamefont {Balabdaoui}},\ and\
  \bibinfo {author} {\bibfnamefont {M.}~\bibnamefont {Polakowski}},\ }\bibfield
   {title} {\bibinfo {title} {Using bayesian model averaging to calibrate
  forecast ensembles},\ }\href {https://doi.org/10.1175/MWR2906.1} {\bibfield
  {journal} {\bibinfo  {journal} {Monthly weather review}\ }\textbf {\bibinfo
  {volume} {133}},\ \bibinfo {pages} {1155} (\bibinfo {year}
  {2005})}\BibitemShut {NoStop}%
\bibitem [{\citenamefont {B{\"u}cker}\ and\ \citenamefont
  {Corliss}(2005)}]{Bucker2005ABo}%
  \BibitemOpen
  \bibfield  {author} {\bibinfo {author} {\bibfnamefont {H.~M.}\ \bibnamefont
  {B{\"u}cker}}\ and\ \bibinfo {author} {\bibfnamefont {G.~F.}\ \bibnamefont
  {Corliss}},\ }\bibfield  {title} {\bibinfo {title} {A bibliography on
  automatic differentiation},\ }in\ \href
  {https://doi.org/10.1007/3-540-28438-9_28} {\emph {\bibinfo {booktitle}
  {Automatic Differentiation}}},\ \bibinfo {series} {Lecture Notes in
  Computational Science and Engineering}, Vol.~\bibinfo {volume} {50},\
  \bibinfo {editor} {edited by\ \bibinfo {editor} {\bibfnamefont {H.~M.}\
  \bibnamefont {B{\"u}cker}}, \bibinfo {editor} {\bibfnamefont {G.~F.}\
  \bibnamefont {Corliss}}, \bibinfo {editor} {\bibfnamefont {P.~D.}\
  \bibnamefont {Hovland}}, \bibinfo {editor} {\bibfnamefont {U.}~\bibnamefont
  {Naumann}},\ and\ \bibinfo {editor} {\bibfnamefont {B.}~\bibnamefont
  {Norris}}}\ (\bibinfo  {publisher} {Springer},\ \bibinfo {address} {New York,
  NY},\ \bibinfo {year} {2005})\ pp.\ \bibinfo {pages} {321--322}\BibitemShut
  {NoStop}%
\bibitem [{\citenamefont {Courtoy}\ \emph {et~al.}(2022)\citenamefont
  {Courtoy}, \citenamefont {Huston}, \citenamefont {Nadolsky}, \citenamefont
  {Xie}, \citenamefont {Yan},\ and\ \citenamefont {Yuan}}]{Courtoy:2022ocu}%
  \BibitemOpen
  \bibfield  {author} {\bibinfo {author} {\bibfnamefont {A.}~\bibnamefont
  {Courtoy}}, \bibinfo {author} {\bibfnamefont {J.}~\bibnamefont {Huston}},
  \bibinfo {author} {\bibfnamefont {P.}~\bibnamefont {Nadolsky}}, \bibinfo
  {author} {\bibfnamefont {K.}~\bibnamefont {Xie}}, \bibinfo {author}
  {\bibfnamefont {M.}~\bibnamefont {Yan}},\ and\ \bibinfo {author}
  {\bibfnamefont {C.~P.}\ \bibnamefont {Yuan}},\ }\href@noop {} {\bibinfo
  {title} {{Parton distributions need representative sampling}}} (\bibinfo
  {year} {2022}),\ \Eprint {https://arxiv.org/abs/2205.10444} {arXiv:2205.10444
  [hep-ph]} \BibitemShut {NoStop}%
\bibitem [{\citenamefont {Lai}\ \emph {et~al.}(2010)\citenamefont {Lai},
  \citenamefont {Guzzi}, \citenamefont {Huston}, \citenamefont {Li},
  \citenamefont {Nadolsky}, \citenamefont {Pumplin},\ and\ \citenamefont
  {Yuan}}]{Lai:2010vv}%
  \BibitemOpen
  \bibfield  {author} {\bibinfo {author} {\bibfnamefont {H.-L.}\ \bibnamefont
  {Lai}}, \bibinfo {author} {\bibfnamefont {M.}~\bibnamefont {Guzzi}}, \bibinfo
  {author} {\bibfnamefont {J.}~\bibnamefont {Huston}}, \bibinfo {author}
  {\bibfnamefont {Z.}~\bibnamefont {Li}}, \bibinfo {author} {\bibfnamefont
  {P.~M.}\ \bibnamefont {Nadolsky}}, \bibinfo {author} {\bibfnamefont
  {J.}~\bibnamefont {Pumplin}},\ and\ \bibinfo {author} {\bibfnamefont {C.~P.}\
  \bibnamefont {Yuan}},\ }\bibfield  {title} {\bibinfo {title} {{New parton
  distributions for collider physics}},\ }\href
  {https://doi.org/10.1103/PhysRevD.82.074024} {\bibfield  {journal} {\bibinfo
  {journal} {Phys. Rev. D}\ }\textbf {\bibinfo {volume} {82}},\ \bibinfo
  {pages} {074024} (\bibinfo {year} {2010})},\ \Eprint
  {https://arxiv.org/abs/1007.2241} {arXiv:1007.2241 [hep-ph]} \BibitemShut
  {NoStop}%
\bibitem [{\citenamefont {Gao}\ \emph {et~al.}(2014)\citenamefont {Gao},
  \citenamefont {Guzzi}, \citenamefont {Huston}, \citenamefont {Lai},
  \citenamefont {Li}, \citenamefont {Nadolsky}, \citenamefont {Pumplin},
  \citenamefont {Stump},\ and\ \citenamefont {Yuan}}]{Gao:2013xoa}%
  \BibitemOpen
  \bibfield  {author} {\bibinfo {author} {\bibfnamefont {J.}~\bibnamefont
  {Gao}}, \bibinfo {author} {\bibfnamefont {M.}~\bibnamefont {Guzzi}}, \bibinfo
  {author} {\bibfnamefont {J.}~\bibnamefont {Huston}}, \bibinfo {author}
  {\bibfnamefont {H.-L.}\ \bibnamefont {Lai}}, \bibinfo {author} {\bibfnamefont
  {Z.}~\bibnamefont {Li}}, \bibinfo {author} {\bibfnamefont {P.}~\bibnamefont
  {Nadolsky}}, \bibinfo {author} {\bibfnamefont {J.}~\bibnamefont {Pumplin}},
  \bibinfo {author} {\bibfnamefont {D.}~\bibnamefont {Stump}},\ and\ \bibinfo
  {author} {\bibfnamefont {C.~P.}\ \bibnamefont {Yuan}},\ }\bibfield  {title}
  {\bibinfo {title} {{CT10 next-to-next-to-leading order global analysis of
  QCD}},\ }\href {https://doi.org/10.1103/PhysRevD.89.033009} {\bibfield
  {journal} {\bibinfo  {journal} {Phys. Rev. D}\ }\textbf {\bibinfo {volume}
  {89}},\ \bibinfo {pages} {033009} (\bibinfo {year} {2014})},\ \Eprint
  {https://arxiv.org/abs/1302.6246} {arXiv:1302.6246 [hep-ph]} \BibitemShut
  {NoStop}%
\bibitem [{\citenamefont {Martin}\ \emph {et~al.}(2009)\citenamefont {Martin},
  \citenamefont {Stirling}, \citenamefont {Thorne},\ and\ \citenamefont
  {Watt}}]{Martin:2009iq}%
  \BibitemOpen
  \bibfield  {author} {\bibinfo {author} {\bibfnamefont {A.~D.}\ \bibnamefont
  {Martin}}, \bibinfo {author} {\bibfnamefont {W.~J.}\ \bibnamefont
  {Stirling}}, \bibinfo {author} {\bibfnamefont {R.~S.}\ \bibnamefont
  {Thorne}},\ and\ \bibinfo {author} {\bibfnamefont {G.}~\bibnamefont {Watt}},\
  }\bibfield  {title} {\bibinfo {title} {{Parton distributions for the LHC}},\
  }\href {https://doi.org/10.1140/epjc/s10052-009-1072-5} {\bibfield  {journal}
  {\bibinfo  {journal} {Eur. Phys. J. C}\ }\textbf {\bibinfo {volume} {63}},\
  \bibinfo {pages} {189} (\bibinfo {year} {2009})},\ \Eprint
  {https://arxiv.org/abs/0901.0002} {arXiv:0901.0002 [hep-ph]} \BibitemShut
  {NoStop}%
\bibitem [{\citenamefont {Harland-Lang}\ \emph {et~al.}(2015)\citenamefont
  {Harland-Lang}, \citenamefont {Martin}, \citenamefont {Motylinski},\ and\
  \citenamefont {Thorne}}]{Harland-Lang:2014zoa}%
  \BibitemOpen
  \bibfield  {author} {\bibinfo {author} {\bibfnamefont {L.~A.}\ \bibnamefont
  {Harland-Lang}}, \bibinfo {author} {\bibfnamefont {A.~D.}\ \bibnamefont
  {Martin}}, \bibinfo {author} {\bibfnamefont {P.}~\bibnamefont {Motylinski}},\
  and\ \bibinfo {author} {\bibfnamefont {R.~S.}\ \bibnamefont {Thorne}},\
  }\bibfield  {title} {\bibinfo {title} {{Parton distributions in the LHC era:
  MMHT 2014 PDFs}},\ }\href {https://doi.org/10.1140/epjc/s10052-015-3397-6}
  {\bibfield  {journal} {\bibinfo  {journal} {Eur. Phys. J. C}\ }\textbf
  {\bibinfo {volume} {75}},\ \bibinfo {pages} {204} (\bibinfo {year} {2015})},\
  \Eprint {https://arxiv.org/abs/1412.3989} {arXiv:1412.3989 [hep-ph]}
  \BibitemShut {NoStop}%
\bibitem [{\citenamefont {Jing}\ \emph {et~al.}(2023)\citenamefont {Jing} \emph
  {et~al.}}]{Jing:2023isu}%
  \BibitemOpen
  \bibfield  {author} {\bibinfo {author} {\bibfnamefont {X.}~\bibnamefont
  {Jing}} \emph {et~al.},\ }\bibfield  {title} {\bibinfo {title} {{Quantifying
  the interplay of experimental constraints in analyses of parton
  distributions}},\ }\href {https://doi.org/10.1103/PhysRevD.108.034029}
  {\bibfield  {journal} {\bibinfo  {journal} {Phys. Rev. D}\ }\textbf {\bibinfo
  {volume} {108}},\ \bibinfo {pages} {034029} (\bibinfo {year} {2023})},\
  \Eprint {https://arxiv.org/abs/2306.03918} {arXiv:2306.03918 [hep-ph]}
  \BibitemShut {NoStop}%
\bibitem [{\citenamefont {Akaike}(1974)}]{Akaike}%
  \BibitemOpen
  \bibfield  {author} {\bibinfo {author} {\bibfnamefont {H.}~\bibnamefont
  {Akaike}},\ }\bibfield  {title} {\bibinfo {title} {{A new look at the
  statistical model identification}},\ }\href
  {https://doi.org/10.1109/TAC.1974.1100705} {\bibfield  {journal} {\bibinfo
  {journal} {IEEE Transactions on Automatic Control}\ }\textbf {\bibinfo
  {volume} {19}},\ \bibinfo {pages} {716} (\bibinfo {year} {1974})}\BibitemShut
  {NoStop}%
\bibitem [{\citenamefont {Schwarz}(1978)}]{BIC}%
  \BibitemOpen
  \bibfield  {author} {\bibinfo {author} {\bibfnamefont {G.}~\bibnamefont
  {Schwarz}},\ }\bibfield  {title} {\bibinfo {title} {{Estimating the Dimension
  of a Model}},\ }\href {https://doi.org/10.1214/aos/1176344136} {\bibfield
  {journal} {\bibinfo  {journal} {The Annals of Statistics}\ }\textbf {\bibinfo
  {volume} {6}},\ \bibinfo {pages} {461 } (\bibinfo {year} {1978})}\BibitemShut
  {NoStop}%
\bibitem [{\citenamefont {Cowan}(2019)}]{Cowan:2018lhq}%
  \BibitemOpen
  \bibfield  {author} {\bibinfo {author} {\bibfnamefont {G.}~\bibnamefont
  {Cowan}},\ }\bibfield  {title} {\bibinfo {title} {{Statistical Models with
  Uncertain Error Parameters}},\ }\href
  {https://doi.org/10.1140/epjc/s10052-019-6644-4} {\bibfield  {journal}
  {\bibinfo  {journal} {Eur. Phys. J. C}\ }\textbf {\bibinfo {volume} {79}},\
  \bibinfo {pages} {133} (\bibinfo {year} {2019})},\ \Eprint
  {https://arxiv.org/abs/1809.05778} {arXiv:1809.05778 [physics.data-an]}
  \BibitemShut {NoStop}%
\bibitem [{\citenamefont {D'Agostini}(1999)}]{DAgostini:1999niu}%
  \BibitemOpen
  \bibfield  {author} {\bibinfo {author} {\bibfnamefont {G.}~\bibnamefont
  {D'Agostini}},\ }\href@noop {} {\bibinfo {title} {{Sceptical combination of
  experimental results: General considerations and application to epsilon-prime
  / epsilon}}} (\bibinfo {year} {1999}),\ \Eprint
  {https://arxiv.org/abs/hep-ex/9910036} {arXiv:hep-ex/9910036} \BibitemShut
  {NoStop}%
\bibitem [{\citenamefont {Paukkunen}\ and\ \citenamefont
  {Zurita}(2014)}]{Paukkunen:2014zia}%
  \BibitemOpen
  \bibfield  {author} {\bibinfo {author} {\bibfnamefont {H.}~\bibnamefont
  {Paukkunen}}\ and\ \bibinfo {author} {\bibfnamefont {P.}~\bibnamefont
  {Zurita}},\ }\bibfield  {title} {\bibinfo {title} {{PDF reweighting in the
  Hessian matrix approach}},\ }\href {https://doi.org/10.1007/JHEP12(2014)100}
  {\bibfield  {journal} {\bibinfo  {journal} {JHEP}\ }\textbf {\bibinfo
  {volume} {12}}\bibfield  {number} {\bibinfo  {number} { (100)}},\ }\Eprint
  {https://arxiv.org/abs/1402.6623} {arXiv:1402.6623 [hep-ph]} \BibitemShut
  {NoStop}%
\bibitem [{\citenamefont {Wolpert}(1992)}]{wolpert1992stacked}%
  \BibitemOpen
  \bibfield  {author} {\bibinfo {author} {\bibfnamefont {D.~H.}\ \bibnamefont
  {Wolpert}},\ }\bibfield  {title} {\bibinfo {title} {Stacked generalization},\
  }\href {https://doi.org/10.1016/S0893-6080(05)80023-1} {\bibfield  {journal}
  {\bibinfo  {journal} {Neural networks}\ }\textbf {\bibinfo {volume} {5}},\
  \bibinfo {pages} {241} (\bibinfo {year} {1992})}\BibitemShut {NoStop}%
\bibitem [{\citenamefont {Yao}\ \emph {et~al.}(2018)\citenamefont {Yao},
  \citenamefont {Vehtari}, \citenamefont {Simpson},\ and\ \citenamefont
  {Gelman}}]{yao2018using}%
  \BibitemOpen
  \bibfield  {author} {\bibinfo {author} {\bibfnamefont {Y.}~\bibnamefont
  {Yao}}, \bibinfo {author} {\bibfnamefont {A.}~\bibnamefont {Vehtari}},
  \bibinfo {author} {\bibfnamefont {D.}~\bibnamefont {Simpson}},\ and\ \bibinfo
  {author} {\bibfnamefont {A.}~\bibnamefont {Gelman}},\ }\bibfield  {title}
  {\bibinfo {title} {{Using Stacking to Average Bayesian Predictive
  Distributions (with Discussion)}},\ }\href
  {https://doi.org/10.1214/17-BA1091} {\bibfield  {journal} {\bibinfo
  {journal} {Bayesian Analysis}\ }\textbf {\bibinfo {volume} {13}},\ \bibinfo
  {pages} {917 } (\bibinfo {year} {2018})}\BibitemShut {NoStop}%
\bibitem [{\citenamefont {Wasserman}(2000)}]{wasserman2000bayesian}%
  \BibitemOpen
  \bibfield  {author} {\bibinfo {author} {\bibfnamefont {L.}~\bibnamefont
  {Wasserman}},\ }\bibfield  {title} {\bibinfo {title} {Bayesian model
  selection and model averaging},\ }\href
  {https://doi.org/10.1006/jmps.1999.1278} {\bibfield  {journal} {\bibinfo
  {journal} {Journal of mathematical psychology}\ }\textbf {\bibinfo {volume}
  {44}},\ \bibinfo {pages} {92} (\bibinfo {year} {2000})}\BibitemShut {NoStop}%
\bibitem [{\citenamefont {Minka}(2000)}]{minka2000bayesian}%
  \BibitemOpen
  \bibfield  {author} {\bibinfo {author} {\bibfnamefont {T.~P.}\ \bibnamefont
  {Minka}},\ }\href {http://www.stat.cmu.edu/minka/papers/bma.html} {\emph
  {\bibinfo {title} {Bayesian Model Averaging is not Model Combination}}},\
  \bibinfo {type} {Tech. Rep.}\ (\bibinfo  {institution} {Microsoft Research},\
  \bibinfo {year} {2000})\ \bibinfo {note} {accessed: 2024-07-15}\BibitemShut
  {NoStop}%
\bibitem [{\citenamefont {Monteith}\ \emph {et~al.}(2011)\citenamefont
  {Monteith}, \citenamefont {Carroll}, \citenamefont {Seppi},\ and\
  \citenamefont {Martinez}}]{Monteith:2011}%
  \BibitemOpen
  \bibfield  {author} {\bibinfo {author} {\bibfnamefont {K.}~\bibnamefont
  {Monteith}}, \bibinfo {author} {\bibfnamefont {J.~L.}\ \bibnamefont
  {Carroll}}, \bibinfo {author} {\bibfnamefont {K.}~\bibnamefont {Seppi}},\
  and\ \bibinfo {author} {\bibfnamefont {T.}~\bibnamefont {Martinez}},\
  }\bibfield  {title} {\bibinfo {title} {Turning bayesian model averaging into
  bayesian model combination},\ }in\ \href
  {https://doi.org/10.1109/IJCNN.2011.6033566} {\emph {\bibinfo {booktitle}
  {The 2011 International Joint Conference on Neural Networks}}}\ (\bibinfo
  {year} {2011})\ pp.\ \bibinfo {pages} {2657--2663}\BibitemShut {NoStop}%
\bibitem [{\citenamefont {Kamary}\ \emph {et~al.}(2018)\citenamefont {Kamary},
  \citenamefont {Mengersen}, \citenamefont {Robert},\ and\ \citenamefont
  {Rousseau}}]{kamary2014testing}%
  \BibitemOpen
  \bibfield  {author} {\bibinfo {author} {\bibfnamefont {K.}~\bibnamefont
  {Kamary}}, \bibinfo {author} {\bibfnamefont {K.}~\bibnamefont {Mengersen}},
  \bibinfo {author} {\bibfnamefont {C.~P.}\ \bibnamefont {Robert}},\ and\
  \bibinfo {author} {\bibfnamefont {J.}~\bibnamefont {Rousseau}},\ }\href
  {https://arxiv.org/abs/1412.2044} {\bibinfo {title} {Testing hypotheses via a
  mixture estimation model}} (\bibinfo {year} {2018}),\ \Eprint
  {https://arxiv.org/abs/1412.2044} {arXiv:1412.2044 [stat.ME]} \BibitemShut
  {NoStop}%
\bibitem [{\citenamefont {Keller}\ and\ \citenamefont
  {Kamary}(2018)}]{keller2017bayesian}%
  \BibitemOpen
  \bibfield  {author} {\bibinfo {author} {\bibfnamefont {M.}~\bibnamefont
  {Keller}}\ and\ \bibinfo {author} {\bibfnamefont {K.}~\bibnamefont
  {Kamary}},\ }\href {https://arxiv.org/abs/1711.10016} {\bibinfo {title}
  {Bayesian model averaging via mixture model estimation}} (\bibinfo {year}
  {2018}),\ \Eprint {https://arxiv.org/abs/1711.10016} {arXiv:1711.10016
  [stat.ME]} \BibitemShut {NoStop}%
\bibitem [{\citenamefont {Abazov}\ \emph {et~al.}(2002)\citenamefont {Abazov}
  \emph {et~al.}}]{D0:2002fhu}%
  \BibitemOpen
  \bibfield  {author} {\bibinfo {author} {\bibfnamefont {V.~M.}\ \bibnamefont
  {Abazov}} \emph {et~al.} (\bibinfo {collaboration} {D0}),\ }\bibfield
  {title} {\bibinfo {title} {{Improved $W$ Boson Mass Measurement with the D0
  Detector}},\ }\href {https://doi.org/10.1103/PhysRevD.66.012001} {\bibfield
  {journal} {\bibinfo  {journal} {Phys. Rev. D}\ }\textbf {\bibinfo {volume}
  {66}},\ \bibinfo {pages} {012001} (\bibinfo {year} {2002})},\ \Eprint
  {https://arxiv.org/abs/hep-ex/0204014} {arXiv:hep-ex/0204014} \BibitemShut
  {NoStop}%
\bibitem [{\citenamefont {Affolder}\ \emph {et~al.}(2001)\citenamefont
  {Affolder} \emph {et~al.}}]{CDF:2000gwd}%
  \BibitemOpen
  \bibfield  {author} {\bibinfo {author} {\bibfnamefont {T.}~\bibnamefont
  {Affolder}} \emph {et~al.} (\bibinfo {collaboration} {CDF}),\ }\bibfield
  {title} {\bibinfo {title} {{Measurement of the $W$ boson mass with the
  Collider Detector at Fermilab}},\ }\href
  {https://doi.org/10.1103/PhysRevD.64.052001} {\bibfield  {journal} {\bibinfo
  {journal} {Phys. Rev. D}\ }\textbf {\bibinfo {volume} {64}},\ \bibinfo
  {pages} {052001} (\bibinfo {year} {2001})},\ \Eprint
  {https://arxiv.org/abs/hep-ex/0007044} {arXiv:hep-ex/0007044} \BibitemShut
  {NoStop}%
\bibitem [{\citenamefont {Schael}\ \emph {et~al.}(2013)\citenamefont {Schael}
  \emph {et~al.}}]{ALEPH:2013dgf}%
  \BibitemOpen
  \bibfield  {author} {\bibinfo {author} {\bibfnamefont {S.}~\bibnamefont
  {Schael}} \emph {et~al.} (\bibinfo {collaboration} {ALEPH, DELPHI, L3, OPAL,
  LEP Electroweak}),\ }\bibfield  {title} {\bibinfo {title} {{Electroweak
  Measurements in Electron-Positron Collisions at W-Boson-Pair Energies at
  LEP}},\ }\href {https://doi.org/10.1016/j.physrep.2013.07.004} {\bibfield
  {journal} {\bibinfo  {journal} {Phys. Rept.}\ }\textbf {\bibinfo {volume}
  {532}},\ \bibinfo {pages} {119} (\bibinfo {year} {2013})},\ \Eprint
  {https://arxiv.org/abs/1302.3415} {arXiv:1302.3415 [hep-ex]} \BibitemShut
  {NoStop}%
\bibitem [{\citenamefont {Abazov}\ \emph {et~al.}(2012)\citenamefont {Abazov}
  \emph {et~al.}}]{D0:2012kms}%
  \BibitemOpen
  \bibfield  {author} {\bibinfo {author} {\bibfnamefont {V.~M.}\ \bibnamefont
  {Abazov}} \emph {et~al.} (\bibinfo {collaboration} {D0}),\ }\bibfield
  {title} {\bibinfo {title} {{Measurement of the W Boson Mass with the D0
  Detector}},\ }\href {https://doi.org/10.1103/PhysRevLett.108.151804}
  {\bibfield  {journal} {\bibinfo  {journal} {Phys. Rev. Lett.}\ }\textbf
  {\bibinfo {volume} {108}},\ \bibinfo {pages} {151804} (\bibinfo {year}
  {2012})},\ \Eprint {https://arxiv.org/abs/1203.0293} {arXiv:1203.0293
  [hep-ex]} \BibitemShut {NoStop}%
\bibitem [{\citenamefont {Aaij}\ \emph {et~al.}(2022)\citenamefont {Aaij} \emph
  {et~al.}}]{LHCb:2021bjt}%
  \BibitemOpen
  \bibfield  {author} {\bibinfo {author} {\bibfnamefont {R.}~\bibnamefont
  {Aaij}} \emph {et~al.} (\bibinfo {collaboration} {LHCb}),\ }\bibfield
  {title} {\bibinfo {title} {{Measurement of the W boson mass}},\ }\href
  {https://doi.org/10.1007/JHEP01(2022)036} {\bibfield  {journal} {\bibinfo
  {journal} {JHEP}\ }\textbf {\bibinfo {volume} {01}}\bibfield  {number}
  {\bibinfo  {number} { (36)},\ \bibinfo {pages} {036}},\ }\Eprint
  {https://arxiv.org/abs/2109.01113} {arXiv:2109.01113 [hep-ex]} \BibitemShut
  {NoStop}%
\bibitem [{\citenamefont {Aad}\ \emph {et~al.}(2024)\citenamefont {Aad} \emph
  {et~al.}}]{ATLAS:2024erm}%
  \BibitemOpen
  \bibfield  {author} {\bibinfo {author} {\bibfnamefont {G.}~\bibnamefont
  {Aad}} \emph {et~al.} (\bibinfo {collaboration} {ATLAS}),\ }\bibfield
  {title} {\bibinfo {title} {{Measurement of the W-boson mass and width with
  the ATLAS detector using proton\textendash{}proton collisions at $\sqrt{s}=7$
  TeV}},\ }\href {https://doi.org/10.1140/epjc/s10052-024-13190-x} {\bibfield
  {journal} {\bibinfo  {journal} {Eur. Phys. J. C}\ }\textbf {\bibinfo {volume}
  {84}},\ \bibinfo {pages} {1309} (\bibinfo {year} {2024})},\ \Eprint
  {https://arxiv.org/abs/2403.15085} {arXiv:2403.15085 [hep-ex]} \BibitemShut
  {NoStop}%
\bibitem [{\citenamefont {Kotwal}(2025)}]{Kotwal:2025xsy}%
  \BibitemOpen
  \bibfield  {author} {\bibinfo {author} {\bibfnamefont {A.~V.}\ \bibnamefont
  {Kotwal}},\ }\bibfield  {title} {\bibinfo {title} {{Model for the curvature
  response of the CDF II drift chamber}},\ }\href
  {https://doi.org/10.1103/PhysRevResearch.7.013128} {\bibfield  {journal}
  {\bibinfo  {journal} {Phys. Rev. Res.}\ }\textbf {\bibinfo {volume} {7}},\
  \bibinfo {pages} {013128} (\bibinfo {year} {2025})}\BibitemShut {NoStop}%
\bibitem [{\citenamefont {Navas}\ \emph {et~al.}(2024)\citenamefont {Navas}
  \emph {et~al.}}]{ParticleDataGroup:2024cfk}%
  \BibitemOpen
  \bibfield  {author} {\bibinfo {author} {\bibfnamefont {S.}~\bibnamefont
  {Navas}} \emph {et~al.} (\bibinfo {collaboration} {Particle Data Group}),\
  }\bibfield  {title} {\bibinfo {title} {{Review of particle physics}},\ }\href
  {https://doi.org/10.1103/PhysRevD.110.030001} {\bibfield  {journal} {\bibinfo
   {journal} {Phys. Rev. D}\ }\textbf {\bibinfo {volume} {110}},\ \bibinfo
  {pages} {030001} (\bibinfo {year} {2024})}\BibitemShut {NoStop}%
\bibitem [{\citenamefont {Amoroso}\ \emph {et~al.}(2024)\citenamefont {Amoroso}
  \emph {et~al.}}]{LHC-TeVMWWorkingGroup:2023zkn}%
  \BibitemOpen
  \bibfield  {author} {\bibinfo {author} {\bibfnamefont {S.}~\bibnamefont
  {Amoroso}} \emph {et~al.} (\bibinfo {collaboration}
  {LHC-TeV~MW~Working~Group}),\ }\bibfield  {title} {\bibinfo {title}
  {{Compatibility and combination of world W-boson mass measurements}},\ }\href
  {https://doi.org/10.1140/epjc/s10052-024-12532-z} {\bibfield  {journal}
  {\bibinfo  {journal} {Eur. Phys. J. C}\ }\textbf {\bibinfo {volume} {84}},\
  \bibinfo {pages} {451} (\bibinfo {year} {2024})},\ \Eprint
  {https://arxiv.org/abs/2308.09417} {arXiv:2308.09417 [hep-ex]} \BibitemShut
  {NoStop}%
\bibitem [{\citenamefont {de~Blas}\ \emph {et~al.}(2022)\citenamefont
  {de~Blas}, \citenamefont {Ciuchini}, \citenamefont {Franco}, \citenamefont
  {Goncalves}, \citenamefont {Mishima}, \citenamefont {Pierini}, \citenamefont
  {Reina},\ and\ \citenamefont {Silvestrini}}]{deBlas:2021wap}%
  \BibitemOpen
  \bibfield  {author} {\bibinfo {author} {\bibfnamefont {J.}~\bibnamefont
  {de~Blas}}, \bibinfo {author} {\bibfnamefont {M.}~\bibnamefont {Ciuchini}},
  \bibinfo {author} {\bibfnamefont {E.}~\bibnamefont {Franco}}, \bibinfo
  {author} {\bibfnamefont {A.}~\bibnamefont {Goncalves}}, \bibinfo {author}
  {\bibfnamefont {S.}~\bibnamefont {Mishima}}, \bibinfo {author} {\bibfnamefont
  {M.}~\bibnamefont {Pierini}}, \bibinfo {author} {\bibfnamefont
  {L.}~\bibnamefont {Reina}},\ and\ \bibinfo {author} {\bibfnamefont
  {L.}~\bibnamefont {Silvestrini}},\ }\bibfield  {title} {\bibinfo {title}
  {{Global analysis of electroweak data in the Standard Model}},\ }\href
  {https://doi.org/10.1103/PhysRevD.106.033003} {\bibfield  {journal} {\bibinfo
   {journal} {Phys. Rev. D}\ }\textbf {\bibinfo {volume} {106}},\ \bibinfo
  {pages} {033003} (\bibinfo {year} {2022})},\ \Eprint
  {https://arxiv.org/abs/2112.07274} {arXiv:2112.07274 [hep-ph]} \BibitemShut
  {NoStop}%
\bibitem [{\citenamefont {Haller}\ \emph {et~al.}(2022)\citenamefont {Haller},
  \citenamefont {Hoecker}, \citenamefont {Kogler}, \citenamefont {M\"onig},\
  and\ \citenamefont {Stelzer}}]{Haller:2022eyb}%
  \BibitemOpen
  \bibfield  {author} {\bibinfo {author} {\bibfnamefont {J.}~\bibnamefont
  {Haller}}, \bibinfo {author} {\bibfnamefont {A.}~\bibnamefont {Hoecker}},
  \bibinfo {author} {\bibfnamefont {R.}~\bibnamefont {Kogler}}, \bibinfo
  {author} {\bibfnamefont {K.}~\bibnamefont {M\"onig}},\ and\ \bibinfo {author}
  {\bibfnamefont {J.}~\bibnamefont {Stelzer}},\ }\bibfield  {title} {\bibinfo
  {title} {{Status of the global electroweak fit with Gfitter in the light of
  new precision measurements}},\ }\href {https://doi.org/10.22323/1.414.0897}
  {\bibfield  {journal} {\bibinfo  {journal} {PoS}\ }\textbf {\bibinfo {volume}
  {ICHEP2022}},\ \bibinfo {pages} {897} (\bibinfo {year} {2022})},\ \Eprint
  {https://arxiv.org/abs/2211.07665} {arXiv:2211.07665 [hep-ph]} \BibitemShut
  {NoStop}%
\bibitem [{\citenamefont {Lin}\ \emph {et~al.}(2018)\citenamefont {Lin} \emph
  {et~al.}}]{Lin:2017snn}%
  \BibitemOpen
  \bibfield  {author} {\bibinfo {author} {\bibfnamefont {H.-W.}\ \bibnamefont
  {Lin}} \emph {et~al.},\ }\bibfield  {title} {\bibinfo {title} {{Parton
  distributions and lattice QCD calculations: a community white paper}},\
  }\href {https://doi.org/10.1016/j.ppnp.2018.01.007} {\bibfield  {journal}
  {\bibinfo  {journal} {Prog. Part. Nucl. Phys.}\ }\textbf {\bibinfo {volume}
  {100}},\ \bibinfo {pages} {107} (\bibinfo {year} {2018})},\ \Eprint
  {https://arxiv.org/abs/1711.07916} {arXiv:1711.07916 [hep-ph]} \BibitemShut
  {NoStop}%
\bibitem [{\citenamefont {Baak}\ \emph {et~al.}(2014)\citenamefont {Baak},
  \citenamefont {C\'uth}, \citenamefont {Haller}, \citenamefont {Hoecker},
  \citenamefont {Kogler}, \citenamefont {M\"onig}, \citenamefont {Schott},\
  and\ \citenamefont {Stelzer}}]{Baak:2014ora}%
  \BibitemOpen
  \bibfield  {author} {\bibinfo {author} {\bibfnamefont {M.}~\bibnamefont
  {Baak}}, \bibinfo {author} {\bibfnamefont {J.}~\bibnamefont {C\'uth}},
  \bibinfo {author} {\bibfnamefont {J.}~\bibnamefont {Haller}}, \bibinfo
  {author} {\bibfnamefont {A.}~\bibnamefont {Hoecker}}, \bibinfo {author}
  {\bibfnamefont {R.}~\bibnamefont {Kogler}}, \bibinfo {author} {\bibfnamefont
  {K.}~\bibnamefont {M\"onig}}, \bibinfo {author} {\bibfnamefont
  {M.}~\bibnamefont {Schott}},\ and\ \bibinfo {author} {\bibfnamefont
  {J.}~\bibnamefont {Stelzer}} (\bibinfo {collaboration} {Gfitter Group}),\
  }\bibfield  {title} {\bibinfo {title} {{The global electroweak fit at NNLO
  and prospects for the LHC and ILC}},\ }\href
  {https://doi.org/10.1140/epjc/s10052-014-3046-5} {\bibfield  {journal}
  {\bibinfo  {journal} {Eur. Phys. J. C}\ }\textbf {\bibinfo {volume} {74}},\
  \bibinfo {pages} {3046} (\bibinfo {year} {2014})},\ \Eprint
  {https://arxiv.org/abs/1407.3792} {arXiv:1407.3792 [hep-ph]} \BibitemShut
  {NoStop}%
\end{thebibliography}%
